\documentclass[pre,showpacs,notitlepage]{revtex4-1}
\usepackage{graphicx}
\usepackage{dcolumn}
\usepackage{bm}
\usepackage{natbib}
\usepackage{color}

\begin{document}

\title{Interplay of anisotropy in shape and interactions in charged platelet suspensions  }


\author{Sara Jabbari-Farouji}
\affiliation{
LPTMS, CNRS and Universit$\acute{e}$ Paris-Sud, UMR8626, Bat. 100, 91405 Orsay, France
}%
\affiliation{
Laboratoire interdisciplinaire de Physique, UMR 5588, F-38041 Grenoble, France
}%

\author{Jean-Jacques Weis}%
\affiliation{
Universit\'{e} Paris-Sud, Laboratoire de Physique Th\'{e}orique, UMR 8627\\
B\^{a}timent 210, 91405 Orsay Cedex, France
}%

\author{Patrick Davidson}%
\affiliation{
Universit\'{e} Paris-Sud, Laboratoire de Physique des Solides, UMR 8502\\
B\^{a}timent 510, 91405 Orsay Cedex, France
}%

\author{Pierre Levitz}
\affiliation{
Laboratoire PECSA, UMR 7195, Universit\'{e} Pierre et Marie Curie \\
 Case Courrier 51, 4 place Jussieu 72522 Paris Cedex 5, France}

\author{Emmanuel Trizac}
\affiliation{LPTMS, CNRS and Universit$\acute{e}$ Paris-Sud, UMR8626, Bat. 100, 91405 Orsay, France}%


\date{\today}

\begin{abstract}
Motivated by  the intriguing phase behavior of charged colloidal platelets, we investigate the  structure and dynamics of charged repulsive disks  by means of Monte-Carlo simulations.  The electrostatic interactions are taken into account through an effective two-body  potential, obtained within the non-linear Poisson-Boltzmann formalism, which has the form of  anisotropic screened Coulomb potential. Recently, we showed that the original intrinsic anisotropy of the electrostatic potential in competition with excluded volume effects leads to a rich phase behavior that not only includes various liquid-crsytalline phases but also predicts 
the existence of novel structures  composed of alternating  nematic-antinematic sheets.  Here,  we examine the structural and dynamical  signatures of each of the observed structures for both translational and rotational degrees of freedom. Finally, we discuss the influence of effective charge value and  our results in relation to experimental findings on charged platelet suspensions.
\end{abstract}

\maketitle

\tableofcontents
\addcontentsline{toc}{section}{}

\newpage

\section{\label{sec:intro} Introduction}

Charged platelet suspensions, such as swelling clays \cite{Beidellite,nontronite, Laponite,Sara-PRE2008,bentonite,clay1,clay2,clay3}, disk-like mineral crystallites \cite{Gibphase, zirconium,gib2008,Gibphase,Gibbsite,Lekker,crys1,crys2,crys3}
or exfoliated nanosheets \cite{nanosheets,nanosheets1,nano1,nano2,nano3,nano4} abound in natural as well as industrial environments. Such minerals can be found in various particle sizes  varying between 10 nm up to a few micrometers and 
the majority of them have a platelet-like shape with high aspect ratio, typically ranging between 20 and 1000. Apart from their natural relevance in geology as agents conferring
solidity to soils, they are exploited in various  industrial applications such as  thickeners, fillers and gels in  agriculture or pharmaceutics.  Despite  widespread applications of charged platelets suspensions, their 
phase behavior  is still elusive.  They exhibit  nonintuitive phase behaviors: while some charged platelet  suspensions \cite{Laponite,bentonite} form arrested states at 
low densities, others \cite{Beidellite,Gibphase}  exhibit an equilibrium isotropic-nematic transition at moderate densities.   The highly charged nature of the platelets 
suggests that the electrostatic interactions 
should  definitely play a major part, although in some cases other   specific interactions seem important for fathoming  the complex phase behavior of such systems.   
Therefore, the  first important step for
addressing the phase behavior of charged platelet suspensions is to elucidate the  role of electrostatic interactions.  
 
 Understanding the influence of long-ranged electrostatic interactions on phase behavior of charged anisotropic colloids is  a non-trivial many-body and challenging problem by itself.  
 The complexity  stems from the multi-component nature of these systems which comprise asymmetric constituents  of macro ions (charged platelets) and ionic species.  
 The high surface charge density of platelets (0.1--0.5 C/m$^{-2}$)  implies that a  large number of counter-ions, not speaking of added salt ions,  need to be accounted for in any primitive model
 treatment. Computationally such a problem becomes cumbersome if not prohibitive with our current computational facilities. Therefore, resorting to an approach where the effects of ions
 are considered through an effective interaction between charged platelets allows us to simplify the problem and proceed further.

Recently, we developed a coarse-grained model where the ionic degrees of freedom are integrated out and  disks interact with an effective pair potential and we charted out the full phase 
diagram of repulsive charged disks  \cite{Scirep}.   Our  effective pair potential, obtained within non-linear Poisson-Boltzmann formalism and therefore repulsive, has the form  of a screened Coulomb potential 
(Yukawa) multiplied by an angular function of the orientations of the two disks, which embodies the \emph{anisotropy} of the interactions  \cite{Trizac,Carlos}. This anisotropy 
function depends on the ionic strength. Our carefully derived   pair potential  has the advantage that it overcomes  the shortcomings of prior models where the renormalized effective 
charge and full orientational dependence of angular potential were not taken into account \cite{Rowan00,Dijkstra,Wensink}.  

We explored the structure and dynamics of charged disks interacting with anisotropic screened Coulomb potential in conjunction with hard core interactions by means of Monte-Carlo (MC) simulations, for a wide range of
 densities and ionic strengths.   Interplay of anisotropy in shape and interactions of charged repulsive platelets leads to  a rich phase diagram that encompasses various crystalline (bcc and hexagonal and plastic crystals) and  
 liquid crystalline phases (nematic and columnar hexagonal), that rationalizes generic features of the complex phase diagram of  charged colloidal platelets such as   Beidellite clay and Gibbsite.  Additionally, we found that the original intrinsic anisotropy of the electrostatic potential  between charged platelets leads to formation of  a novel structure  which consists of  alternating nematic and anti-nematic layers.  We coined this new phase \emph{intergrowth texture}.  Investigating translational and rotational  dynamics of charged disks as a function of density, we found evidences of a strong slowing down  of the dynamics  upon approaching the orientational disorder-order line.  However, we did not present the structural and dynamical signatures of each structure. Here, we provide a complete description  of each phase by characterizing  the positional and orientational pair correlation functions  (statics) and  self-intermediate scattering  and time orientational correlation functions (dynamics) for disks. Additionally, we inspect the dependence of nematic order parameter and hexagonal bond orientational order parameter as function of density at different ionic strengths.

 The article is organized as follows. In section \ref{sec:effInt}, we review briefly the arguments leading to the effective interactions. We present the angular dependence of the anisotropy function 
 for charged disks at several values of ionic strength and  we discuss the corresponding variation of effective charges.
 In section \ref{sec:method}, our simulation method  and the details of equilibration procedure are  presented.
In section \ref{sec:results}, we  represent our  phase diagram and then discuss in detail the static and dynamical features of each of the structures. We also analyze 
the influence of effective charge on structure.  Concluding remarks are drawn in  section \ref{sec:conclusion}.

\section{Effective interactions for charged disks: Anisotropic Yukawa potential }
\label{sec:effInt}

\begin{figure}[h]
\includegraphics[scale=0.4]{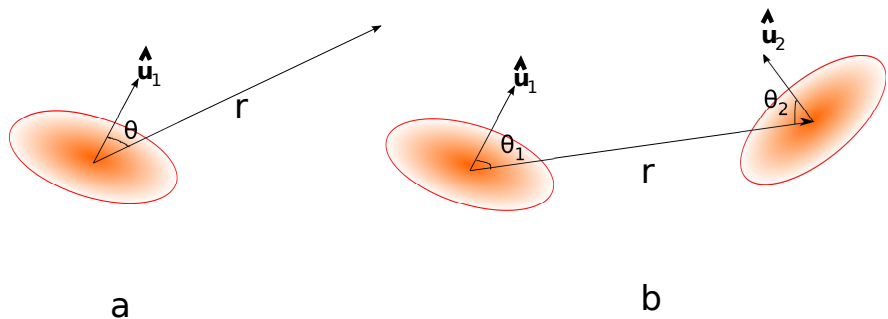}
\caption{Schematic drawing showing  a) one disk of radius $R$ at a distance $r$ from an observation point and its relative orientation b) two disks at a separation $r$ with their relative orientations with respect to their connecting line.}
\label{fig1}
\end{figure}
In order to obtain a tractable   effective interaction potential between two charged platelets, we need to integrate out the ionic degrees of freedom. Such an effective potential can be derived within Poisson-Boltzmann formalism 
 where we consider thin  uniformly charged disks  of   density $Q_s$ and  a continuous  charge density profile for  the  ions. 
The far-field behavior of the electrostatic  potential for  highly charged disks (valid in relatively dilute colloidal suspensions)   
is exactly the same as the one obtained within the linearized PB theory provided that the boundary condition of  constant surface potential is imposed on the surface of  the colloid \cite{TrBA02,Trizac,Carlos}. 
The value of the surface potential  on the disk depends on its charge density $Q_s$ and ionic strength, that sets $\kappa=(4 \pi \lambda_B \sum _i n_i z_i^2)^{1/2}$, where 
$\lambda_B= e^2/(\varepsilon  k_B T)$ is the Bjerrum length \cite{Trizac}.
Coarse-graining thereby leads to an effective charge $Z_{eff}$ for the disks that is considerably lower   
than their real charge, and its value depends on the ionic strength.

\begin{figure}[h]
\includegraphics[scale=0.25]{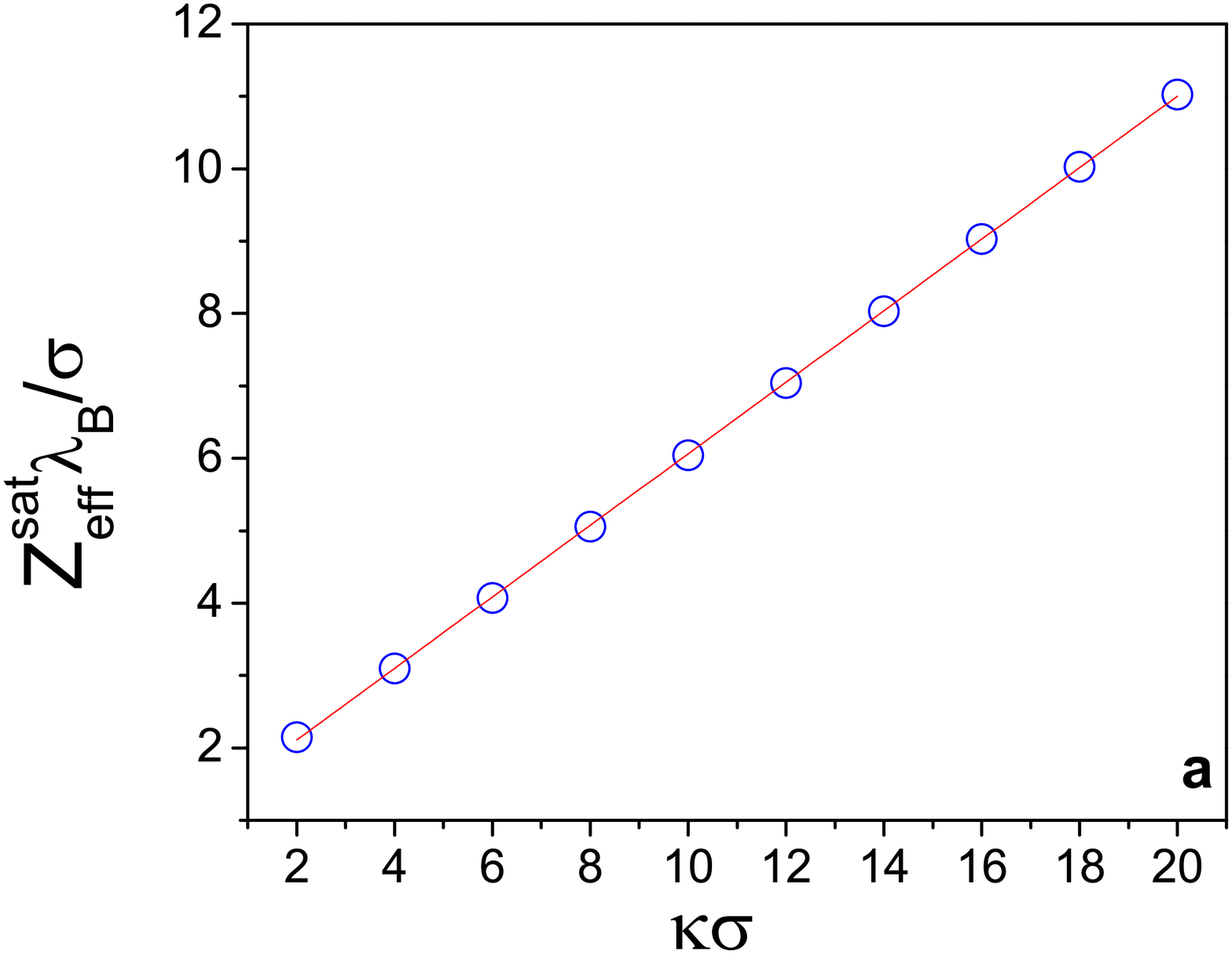}
\includegraphics[scale=0.25]{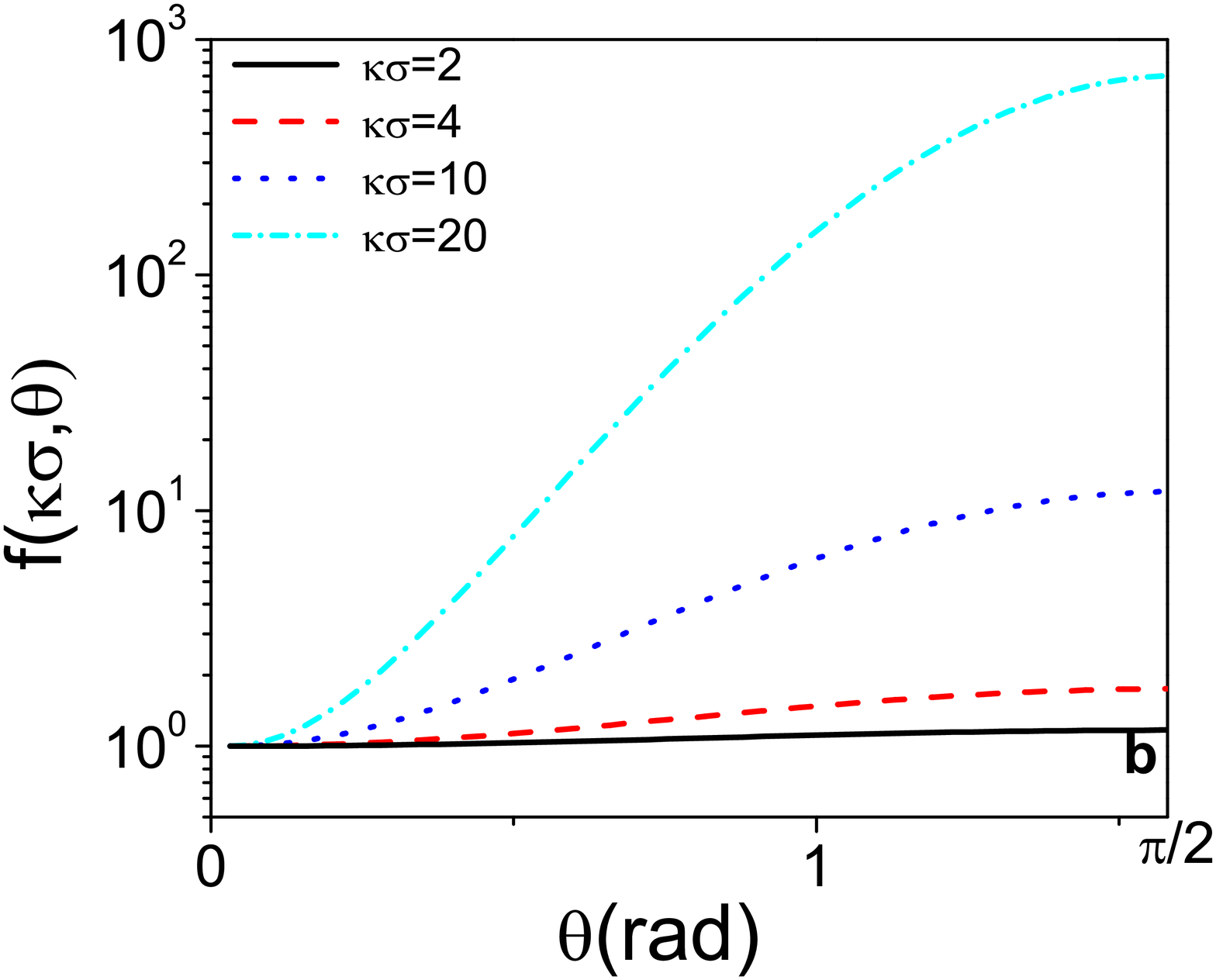}
\caption{ a) The effective charge at saturation as a function of ionic strength grows  linearly $Z_{eff}^{sat} \lambda_B/\sigma= 0.494 \kappa \sigma+1.12$.
b) Anisotropy function versus angle for different values of $\kappa \sigma$}
\label{figZ}
\end{figure}

   The spatial form of the resulting far-field electrostatic potential is given by the product of a screened Coulomb
    potential and an anisotropy function that depends on  the azimuthal angle under which the disk is seen. More precisely, this potential at
   a  distance $r$ from a disk center with diameter $\sigma$ and at an angle $\theta$  with its orientation is given by (see Fig. \ref{fig1}a)
     \cite{Trizac,Carlos}:
\begin{equation}\label{eq:pot1}
\frac{ e \phi (r,\theta)}{k_B T}=  Z_{eff}  \lambda_B  f(\kappa \sigma, \theta) \frac{\exp(- \kappa r)}{r} . 
\end{equation}
The anisotropy function $f(\kappa \sigma, \theta)$
takes into account the orientational dependence of the potential  and can be expressed in terms of spheroidal wave
functions \cite{Carlos}.  For highly charged plates, the value of  the effective charge $Z_{eff}$  somewhat simplifies, and only depends on
the ionic strength (together with disc radius). This is the saturation phenomenon \cite{TrBA02,BoTA02}.
The charge is shown  in Fig. \ref{figZ}, where it can be seen that the
  effective charge increases linearly with ionic strength $\kappa  \sigma$, while the anisotropy function shows an angular
  dependence which increases with  $\kappa  \sigma$. 
  Eq. \ref{eq:pot1} shows that the effective potential of a charged disk in the presence of ions remains anisotropic at all lengthscales unlike the Coulomb potential of a charged
 disk in vacuum. It should be emphasized that the above results hold provided that $\kappa\sigma$ is not too small \cite{TrBA02,Trizac}. 
 For $\kappa \sigma < 2$, the anisotropy factor  approaches unity and the value of effective charge at saturation  is  determined from the solution of full non-linear
 Poisson-Boltzmann equation. The corresponding charge value at $\kappa \sigma=1$ that we used in our simulations is $ Z_{eff}\lambda_B/\sigma=2.5 $ \cite{Trizac}.

Having obtained the effective potential for a charged disk in an electrolyte, we can
  construct the two-body interaction potential for two charged disks, in an electrolyte medium, whose centers are a distance $r$ apart and whose normals
    make an angle $\theta_i$ with the line connecting their centers (see Fig. \ref{fig1}b),  as outlined in reference \cite{Trizac}:
\begin{equation} \label{eq:pot2}
\frac{ U_{12}(r/\sigma,\theta_1, \theta_2)}{k_B T}=  \frac{\sigma}{\lambda_B} Z^{\prime 2} (\kappa \sigma)   f(\kappa \sigma, \theta_1) f(\kappa \sigma, \theta_2) \frac{\exp(- \kappa r)}{r/\sigma},
\end{equation}
where $Z' = Z_{eff}\lambda_B/\sigma$.
The presence of the anisotropy functions in the pair  interaction potential implies a strong asymmetry between coplanar and stacked configurations. 
Given the form sketched in Fig. \ref{figZ}, for a fixed center to center distance, the stacked configuration  corresponds to the minimum of interaction potential while the coplanar arrangement gives rise to  the unfavorable situation of  maximum repulsion. Although   both arrangements correspond to non-overlapping disks configurations and  are equivalent from the excluded volume point of view, electrostatic interactions robustly favor  stacked ordering  against coplanar arrangement. Hence, for charged disks and more generally in charged oblate spheroids, a competition between the anisotropic excluded volume and electrostatic effects ensues.  
 The interplay between anisotropy in shape and interactions  has important consequences on the phase behavior and dynamics as will be the topic of discussion in the following.  However, before discussing the phase behavior we first outline the method and  details of our simulations in the next section.

%


%
%

\section{ Simulation methods and details }
\label{sec:method}

  We carried out canonical ($NVT$) Monte-Carlo simulations \cite{Allen,Frenkel} on a 
system of $N$ infinitely thin hard disks, i.e. zero thickness,  with
diameter $\sigma$ in a cubic simulation box of side $L$ with periodic boundary conditions using the anisotropic
two-body potential of Eq. (\ref{eq:pot2}).  For reasons of efficiency, the potential and the anisotropy function
  were tabulated as functions of  radial and angular variables.  The simulations were performed for a wide range of densities $\rho^*=N \sigma^3/L^3=0.1-8$ and  ionic strengths giving rise  to  $\kappa
  \sigma=1-20$. The temperature used in our simulations is  $T=300$ K which corresponds to a Bjerrum length  $\lambda_B \simeq 7 $ A.
The dimensionless diameter $\sigma/\lambda_B$ was taken to be 43 in the simulations, although we shall argue that its precise value, in a parameter range to be specified, is irrelevant. The number of particles varied 
between $10^3 \leq N \leq 10^4 $.  However, most of simulations were performed with $N=1372$ disks corresponding to 
box sizes ranging from $L/ \sigma=17.0$ for $\rho^*=0.1$ to $L/ \sigma=5.5$ for $\rho^*=8$.
For  $\kappa \sigma \leq 2 $, we adopted an Ewald-like scheme (i.e. with neglect
of the small Fourier space contribution) to take into account the relatively
long range of the potential \cite{salin:00}. More details on the Ewald-like summation method for anisotropic Yukawa potential can be found  in Appendix A. One should note, however, that due to the presence of the anisotropy
function $f(\kappa \sigma, \theta)$, the screened potential depends  on both $r$ and $\theta$. Therefore, this scheme can only be applied in an approximate way (see Appendix A).  For  $\kappa \sigma=2$ in the density range $\rho^*= 1-4 $, we confirmed that Ewald-like simulations performed with  $ N=1024 $ particles lead to the same results as  with larger box sizes ($N=4000-10976$, $ L/ \sigma \approx 15 $) without Ewald sums.

Due to relatively large values of the effective charge leading to a
complex free energy landscape, we sometimes encountered, at large
densities, $\rho^* \geq 3$, a dependence of the final configuration on
initial conditions. We circumvented this problem by  simulated annealing \cite{SimuAnneal} i.e. we started from an initial equilibrium configuration of (uncharged) disks at the desired density, then increased the charge gradually from
zero to the final value $Z_{eff}^{sat} (\kappa \sigma) $.  We  performed (depending on the density and $\kappa \sigma$)
$0.3-1 \times 10^6$ MC cycles, (a cycle consisting of translation and
rotation of the $N$ particles) at each equilibration step before increasing the charge value. Depending on the density and ionic strength, we varied the amplitudes of translational  and rotational moves in the range $0.001-0.1$  to obtain an acceptance ratio of about 30-40\%. It turned out essential to guarantee equilibrium of the system in the
initial stages (low charge values) of the process. This gradual charge increment, similar to gradual cooling of the
system for a fixed value of the effective charge, leads to reproducible
results for independent simulation runs performed with different initial
conditions, provided the charge increment between two subsequent steps is small enough, typically
$\Delta Z_{eff} \lambda_B / \sigma \leq 0.5 $.  It may be worthwhile to point out that common strategies like starting from
an equilibrated high density configuration then lowering the density by
scaling of the simulation box dimensions (at fixed $\kappa\sigma$) or
starting from an equilibrium configuration at some value of
$\kappa\sigma$ and varying the value of  $\kappa\sigma$ (at fixed
density) failed due to pronounced ``hysteresis'' effects.

To characterize the features of each structure, we computed both structural and dynamical correlation functions. The spatial arrangement  of particles was identified  by radial pair correlation function $ g(r) $ and its spatial Fourier-transform, \emph{i.e.}, the structure factor.  The degree of orientational order was characterized by the nematic order parameter
$ S\equiv \left\langle P_2(\cos (\psi)\right\rangle=\left\langle\frac{1}{2} (3 \cos^2 \psi -1) \right\rangle$ where $\psi$ is the angle of a platelet normal with the director $\widehat{n}$ and the brackets mean
averaging over all particles  "ensemble averaging". The director is obtained by finding the eigenvector corresponding to the  eigenvalue  of nematic order tensor with the largest 
magnitude. By definition, $-1/2<S<1$ with $S = 0$ for the disordered isotropic phase. When $0 < S < 1$, the platelet normals point on average along the director $\widehat{n}$ while in the more unconventional $S <0$ regime,
the platelet normals are on average  perpendicular to $\widehat{n}$ ("antinematic" order). Likewise, the extent of  orientational order in space was quantified by orientational pair correlation 
function $g_{or}(r)= \left\langle P_2(\cos (\theta(r))\right\rangle $ where $\theta $ is the angle between the orientations of two particles which are a distance $r$ apart.

We further investigated the dynamics of the orientationally disordered structures with vanishing $S=0$ by means of
 dynamic Monte-Carlo (DMC)  that we have developed  for anisotropic colloids  \cite{DMCsara}.  In this case, very small values of displacements were used and we performed $0.1-2 \times 10^7$ MC cycles to obtain  the dynamical quantities. 
 The dynamics of translational degrees of freedom was characterized by self intermediate scattering function 
 $F_s(q,t)=\frac{1}{N}  \left\langle \sum_{i=1} ^N \exp(i\vec{q} \cdot (\vec{r}_i(t)-\vec{r}_i(0))\right\rangle$ and mean-squared displacement $\left\langle \Delta r^2 (t) \right\rangle$. We quantified the rotational dynamics by computing  orientational time correlation function  $\left\langle P_2 (\widehat{u}_i (t)\cdot \widehat{u}_i(0)) \right \rangle$.   In  Appendix B,    we   provide a complete list of all  quantities  computed to characterize the structure and dynamics of  the observed structures as well as their definitions.

A few lines are here in order concerning the methodology. Our goal is to chart out the phase diagram 
of the platelet system under study. Ideally, this proceeds by computing the free energies of a number of candidate phases. 
Of course, due inclusion of micro-ions in the simulations, both counter-ions and co-ions, opens the way to proper free energy calculations,
but such a route seems prohibitively expensive in computational resources \cite{commentJJ}, which explains why we resorted to an effective
description, integrating out the micro-species from the description.
It should be realized next that knowing the effective pair interactions between the colloids, as is the starting point for us, 
is not sufficient to compute the free energy. It would be incorrect to use the corresponding effective Hamiltonian, which only depends on the colloidal degrees of freedom, 
to compute the free energy in a one component model view. This has been explained in the literature, see e.g. \cite{vanRoij,Belloni,Dobnikar}. This is due to what is often 
called 'volume terms' which are state-dependent and depend on the volume fraction as well. They are not known for aspherical objects, for which charge renormalization effects, of paramount importance here, have not been studied so far \cite{Denton}. 
This precludes the use of the effective Hamiltonian to get the total free energy of the system. The key point here is that, nevertheless, the effective potential, in the one component view, 
yields the correct colloidal {\em structure}.  This is precisely what we do in our study (at Monte Carlo level to sample phase space). As a consequence, we cannot reach the same level of accuracy 
in location of phase boundaries as with proper free energy calculations, but we nevertheless provide a trustworthy exploration of the phase diagram. 

\section{ Results }
\label{sec:results}

\begin{figure}[h]
\includegraphics[scale=0.6]{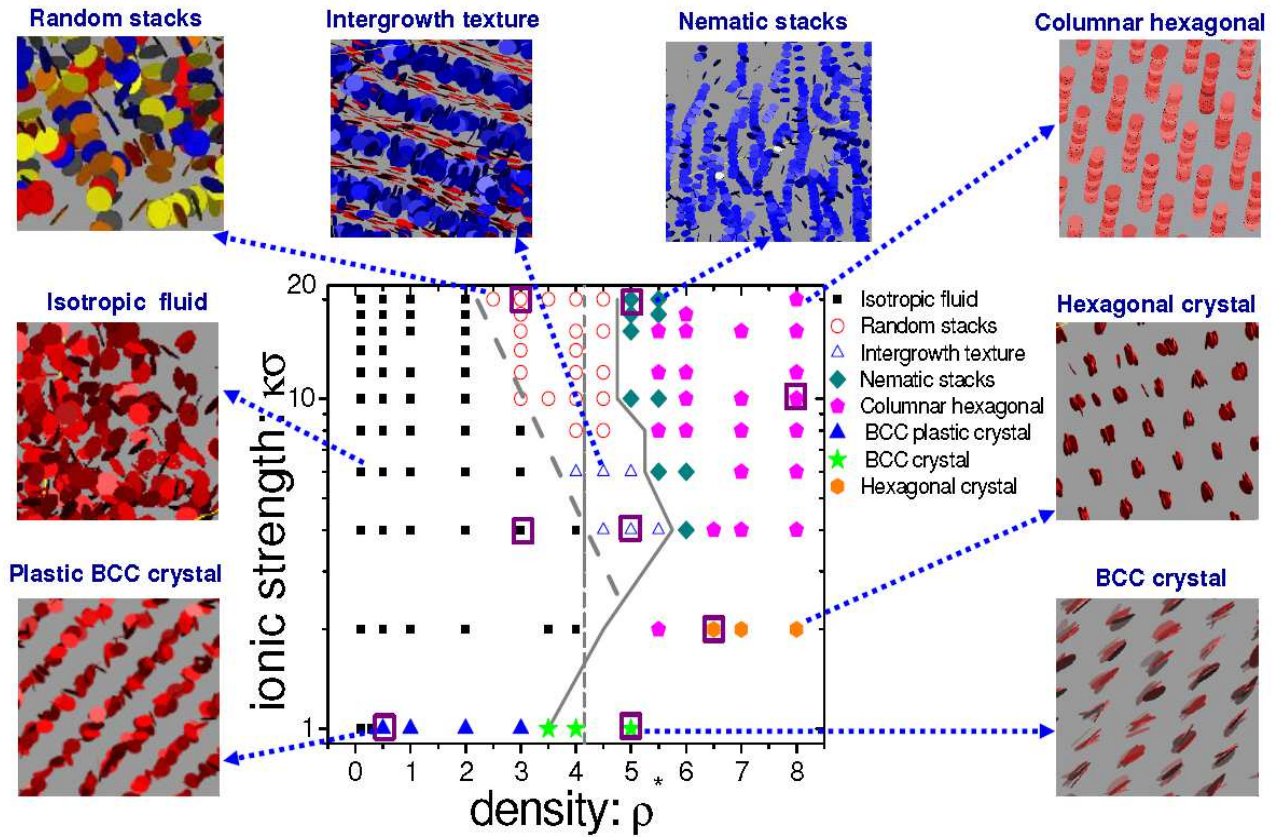}
\caption{ Phase diagram of  charged disks as a function of dimensionless density $\rho^*$ and screening parameter $\kappa \sigma$. We can recognize
 8 distinct structures: isotropic fluid,  stacks, intergrowth texture, nematic stacks,  columnar hexagonal and three different crystals.  
 The straight dotted line shows the density beyond which the nematic phase for hard disks \cite{Daan-disks}, ( i.e.,  no 
 electrostatic interactions),  appears. The solid line separates the orientationally ordered and disordered phases of charged disks.  We have discussed in details  the structural and dynamical  features of the points marked by
big squares in the text. 
}
\label{phase}
\end{figure}

Having introduced the effective interactions between charged disks and our simulation method, we investigate the phase behavior as a function of ionic strength  and density at  saturation 
value of effective charge, a case which pertains to highly charged 
disks. In the following, we first provide an overview of the  structures observed and present our concluding phase diagram  shown in Fig. \ref{phase}. Then we discuss in more detail the features of each structure. Subsequently, we examine 
 the  influence of effective charge magnitude on the formation of structures in both  orientationally ordered and disordered regions of the phase diagram.
\subsection{\label{sec:phase} Phase behavior }

Before going into  details about the characteristics of each
structure in the following subsection, we discuss our diagram (see Fig. \ref{phase}) as a function of density and ionic strength (proportional to $ (\kappa \sigma) ^2$).  
Through a detailed analysis of the observed structures by various static and dynamical correlation
functions, snapshots, etc  (see below), we have identified eight distinct structures.  Four of them have a vanishing overall
nematic order parameter, $S$ and are formed at low and moderate densities. The first  structure is that of  an isotropic fluid with both short-ranged positional and orientational order. The second one is 
a plastic crystal with long-ranged positional order and random orientations of particles. The third one is a  set of  disk stacks whose orientations lack any long-range orientational order. This structure appears at  
high ionic strengths and densities preceding the nematic phase.  The fourth  structure, also with vanishing $S$, is a more  "exotic" arrangement that consists of intertwined nematic and anti-nematic layers whose directors are
perpendicular to each other and  we coined it  \emph{nematic-antinematic intergrowth texture} \cite{Scirep}. 
The orientationally ordered phases with $S> 0.4$  appear at higher densities. They include two liquid-crystalline phases, i.e. nematic phase of platelet stacks (sometimes called "columnar nematic") and hexagonal
 columnar phase, which appear at moderate to high ionic strength. The other two ordered structures  are  crystalline phases with bcc-like and hexagonal symmetry which exhibit  both positional and orientational long-ranged  order.

\begin{figure}[h]
\includegraphics[scale=0.25]{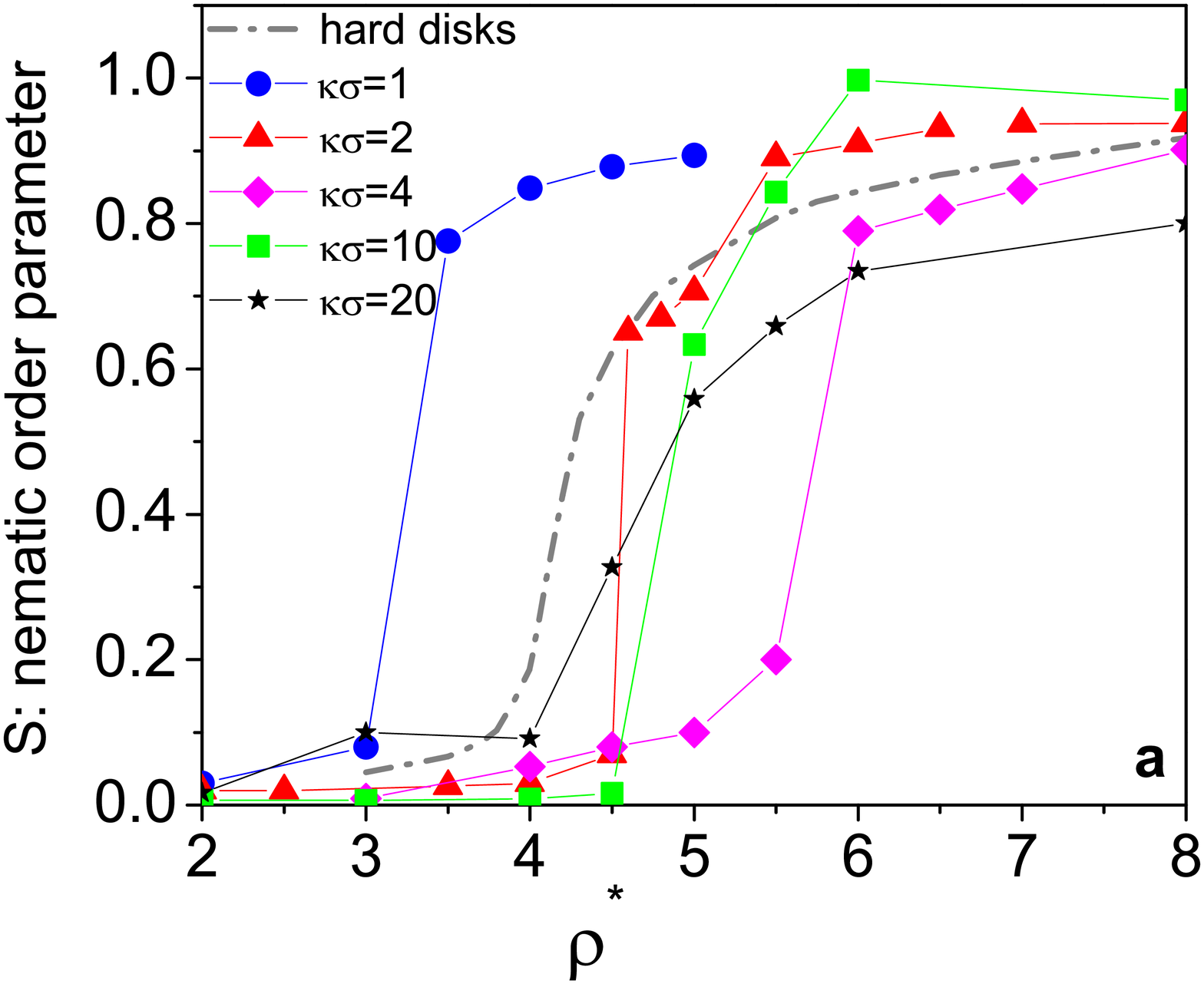}
\includegraphics[scale=0.25]{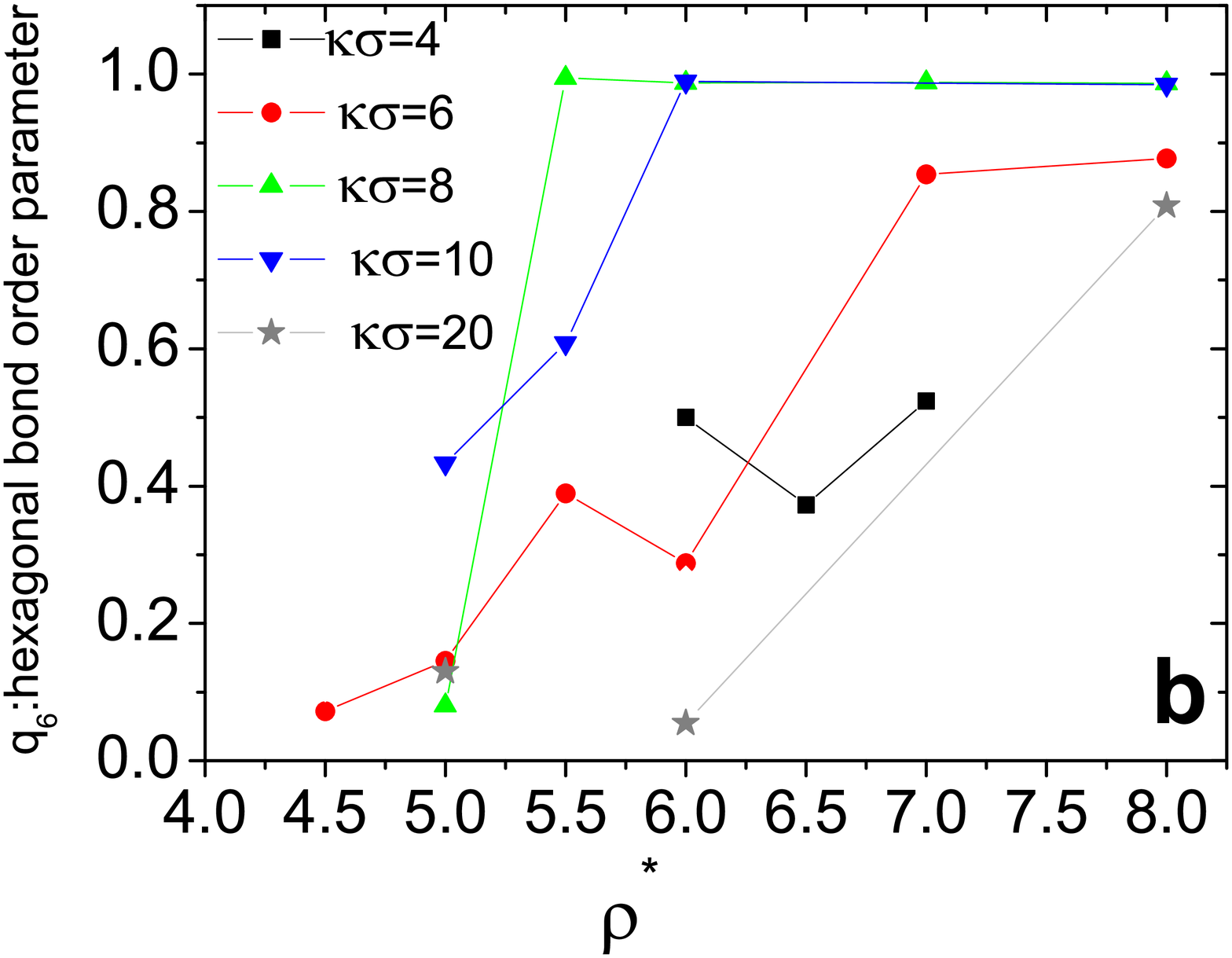}
\caption{  a) The nematic order parameter  b) the hexagonal bond-orientational order parameter as a function of density for different values of $\kappa\sigma$.
Note the non monotonous behaviour of the function $S(\rho^*)$ as a function of $\kappa \sigma$. 
  }
\label{nematic}
\end{figure}

We summarize  briefly  some important aspects of the phase diagram (Fig. \ref{phase}) in connection
with the features of our pair potential. A distinctive feature  is the non-monotonic
behavior of orientational disorder-order transition with $\kappa \sigma$.  While for the fairly isotropic case with $\kappa \sigma =1$, the orientationally ordered phases appear at lower
 densities than those of hard disks \cite{Daan-disks,Harnau},  for $\kappa \sigma \geq 2$ the liquid-crystalline phases of charged disks
appear at densities $\rho^* \geq 4.5 $.   In Fig. \ref{nematic}a, we have shown the nematic order parameter versus density
for several $\kappa \sigma$. The curves show that the overall nematic order parameter is a
 sensitive function of ionic strength. For the orientationally ordered phases, in Fig. \ref{nematic}b, we have plotted  the hexagonal bond orientational order parameter $q_6$ as a function of density 
 for several values of $\kappa \sigma$ . $q_6$ characterizes the positional ordering of particles on planes perpendicular to the  nematic director.   We notice that $q_6$ also exhibits  a non-monotonic  
 dependence versus $\kappa \sigma$. These non-monotonic behaviors can be understood as opposing effects of decreasing  screening length (favoring hard disk-like behavior)  and  enhanced anisotropy function (disfavoring nematic ordering)   upon increase of  the ionic strength.  Note that $Z_{eff}^{sat}$ is also an increasing function of $ \kappa \sigma $ but   we verified that the observed trend is unchanged provided $Z_{eff}\lambda_B/\sigma >2.3$; this issue is  discussed in more detail  in Sec. \ref{sec:charge}. 
This in turn implies that  the particular value of $\sigma/\lambda_B$, which is {\it a priori} a relevant dimensionless parameter, in practice plays little role. This opens the possibility to discuss the results pertaining to different particle sizes  solely in terms of $\rho^*$ and $\kappa \sigma$, as will be done in the following.

\begin{figure}[h]
\includegraphics[scale=0.25]{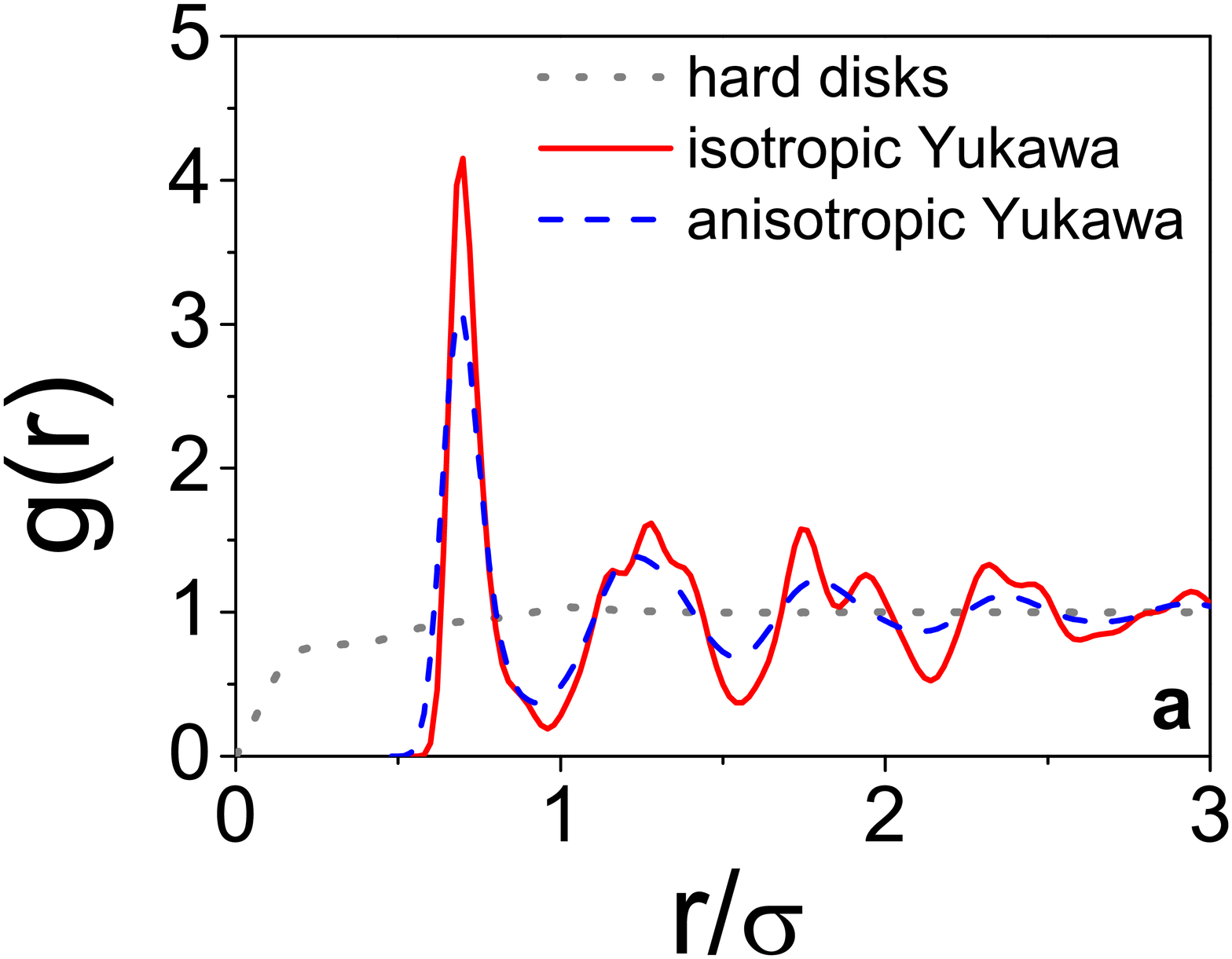}
\includegraphics[scale=0.25]{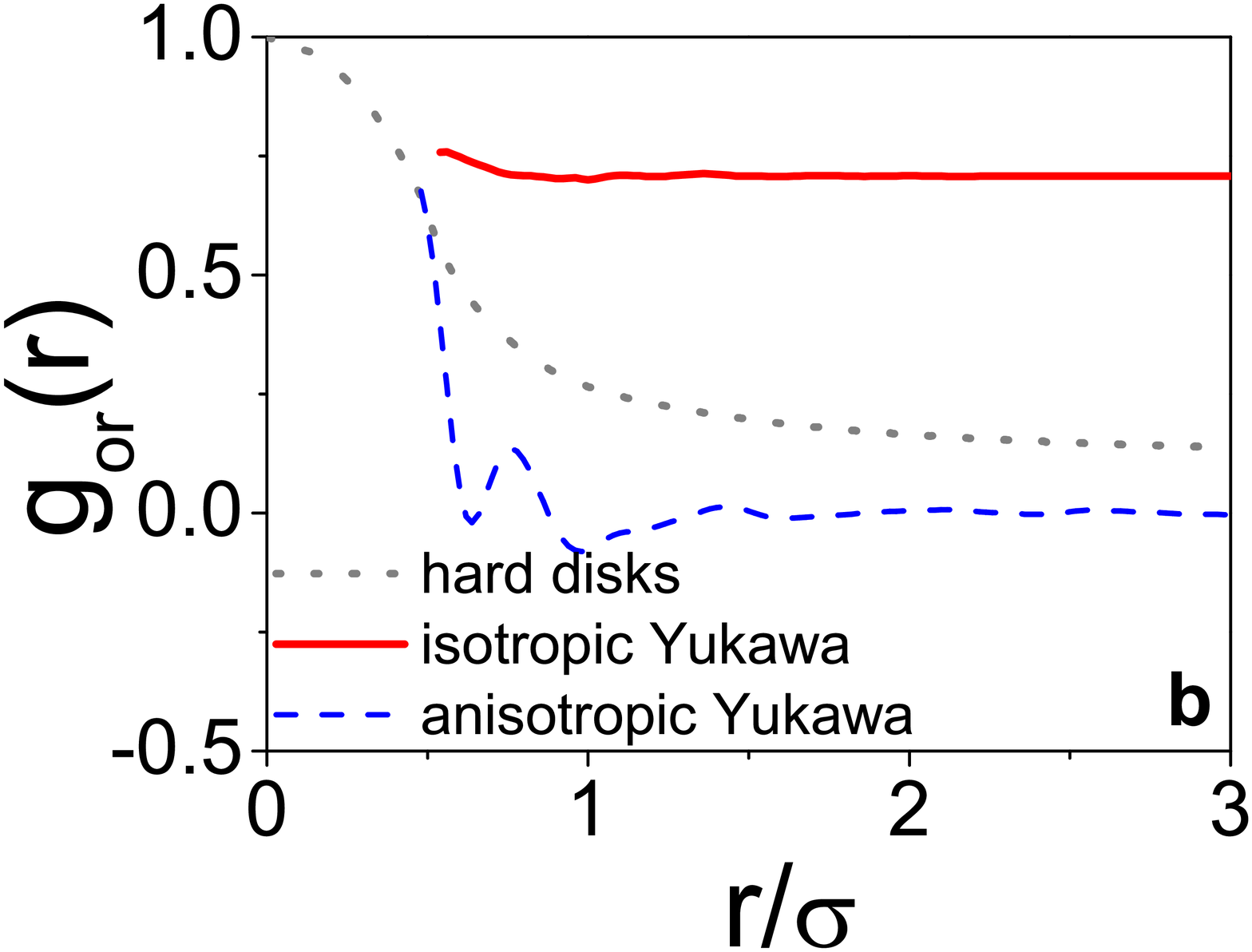}
\caption{a) positional and b) orientational pair correlation function at density  $\rho^*=4$ displayed for hard disks
and disks interacting with isotropic and anisotropic Yukawa potential, at $\kappa \sigma=2$.}
\label{fig:iso}
\end{figure}

 In the limit of low ionic strengths corresponding to $\kappa \sigma=1 $, where the potential is rather isotropic and long-ranged,
bcc-like structures are formed, as observed with charged spheres
at low $\kappa \sigma$ \cite{Dijkstra}. At low densities, disks are orientationally disordered (plastic crystal)
and are reminiscent of the Wigner crystals
 observed for low volume fractions of charged spheres \cite{chaikin}. Increasing the ionic strength, at
 $\kappa \sigma=2$ where the amplitude of the anisotropy function is less than 1.2, the plastic crystal is replaced by an isotropic
 fluid at low and moderate densities. The crystal disappearance for a slightly larger
 $\kappa \sigma$ and shorter range of potential highlights the effect of anisotropy. Performing simulations for a system of hard disks interacting with
 an isotropic Yukawa potential at the same value of $Z_{eff}$ and $\kappa \sigma$, we verified that the observed geometrical frustration is indeed due to anisotropic   nature of potential 
 which leads to melting of crystalline structures at moderate densities. In Fig. \ref{fig:iso}, we have plotted positional and orientational correlation functions  at a density 
 close to the isotropic-nematic transition for 
 hard disks, i.e., $\rho^*=4$ for both hard and charged disks at $\kappa \sigma=2$ as well as disks interacting with isotropic Yukawa potential.  Interestingly, we find that an isotropic Yukawa potential 
 enhances orientational order, while the small  anisotropy amplitude of charged disks at  $\kappa \sigma=2$ leads to destruction of the orientational order, 
 thereby illustrating the frustrating effect of the anisotropy function on the ordering of disks.

At $\kappa \sigma=2$, in the high density regime, the long-range positional order is still preserved as crystals with hexagonal
symmetry appear. Interestingly, for large enough ionic strengths corresponding to $4 \leq \kappa \sigma \leq 6$ and at moderate densities, the
new intergrowth texture appears and, for $\kappa \sigma \geq 8$, randomly oriented stacks of disks are observed. 
At still higher densities, further increase of ionic strength equivalent to $\kappa \sigma \geq 4$, leads to weakening of the positional
order. Hexagonal crystals are replaced by hexagonal columnar liquid-crystals with no positional correlations along the nematic director. 

Having provided a general overview of the phase behavior,  we turn,  in
the next subsection, to a detailed description of each of these structures. To help the reader, we have marked by empty squares the state points on the phase diagram whose structural and dynamical features are discussed in Section. \ref{sec:structures}.

\subsection{\label{sec:structures} Description of the structures  }

\begin{itemize}
\item
 \textbf{Isotropic  fluid} \\

\begin{figure}[h]
\includegraphics[scale=0.2]{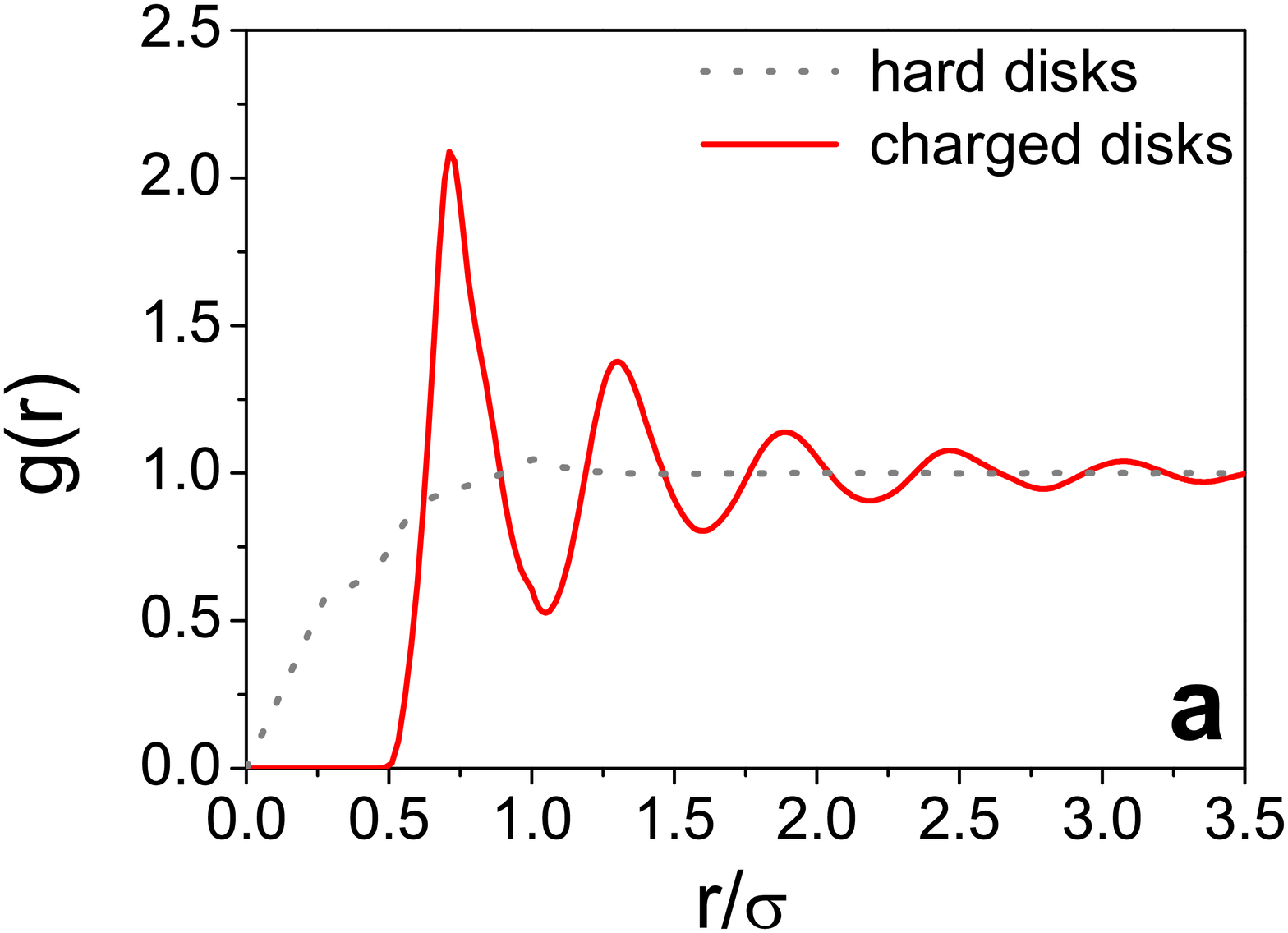}
\includegraphics[scale=0.2]{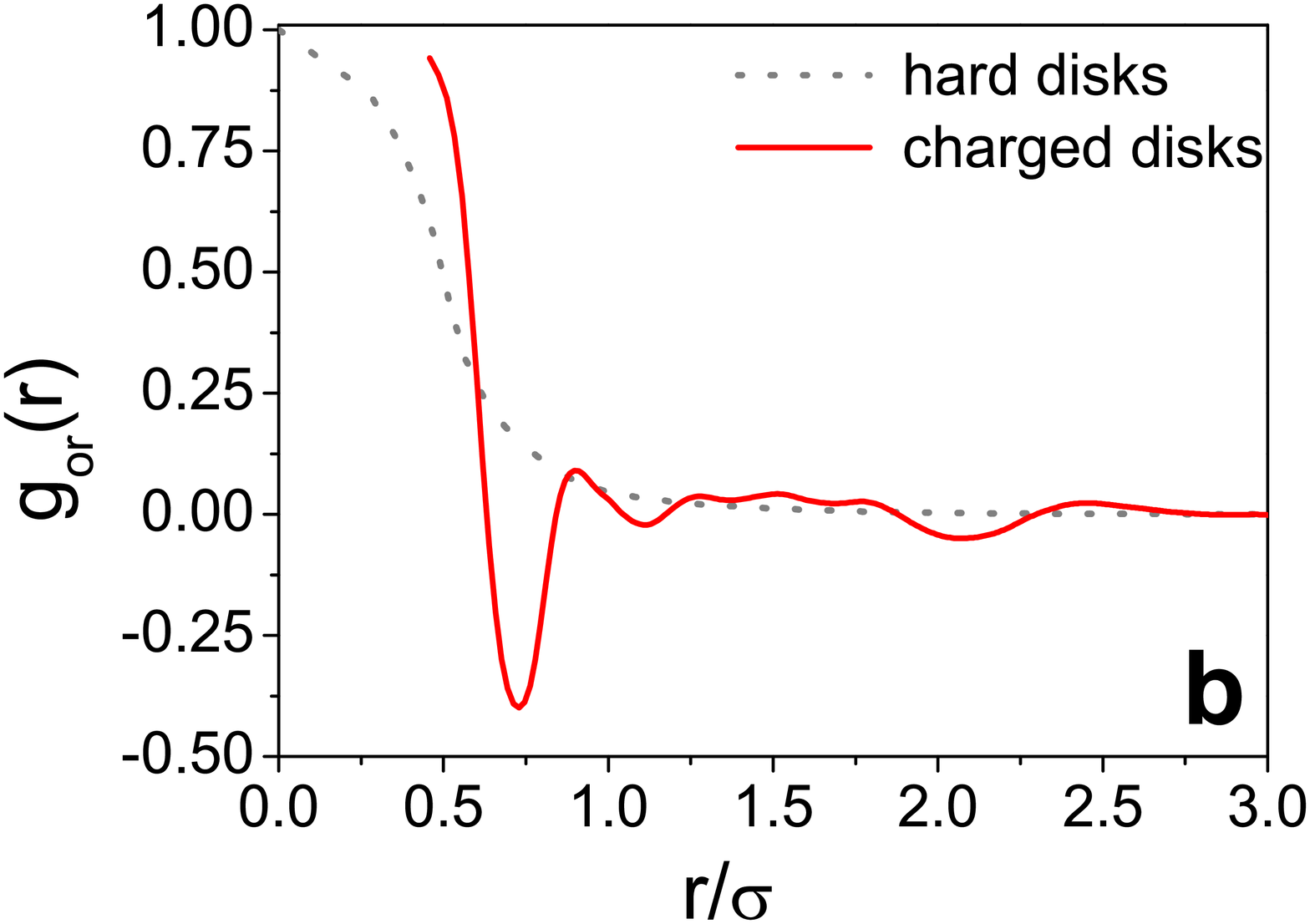}
\includegraphics[scale=0.2]{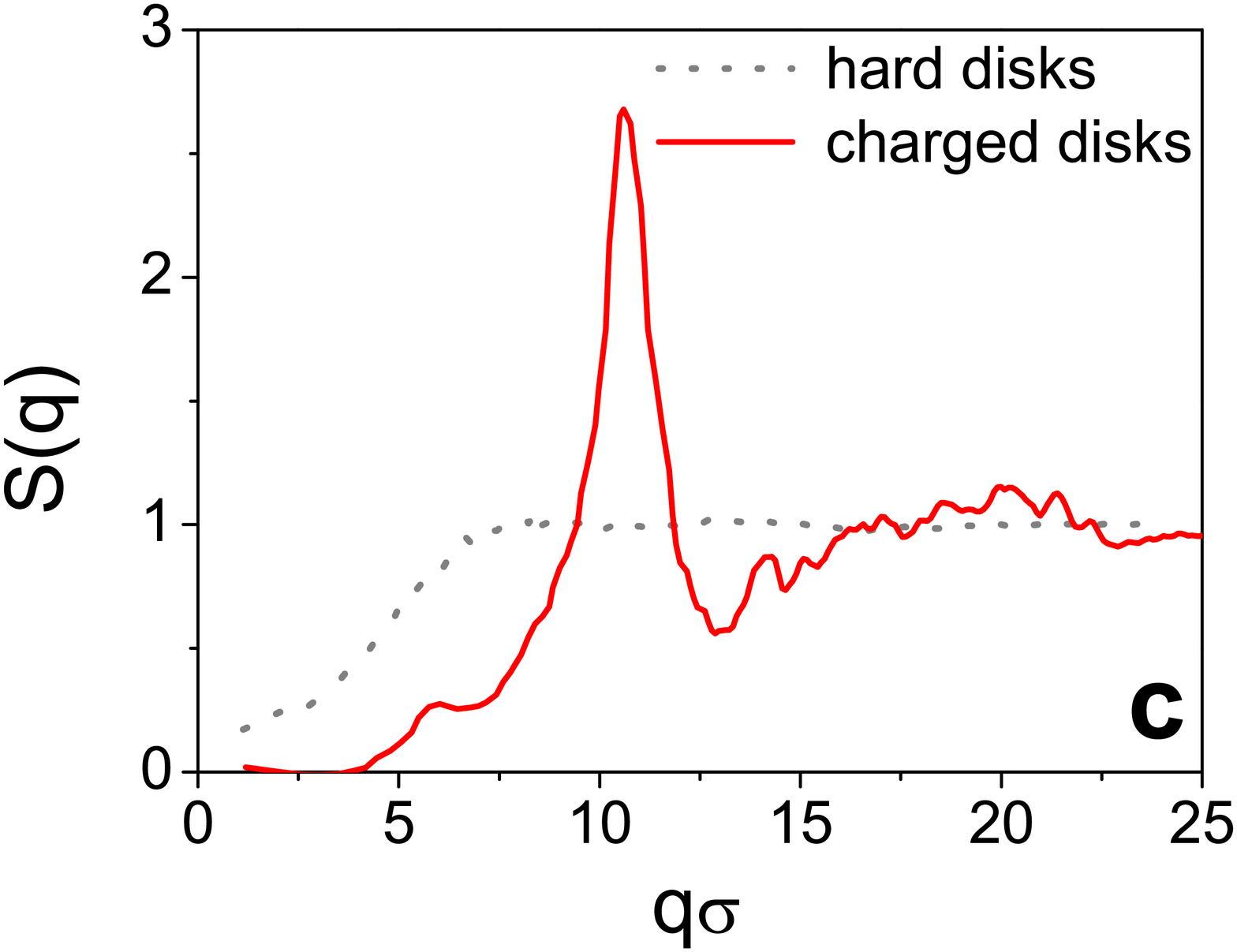}
\includegraphics[scale=0.2]{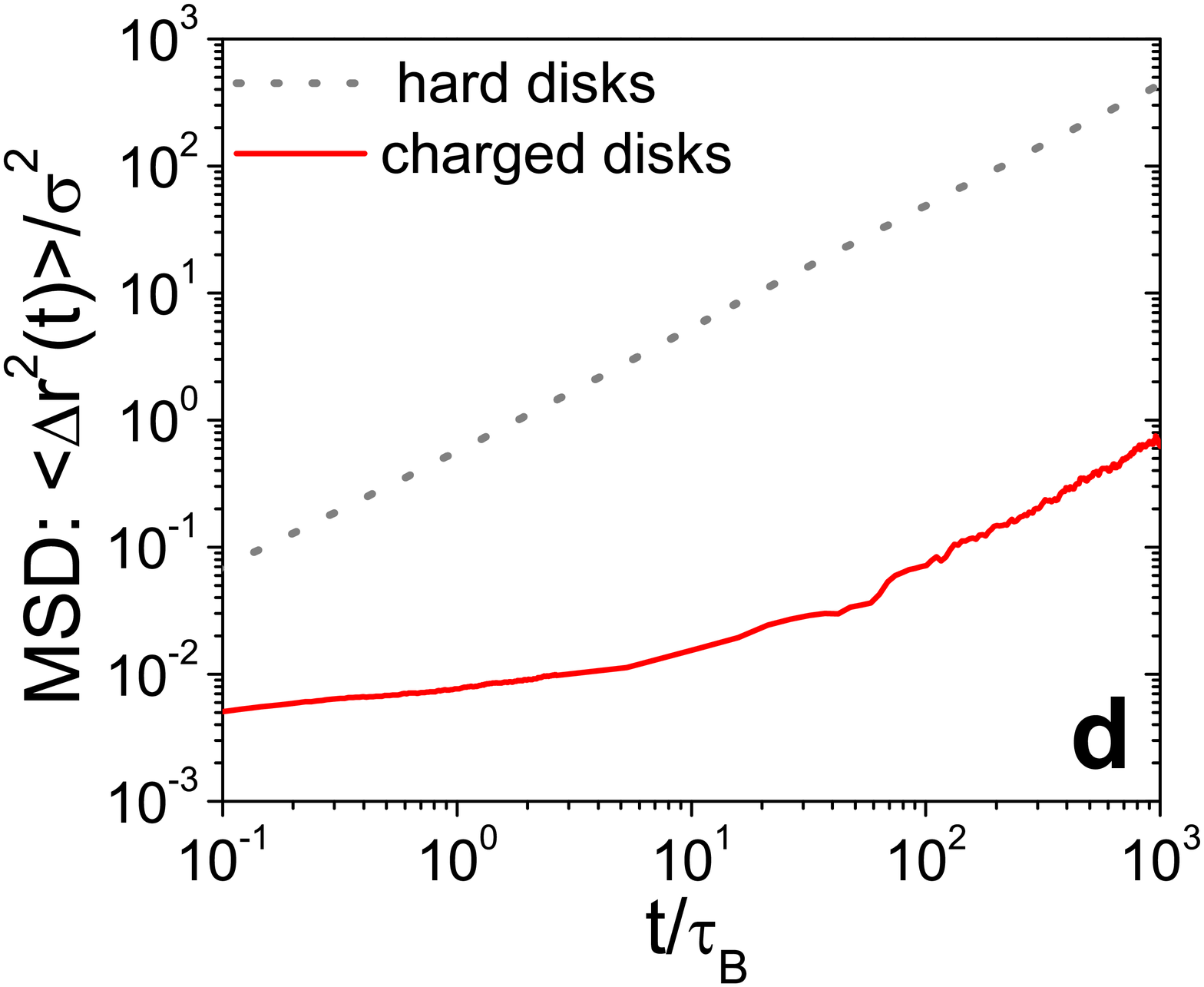}
\includegraphics[scale=0.2]{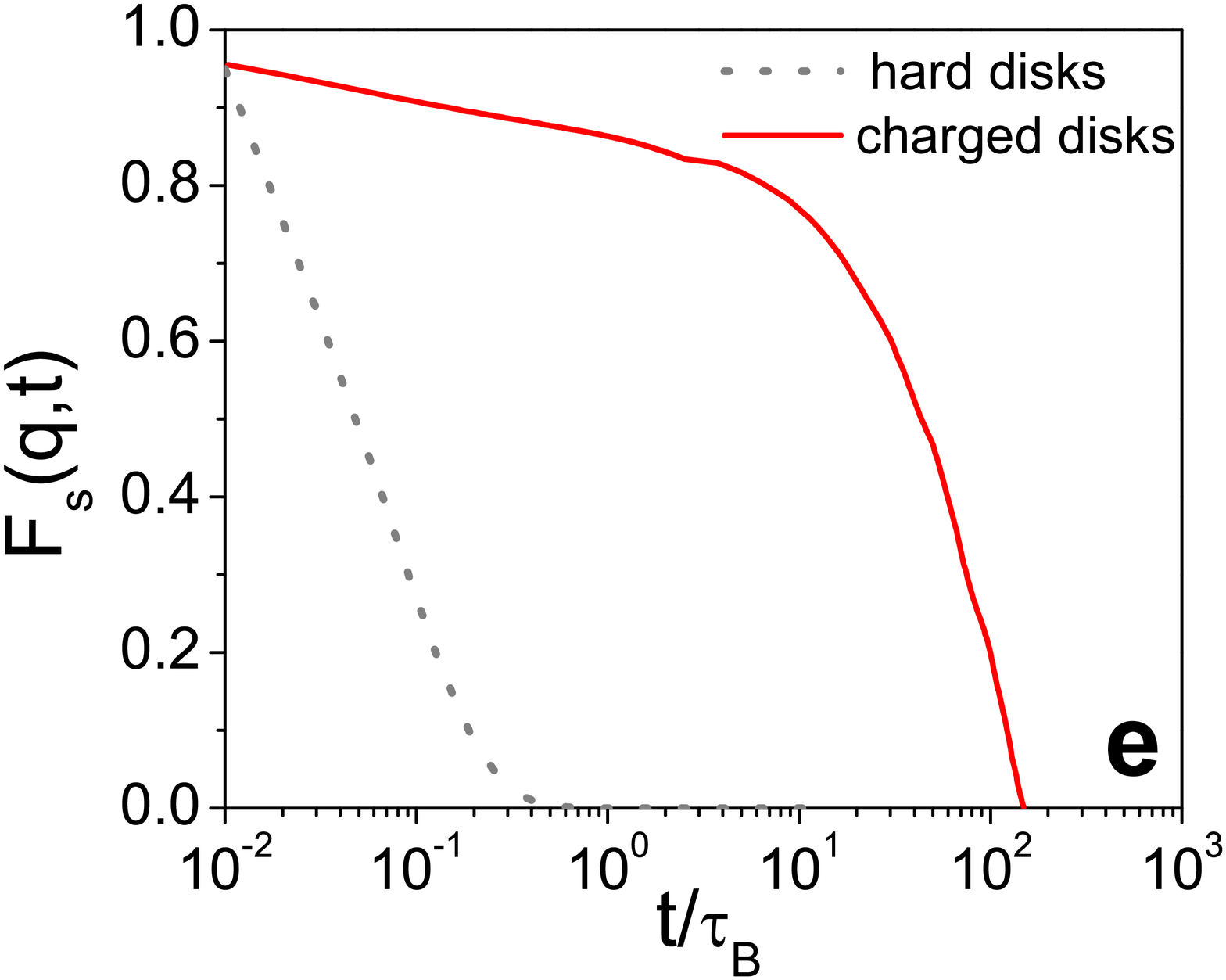}
\includegraphics[scale=0.2]{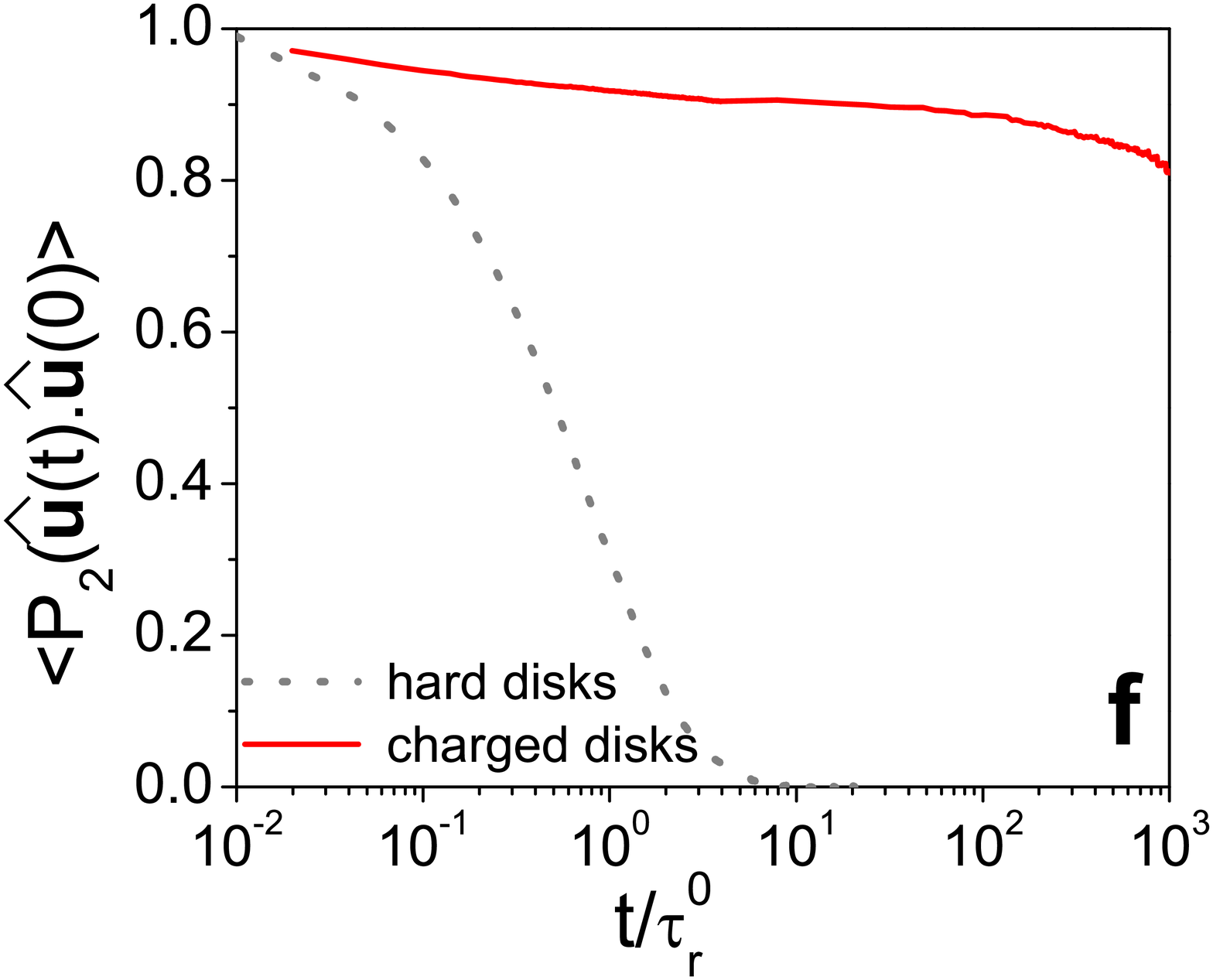}
\caption{ a) the  radial pair distribution function $g(r)$, b) the orientational pair distribution function $g_{or}(r)$
  , c) structure factor $S(q)$  d)  mean squared displacement $\left\langle \Delta r^2(t) \right\rangle$   e) intermediate scattering function $F_s(q,t)$ computed at $q\sigma=10.63$
and   f) orientational time correlation function $\left\langle P_2(\widehat{u}(t).\widehat{u}(0)) \right\rangle$   for an \emph{isotropic fluid} of $\rho^*=3$ and  $\kappa \sigma=4$ . 
}
\label{fig6}
\end{figure}
Similar to the case of hard disks, we  observe an isotropic fluid  at the lowest densities of the charged disks. However, the highest density for which an isotropic fluid is observed is 
a non-monotonous function of $\kappa \sigma$ and the  electrostatic interactions alter the spatial arrangement of charged disks conspicuously. To demonstrate this point, in  Fig. \ref{fig6}a, \ref{fig6}b, and \ref{fig6}c , we
have shown an example of  radial  pair correlation function $g(r)$, orientational correlation function $g_{or}(r)$ and structure factor $S(q)$ for  an isotropic fluid 
of charged disks at $\rho^*=3$ and $\kappa \sigma =4$. In each graph, we  have also plotted the corresponding quantities  for hard disks of the same density. All of these figures  illustrate a marked difference between 
charged and hard disks isotropic fluids and demonstrate that charged disks form a  structured isotropic fluid.  As  can be seen from the radial pair distribution function, the charged disks   are prevented to come too close due to repulsion.  The exclusion of particles from  each others proximity results  in  local ordering (manifested in a strong peak in the structure factor)  but no long-range spatial correlations exist. The average number of nearest neighbors is obtained by integrating over the first peak of $g(r)$ till the first following minimum; for this structure we obtain $N_{nn}=13-14$. This value is comparable to the total number of first and second nearest neighbors in BCC lattices (8+6) and suggests that  locally BCC-like ordering is favored for charged disks. 
The pronounced  peak in the structure factor also confirms the presence of a strong degree of local ordering. Looking into the organization of orientations of charged disks in space, we again notice a prominent deep 
minimum in the orientational pair correlation function at separations slightly higher  than particle radius ($r \approx 0.7 \sigma$). This means that at such short distances T-shaped configurations,  intermediate 
between coplanar and stacked, are favored.

We characterize the dynamical behavior of translational and rotational degrees of freedom  by mean-squared displacement $\left\langle \Delta r^2(t) \right\rangle$, intermediate scattering function and orientational 
time correlation function (see appendix \ref{appendixB} for definitions) as  presented in Fig. \ref{fig6}d, \ref{fig6}e and \ref{fig6}f, respectively. The time for translational motion is rescaled with Brownian time-scale $ \tau_B \equiv \sigma^2/ (6 D^t_{0}) $ and for rotational motion is reduced by the characteristic time scale for free rotational 
diffusion of disks, {\it i. e.}  $\tau_{0}^{r}\equiv1/(2 D_{0}^r) $. Again, we find that the electrostatic interactions influence 
the dynamics of charged disks remarkably.  Both translational and rotational motion of charged disks  are dramatically slowed down compared to that of hard disks at the same density. However, both the
self-intermediate scattering function and the  orientational  time correlation
function  decay on a time scale of the order of $t/\tau_B=1000$. Although we have chosen a relatively high density in the isotropic phase and  relaxation times are large, the system remains ergodic. 



%
\begin{figure}[h]
\includegraphics[scale=0.2]{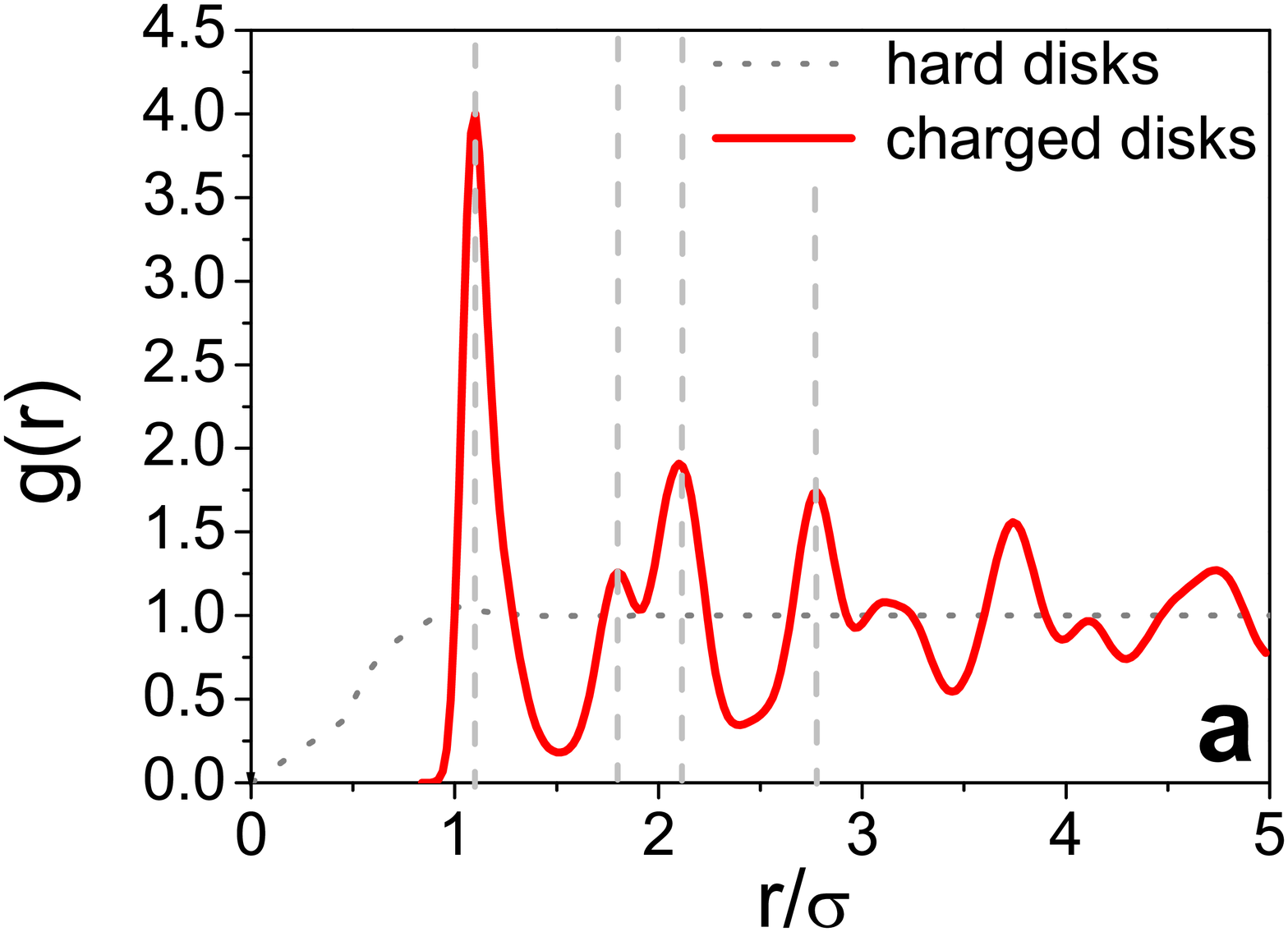}
\includegraphics[scale=0.2]{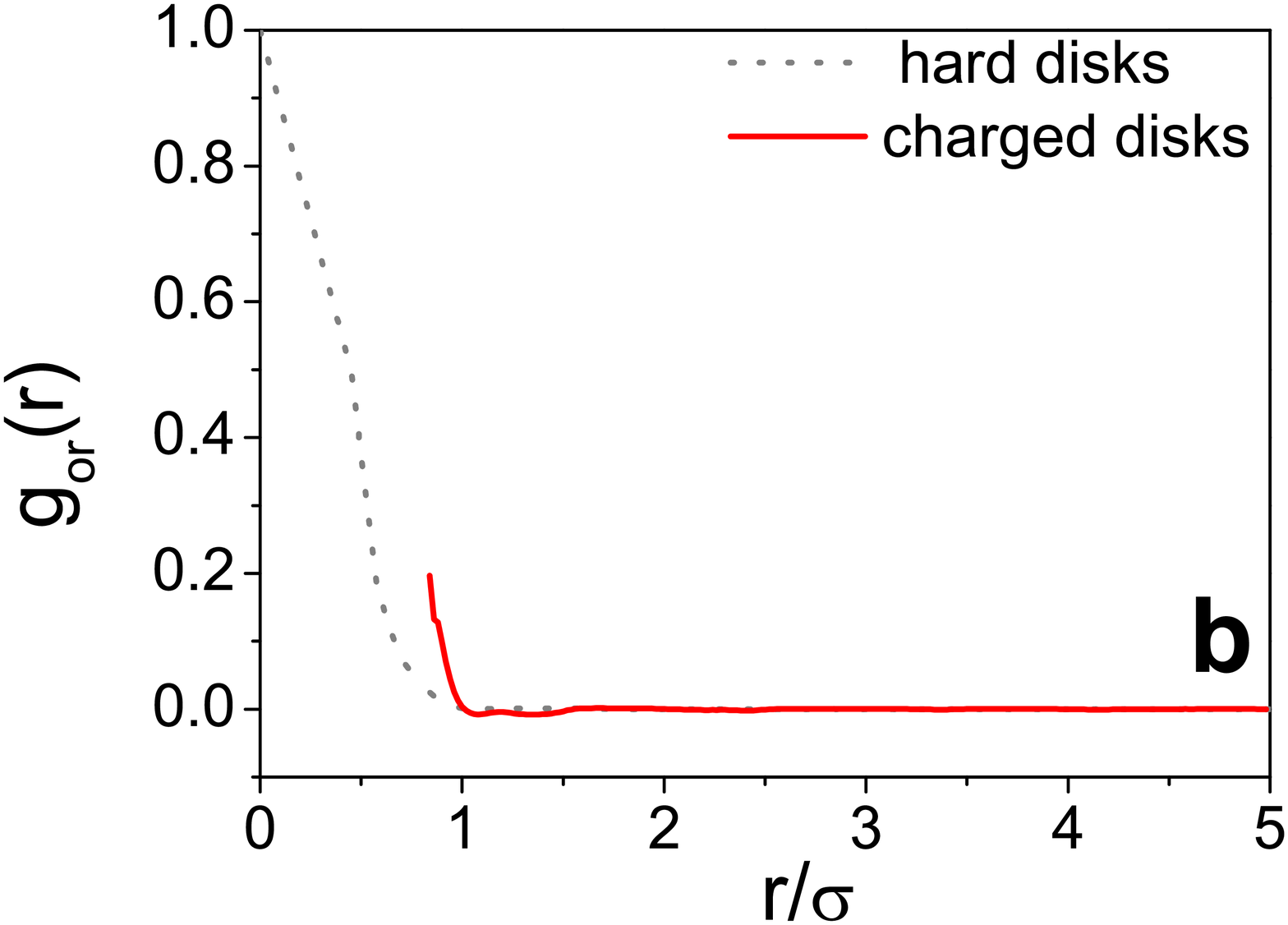}
\includegraphics[scale=0.2]{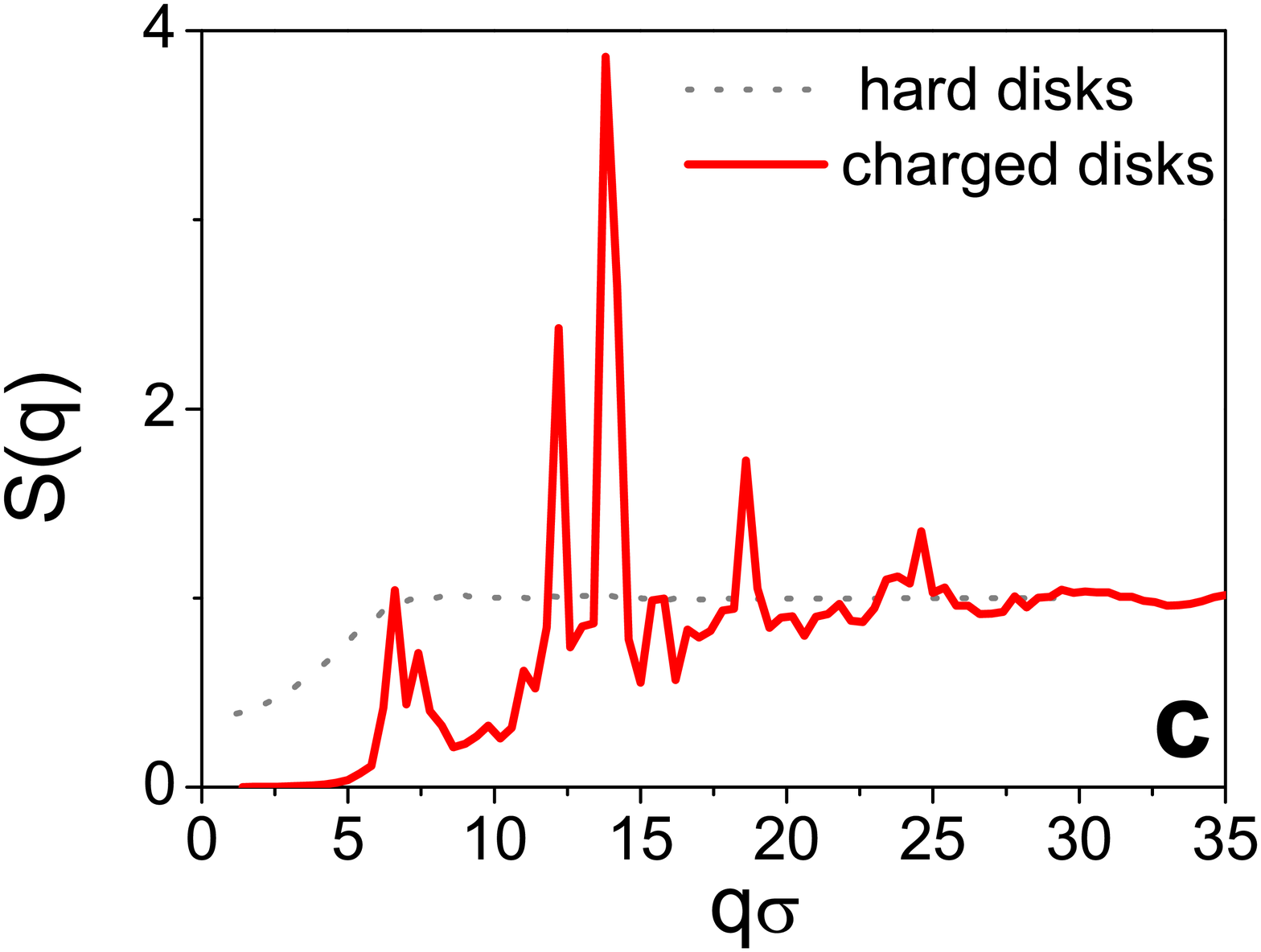}
\includegraphics[scale=0.2]{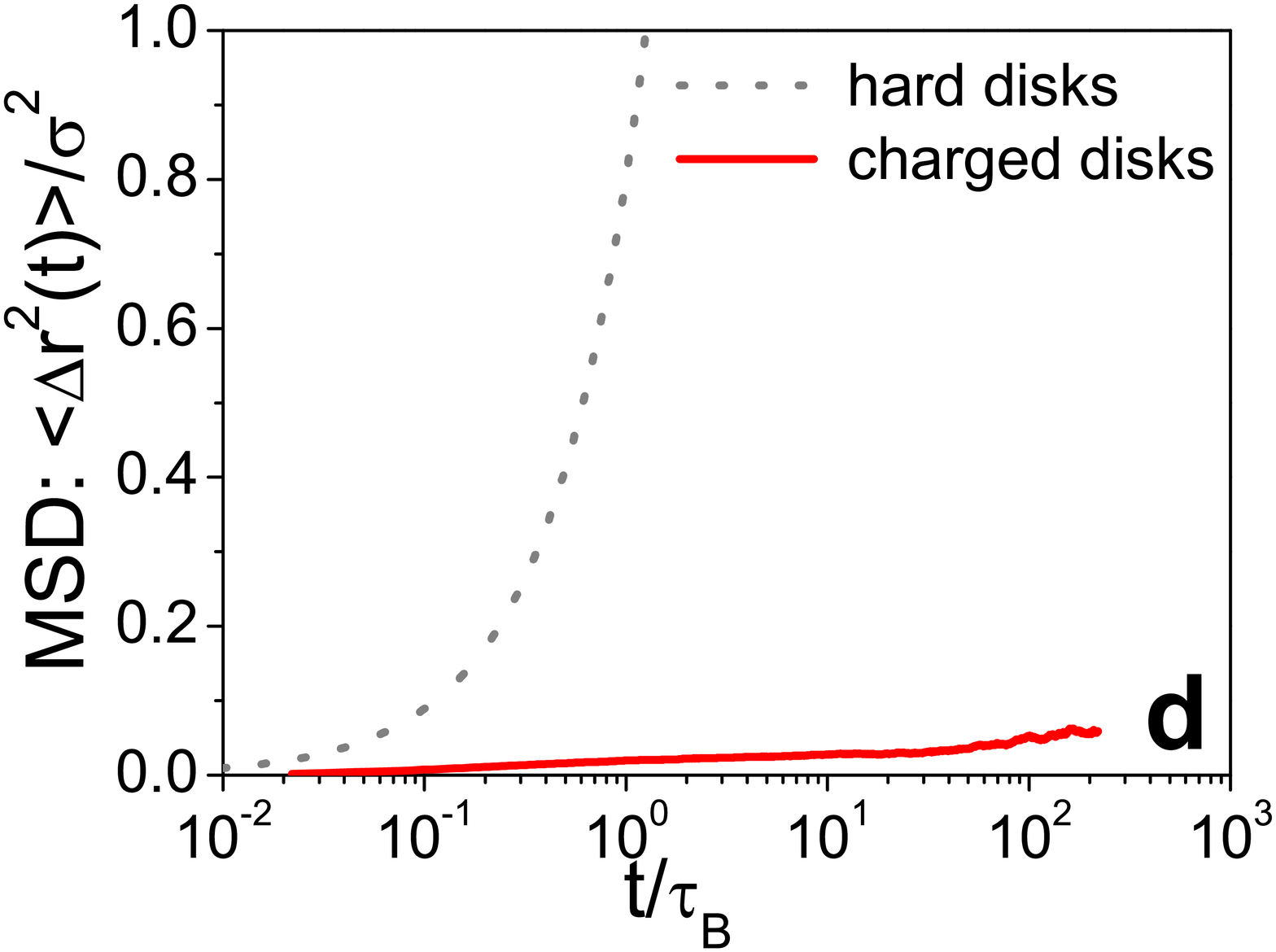}
\includegraphics[scale=0.2]{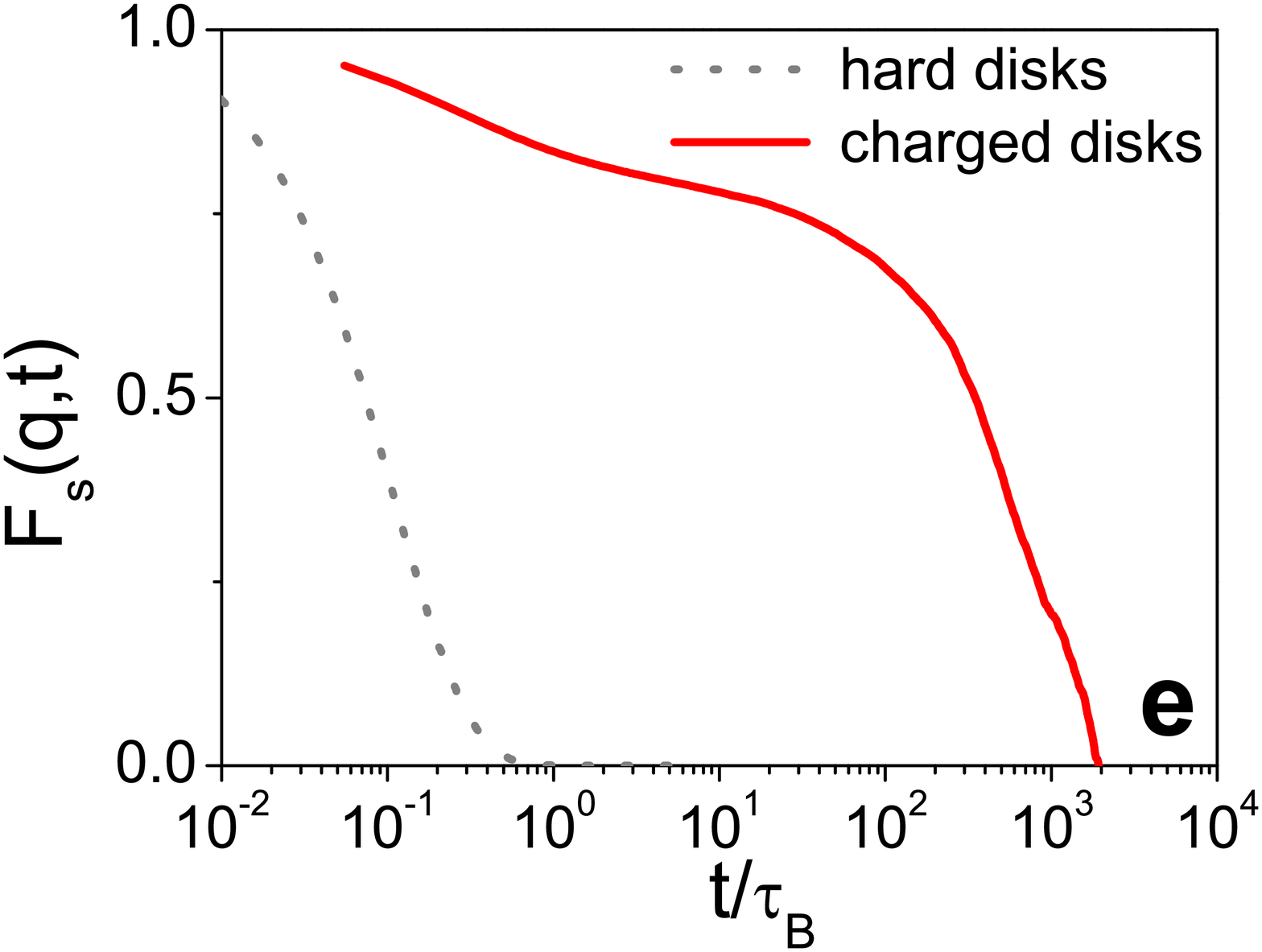}
\includegraphics[scale=0.2]{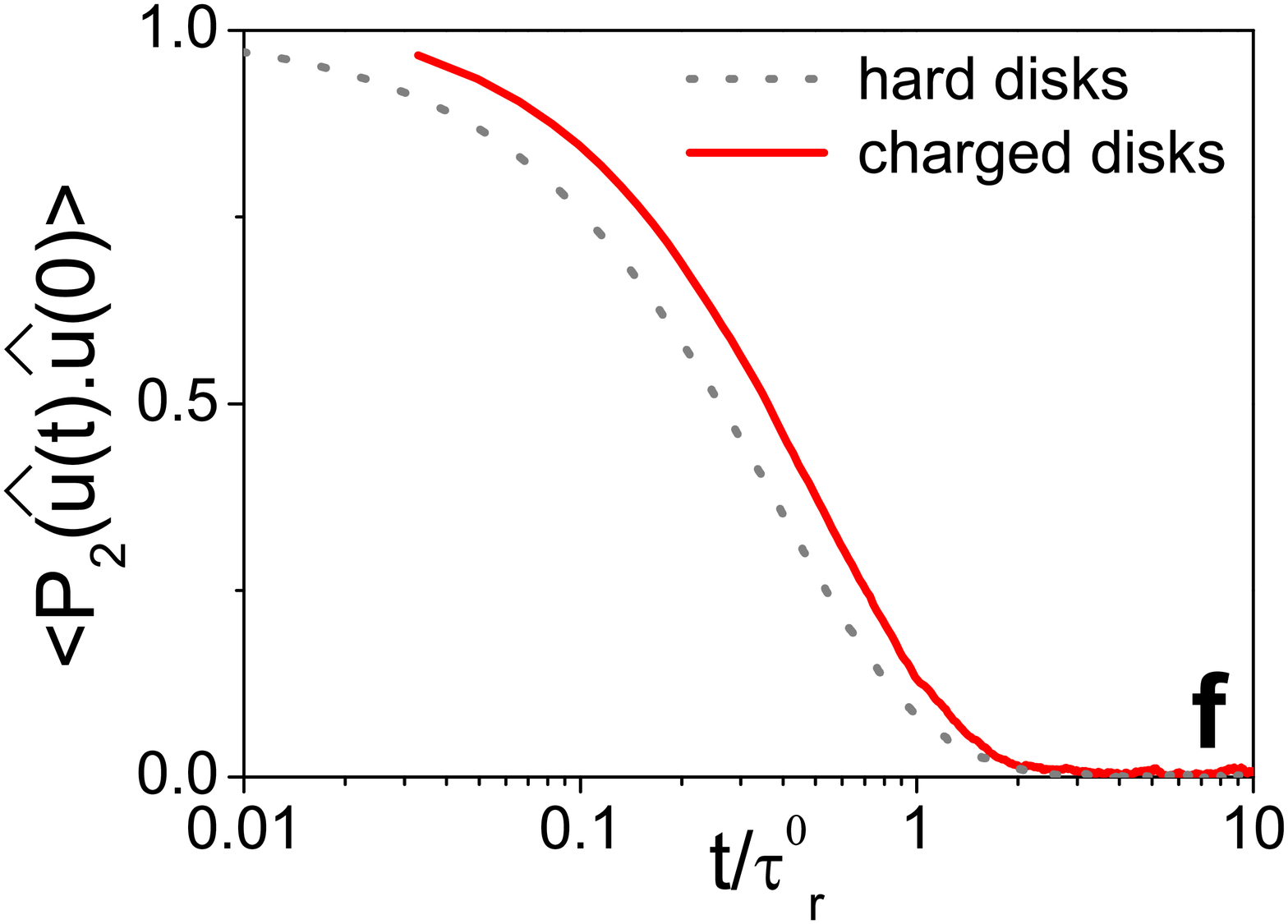}
\caption{ a) The  radial pair distribution function $g(r)$, b) the orientational pair distribution function $g_{or}(r)$
  , c) structure factor $S(q)$, d)  Mean squared displacement $\left\langle \Delta r^2(t) \right\rangle$,   e) intermediate scattering function at 
  $q \sigma=7.9$
and   f) orientational time correlation function $\left\langle P_2(\widehat{u}(t).\widehat{u}(0)) \right\rangle$  for a \emph{plastic crystal} with  $\rho^*=1$ and  $\kappa \sigma=1$.
The dashed vertical lines in Fig a) show the expected sequence of peaks in a bcc structure.}
 \label{fig7}
\end{figure}

\item \textbf{Plastic crystal} \\

Now we turn to the second orientationally disordered structure, {\it i.e.} the plastic crystal, which shows long-ranged positional order despite the lack of long-range orientational order. This structure appears at 
small ionic strengths,  $\kappa \sigma=1$, where the potential is fairly isotropic  at rather low densities $0.5< \rho^*<3.5$. At such low ionic strengths  the range of interaction potential is large enough to
induce arrangement of particles into crystalline order at intermediate range of densities.  However, these  densities are low enough that the disks can have a random orientation without conflicting with excluded volume constraints.

An example of  radial and orientational pair correlation functions and
structure factor for  such a structure are shown in Fig. \ref{fig7}a,
\ref{fig7}b, and \ref{fig7}c for $\rho^*=1$ and $\kappa \sigma =1$. Similar to the case of isotropic fluid the particles are excluded from proximity of each others for separations compared to the range of
potential {\it i.e.} $r<0.9 \sigma$.
As it can be noticed from both radial pair correlation function and structure factor, there is a robust positional ordering in contrast to  hard disks. Examining the positions of
peaks in $g(r)$, we verified that the formed crystal has  BCC-like positional ordering.
The first 8  peak positions expected for an ideal bcc lattice are given by the sequence $ a, 2/\sqrt{3}a, \sqrt{8}/\sqrt{3} a, \sqrt{11}/\sqrt{3} a, 2 a, 4/ \sqrt{3}a, \sqrt{19}/\sqrt{3} a$, and $\sqrt{5} a$ where $a$ is the nearest
neighbor distance. In Fig. \ref{fig7}a, the dashed lines at distances  $ r/\sigma=1.0911, 1.78,  2.09$ and 2.75 correspond to  first, third, fourth and seventh peak. 
The second, fifth, sixth and eighth peaks are
merged in with the adjacent peaks and are not visible. Integrating over the first peak of $g(r)$  merged with its second maximum till the first following minimum, 
we obtain 14.07 for the number of nearest neighbors which 
is consistent with the sum of first and second nearest neighbors for the BCC lattice. In contrast to the long-ranged positional order evidenced by $g(r)$, the orientational pair correlation function decays at a distance of the order of one diameter, testifying  lack of any  long-range orientational order. This
state is similar to the so-called Wigner crystal  observed for charged spheres at low volume fractions \cite{chaikin}.

Now, we turn to the dynamical characteristics  of the plastic crystal as presented in Figs. \ref{fig7}d, \ref{fig7}e  and \ref{fig7}f.  Inspecting the mean-squared displacement,  we find 
that the diffusion of  charged disks in plastic crystal is  greatly slower than for hard disks at the same density. Looking at the self-intermediate scattering functions at the peak of 
structure factor, $q \sigma=7.9$ ,  we also observe a two-step decay of $F_s(q,t)$ with a large  average relaxation time which confirms again that the relatively long-range electrostatic potential leads to a slow dynamics, similar
to what is observed in Laponite suspensions \cite{Abou}.  This strong slowing down of translational dynamics is  anticipated  as a consequence of long-ranged positional order  in plastic crystal. On the contrary, the rotational dynamics of the plastic crystal is not much influenced
by electrostatic interactions and is only slightly slower than that of hard disks. This behavior shows that the disks can freely rotate in the plastic crystal. 


\item \textbf{Random stacks} \\

The third  structure with no long-ranged orientational order is that of random stacks. In this configuration,  the orientations of the stacks are distributed in such  a way that there is no net orientational order.  Fig. \ref{fig8}a, \ref{fig8}b  and \ref{fig8}c  present an example of radial and orientational pair correlation functions and
structure factor at $\rho^*=3$ and $\kappa \sigma =20 $. 
The  presence of stacks manifests itself in a sharply peaked $g(r)$ corresponding to the average spacing
of two disks in a stack. For the example  shown here the stack spacing is about $r_{max}\approx 0.36 \sigma$ which also gives rise to a peak of structure factor obtained in the Fourier space 
$q_{max} \sigma =2 \pi \sigma	/r_{max}=19$. Integrations of $g(r)$ over its first peak  yields $N_{nn}=2$,
confirming the stacked nature of the structure.  We notice the presence of regularly spaced peaks up to distance $2.5 \sigma$ which suggests that the  number of disks in majority of stacks 
is less than 7. To verify this surmise, we determine the  size distribution of stacks. 
Using the position of the first minimum of $g(r)$  as a threshold for spatial correlations between particles, we could identify the different stacks
and obtain their size  distribution function as presented in Fig. \ref{pdf}a. We notice that the number of stacks of $n$ disks decreases exponentially
with stack size, similar to what is observed in linear self-assemblies obtained by living polymerization \cite{LivingPolymer}.


The orientational pair correlation attests the absence of   long-range orientational order in this structure, although we observe a short-range orientational order at disks
separations $r<0.5 \sigma$. This short-ranged order is succeeded by  deep  negative minima at distance $r=0.6$ and $0.8 \sigma$ which are manifestations of  competing effects of  electrostatic potential 
and excluded volume interactions that result in T-shaped configurations of disks at such distances.

 Examining the dynamics of  random stacks shown in Figs. \ref{fig8}d, \ref{fig8}e  and \ref{fig8}f, we find that  both translational and rotational dynamics of charged disks in this structure are strongly slowed down compared to that of hard disks. The translational degrees of freedom  still remain diffusive,  whereas the orientational  time correlation function decays only slightly over a time scale of the order of 1000 times the
 orientational relaxation time of an isolated disk. The observed slowing down of rotational diffusion  is quite spectacular
  and is reminiscent of an orientational  glass \cite{orGlass,schilling}, where  translational degrees of freedom are
 liquid-like and the orientational degrees are frozen.

\begin{figure}[h!]
\includegraphics[scale=0.2]{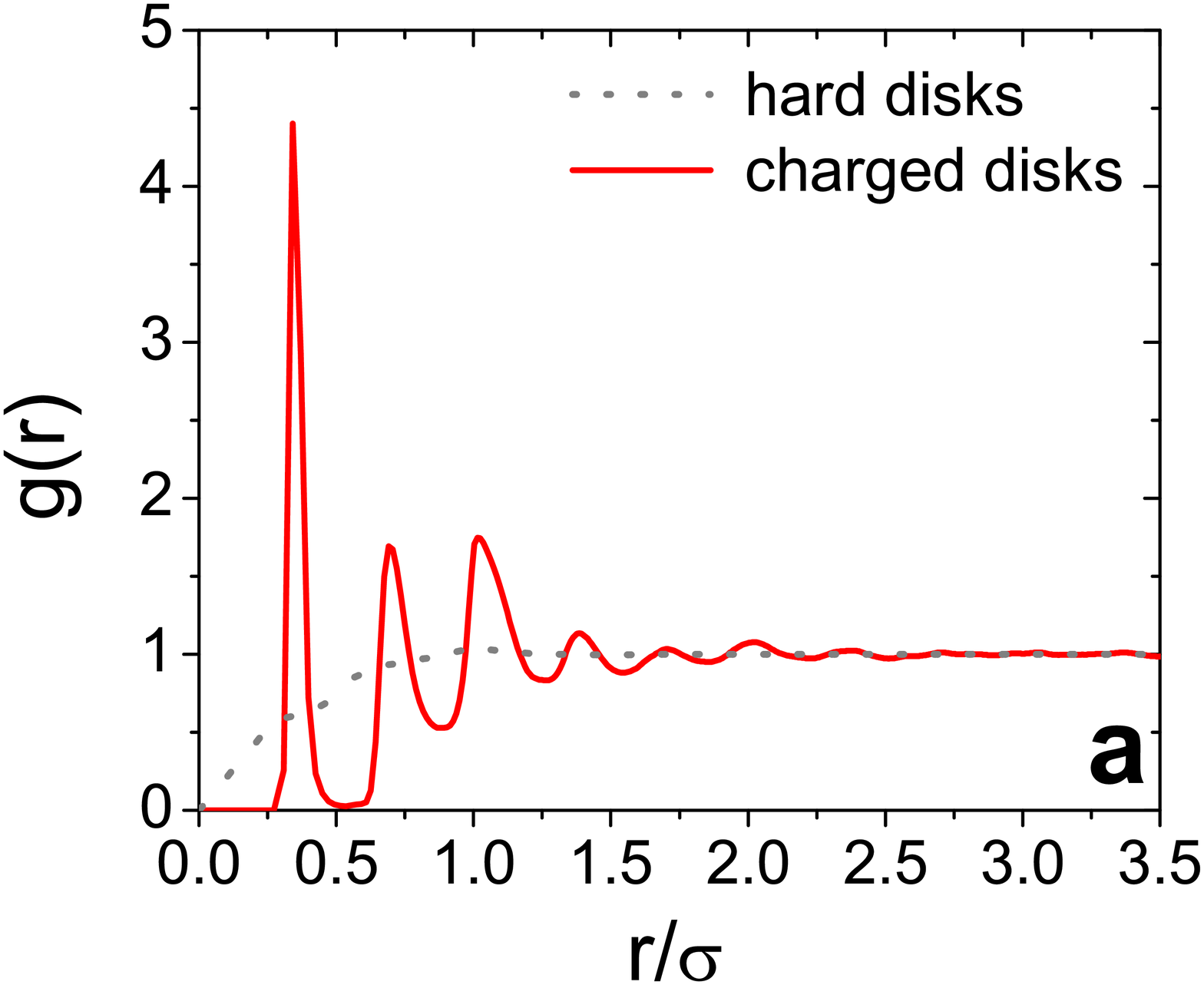}
\includegraphics[scale=0.2]{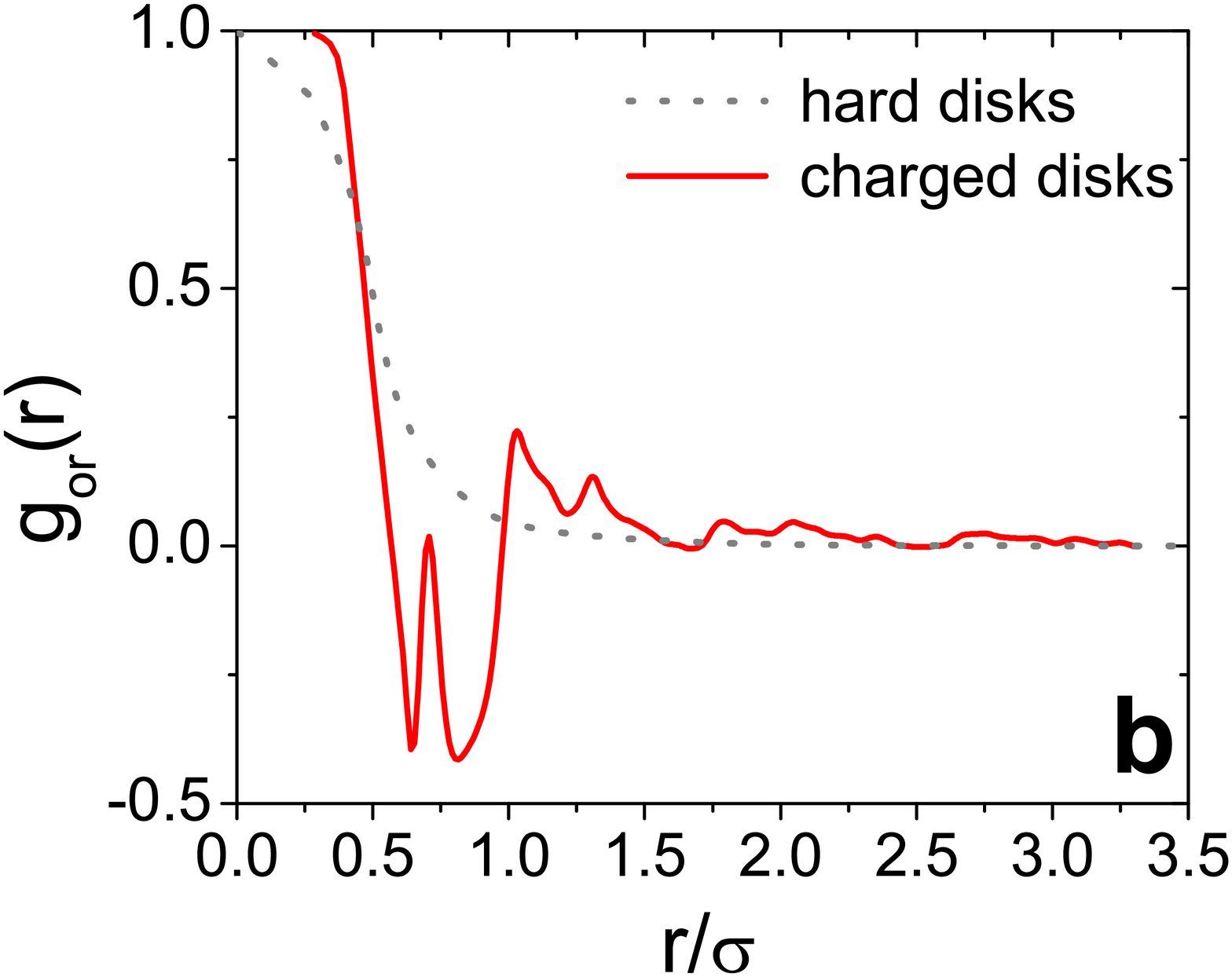}
\includegraphics[scale=0.2]{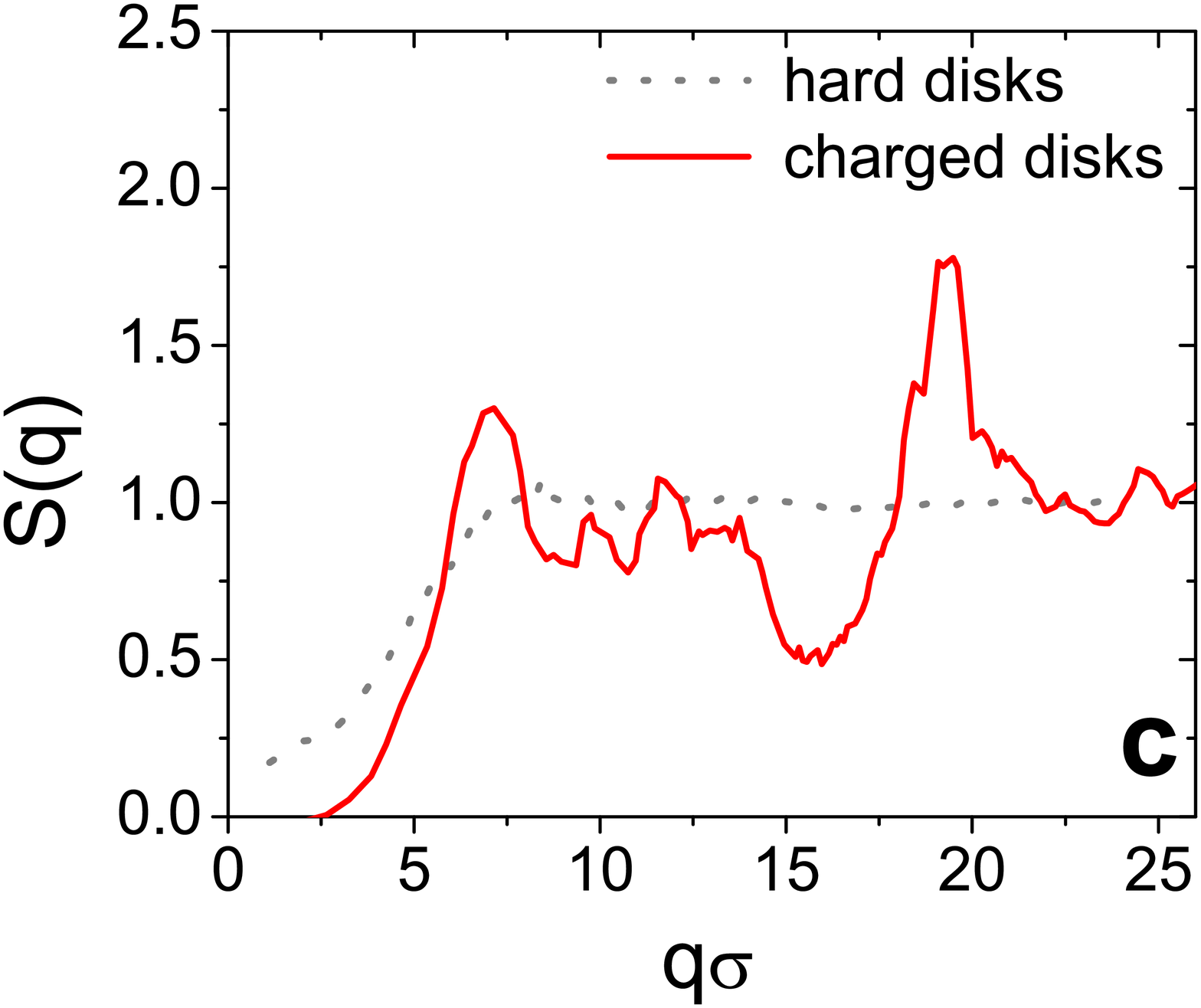}
\includegraphics[scale=0.2]{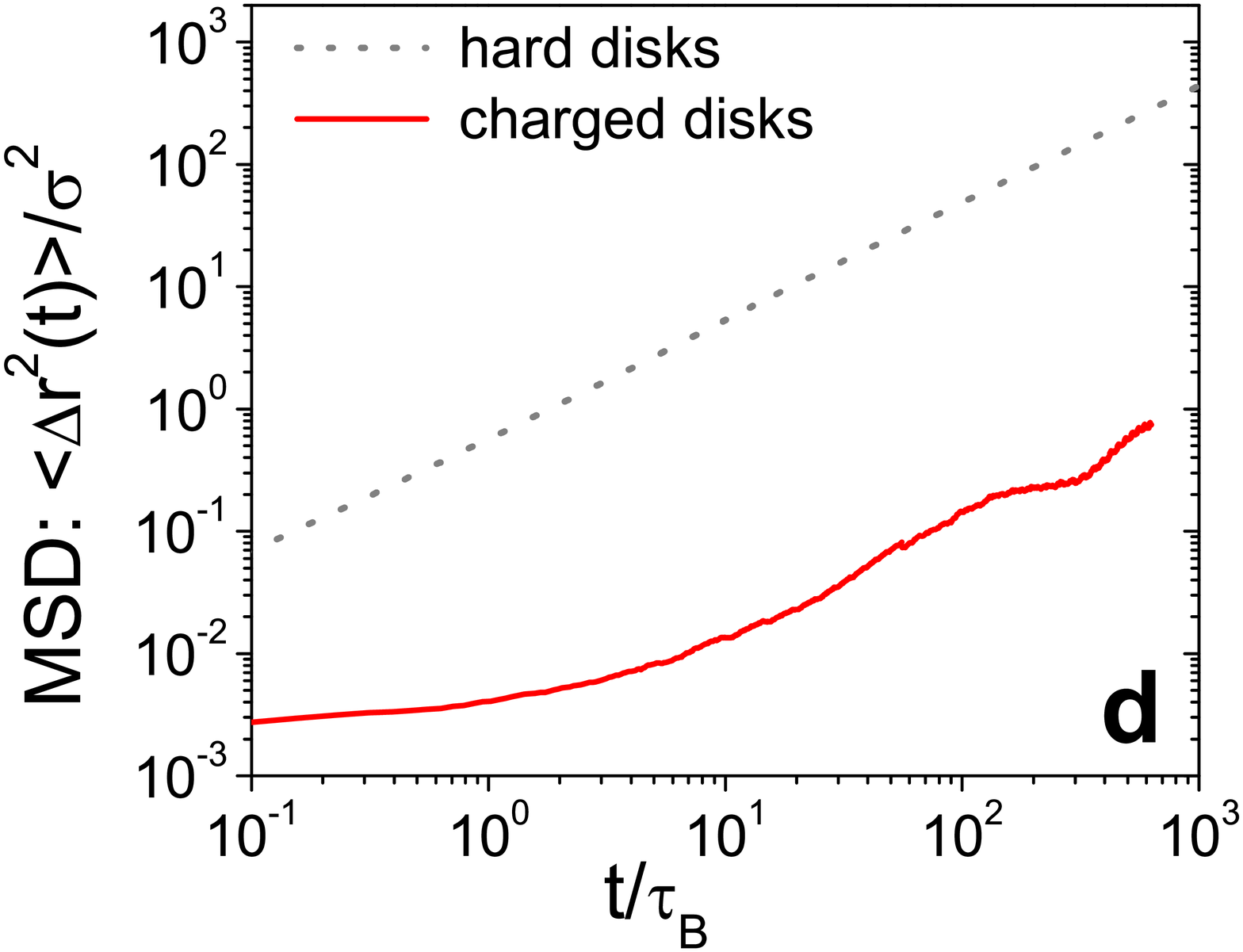}
\includegraphics[scale=0.2]{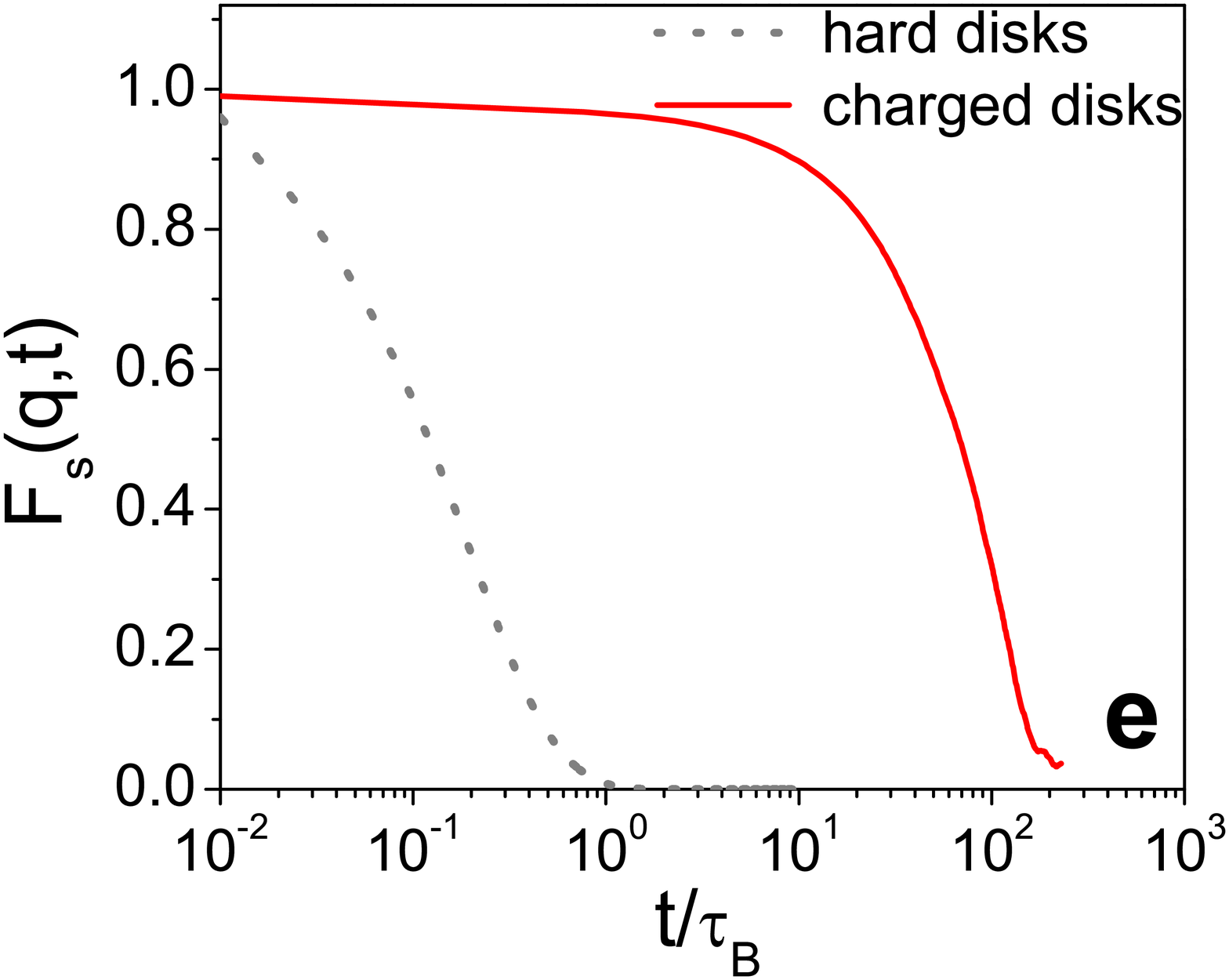}
\includegraphics[scale=0.2]{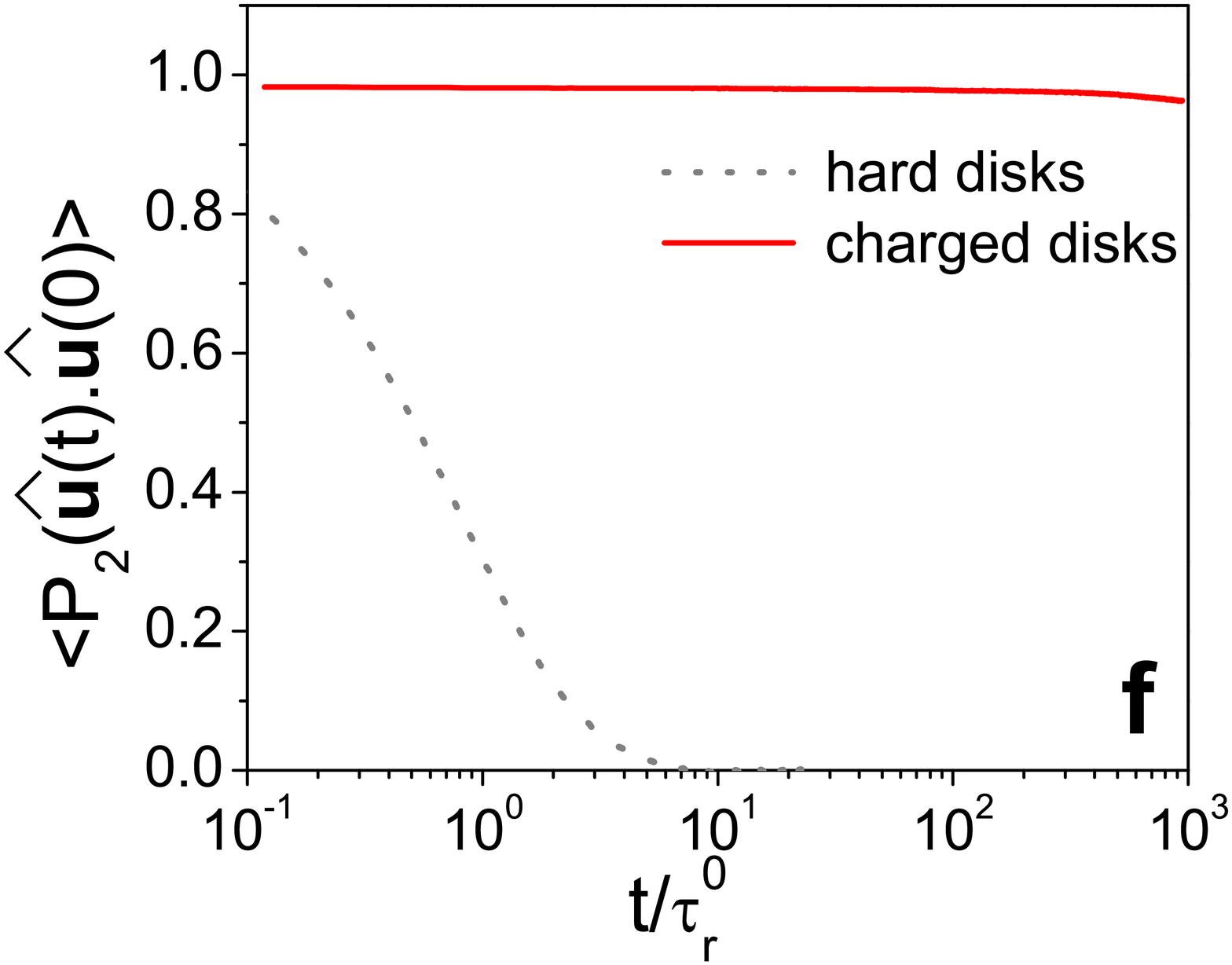}
\caption{ a) The  radial pair distribution function $g(r)$, b) the orientational pair distribution function $g_{or}(r)$ ,  c) structure factor $S(q)$,
 d)  mean squared displacement $\left\langle \Delta r^2(t) \right\rangle$, e) intermediate scattering function $F_s(q,t)$ at $q \sigma=7.1$ f) orientational time correlation
 function $\left\langle P_2(\widehat{u}(t).\widehat{u}(0)) \right\rangle$
shown for  $\rho^*=3$ and $\kappa \sigma=20$, which forms a \emph{random stacks} structure.
}
\label{fig8}
\end{figure}
\item \textbf{Intergrowth texture} \\
Now, we turn to the last structure with no net nematic order parameter, {\it i. e.} the novel structure of \emph{intergrowth texture}. This structure  consists of sets of aligned disks ($0.75 < S < 0.93$) 
interspersed with layers exhibiting anti-nematic order ($-0.45 < S < -0.3$)  \cite{antinem,antinem1}and  both types of layers share the same director.  The intergrowth texture appears at ionic strengths $ 2 < \kappa \sigma < 6$ 
and moderate densities just before orientational disorder-order transition of charged disks in the region where the hard disks are already in the nematic phase. In Fig. \ref{fig9}a, \ref{fig9}b  and 
\ref{fig9}c, we have shown the positional and orientational correlation functions as well as structure 
factor for an example of such a texture  at $\rho^*=5$ and $\kappa \sigma=4$. As can be noticed from these figures, there is no evidence of long-ranged positional and orientational order in this system, although, we can distinguish several overlapping consecutive peaks in $g(r)$ which mark some degree of local ordering. The negative value of $g_{or}$ at $r \approx 0.55$ shows that at such distances particles organize with relative T-shaped configurations.
\begin{figure}[h!]
\includegraphics[scale=0.2]{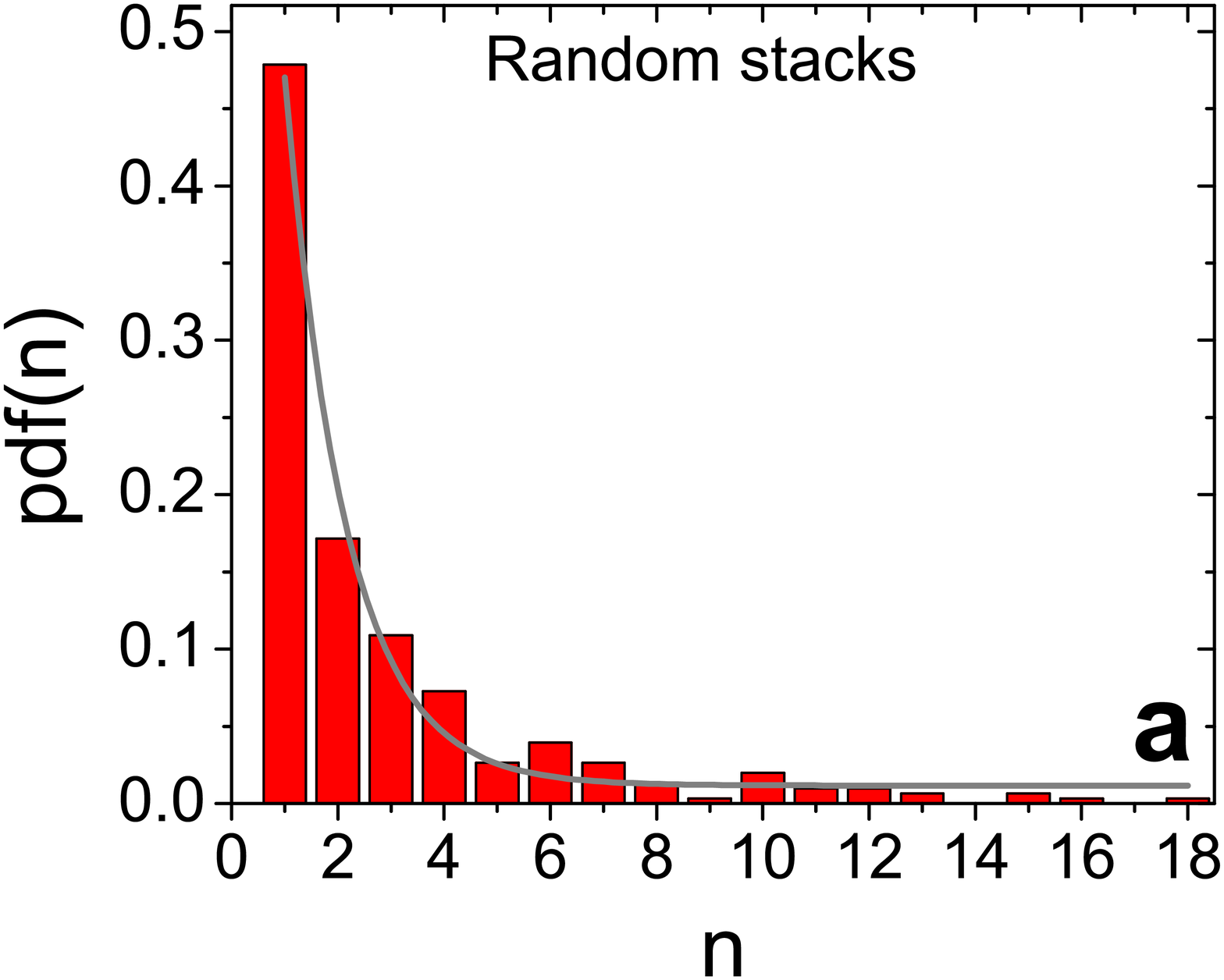}
\includegraphics[scale=0.2]{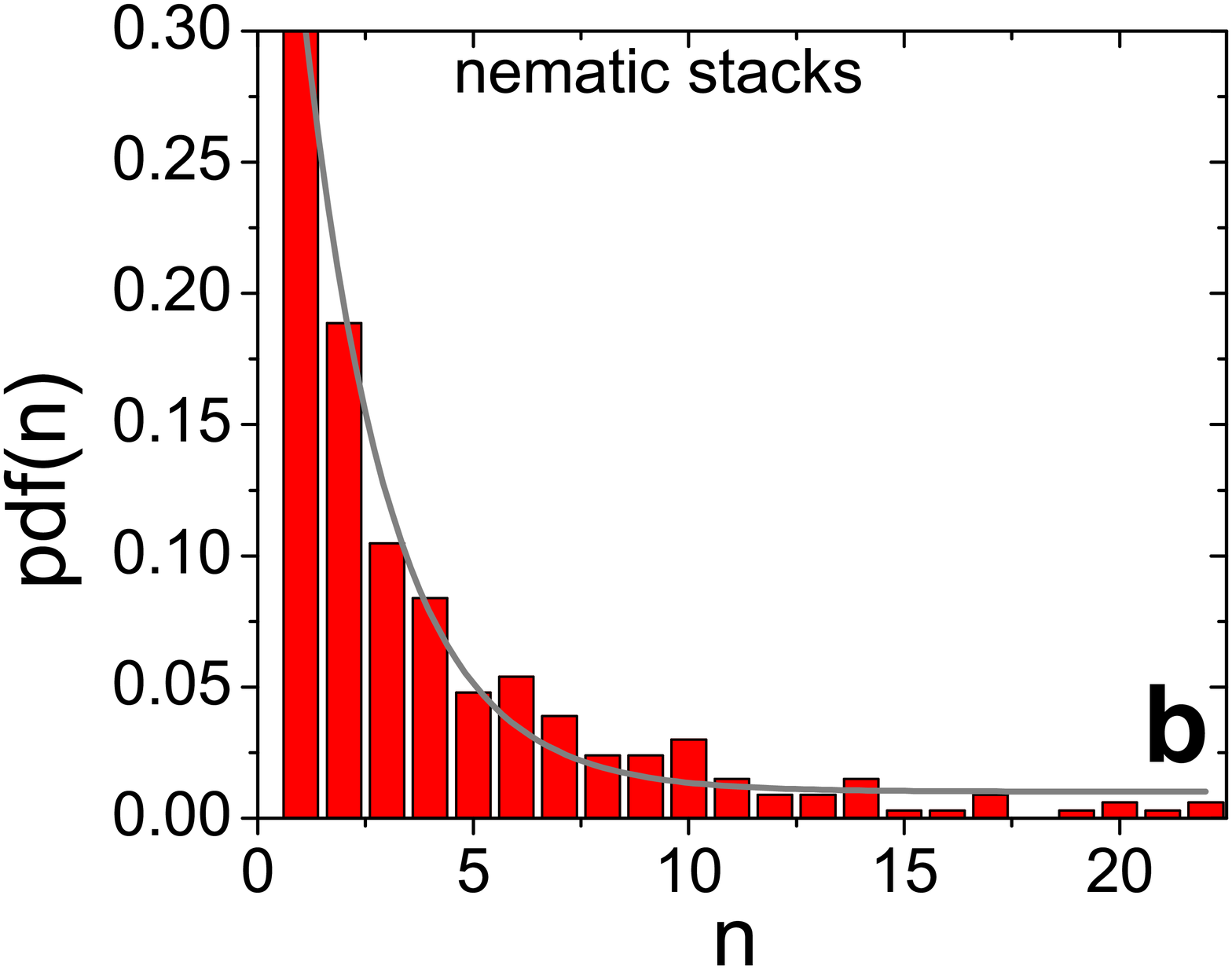}
\includegraphics[scale=0.2]{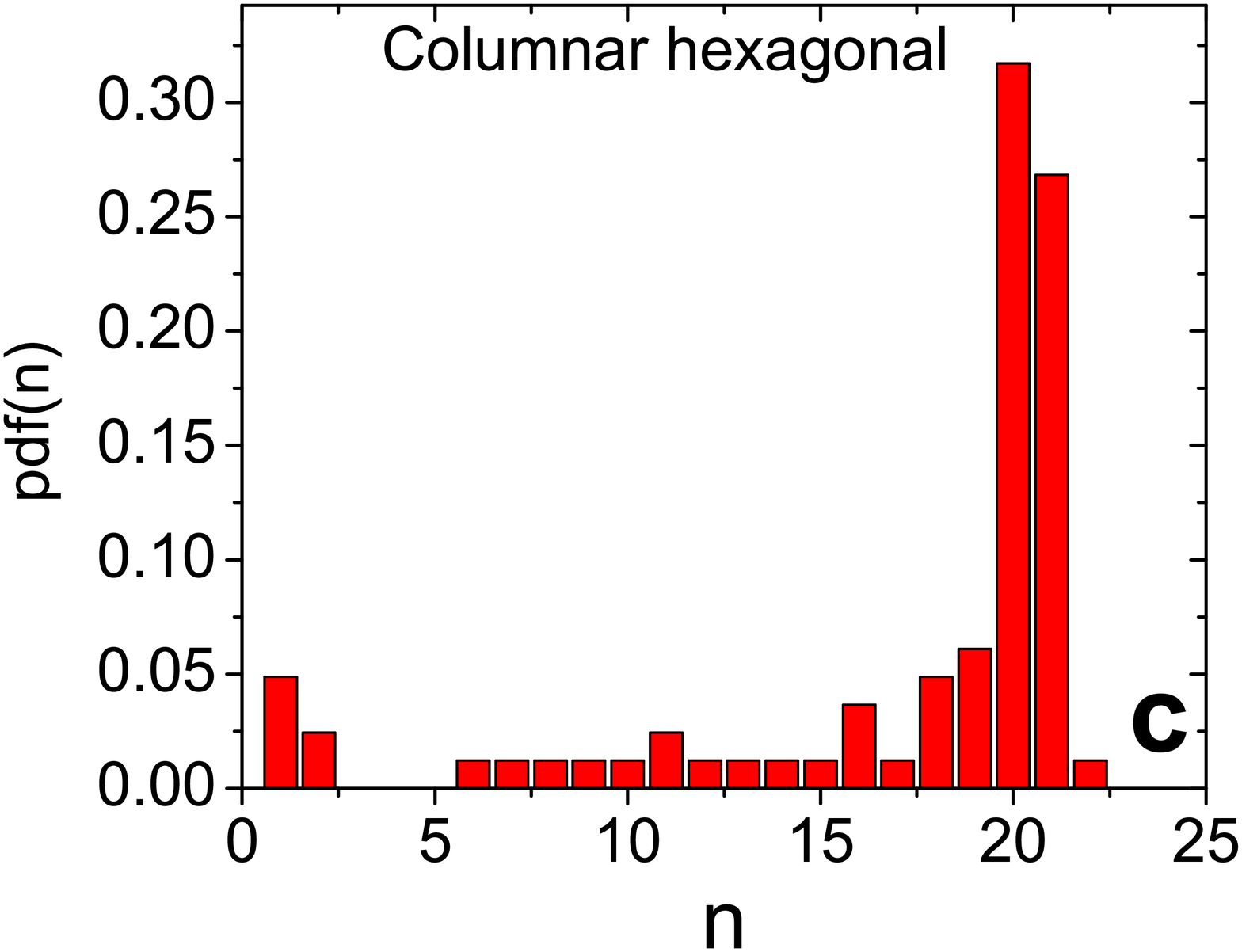}
\caption{ The probability distribution function of number of stacks or columns for a) random stacks structure $\rho^*=3$
and $\kappa \sigma=20$, with average stack size 2.85 b) nematic stacks $\rho^*=5$ and $\kappa \sigma=20$  with average column length 4.1 c)  columnar hexagonal structure
   $\rho^*=8$ and $\kappa \sigma=10$. The lines in panels a) and b) show the fit of the probability distribution function with an exponential form.}
\label{pdf}
\end{figure}
To distinguish between nematic and anti-nematic layers, we identified the sets of aligned disks by finding disks whose orientations were correlated, {\it i.e.}  $ \mathrm{\cos }(\theta_{ij})> 0.8$, where 
$\theta_{ij}$ is the angle between orientations of particles $i$ and $j$. Then we calculated the nematic tensor of the set of aligned disks. The director obtained from the nematic tensor  was used in a second step as a criterion to separate disks  which made an angle $\theta $ with the director so that  $\mathrm{\cos} (\theta)> 0.8$. We repeated this procedure iteratively, until 
the vector obtained for the director from two consequent steps converged. After separating the disks with nematic ordering, we calculated the nematic tensor for the set of remaining disks 
and we found out that these disks exhibit an anti-nematic ordering ($-0.45 < S < -0.3$) and share the same director as the nematic set of disks.   The well-defined separation between the 
nematic and anti-nematic layers is justified by examining  the angular distribution of disks  with respect to the common director  as plotted in Fig. \ref{fig9}f. Here, we find that 
the disks are either aligned with the director (nematic) or their orientations  are perpendicular to the director. This very original structure is indeed a manifestation  of the 
anisotropic nature of  screened electrostatic potential which disfavors coplanar configurations of disks. The disks, instead, opt out for anti-nematic ordering.

\begin{figure}[h]
\includegraphics[scale=0.2]{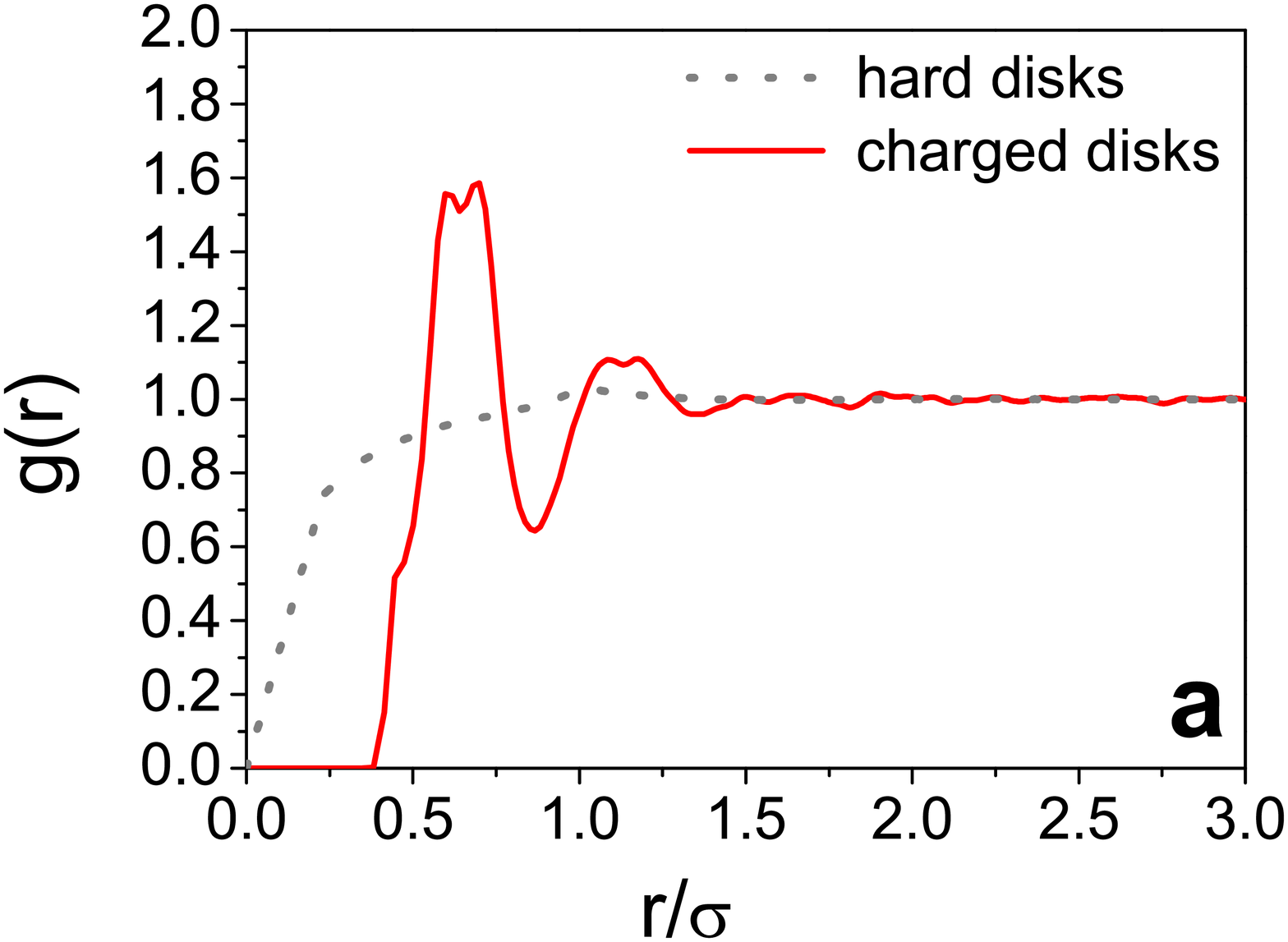}
\includegraphics[scale=0.2]{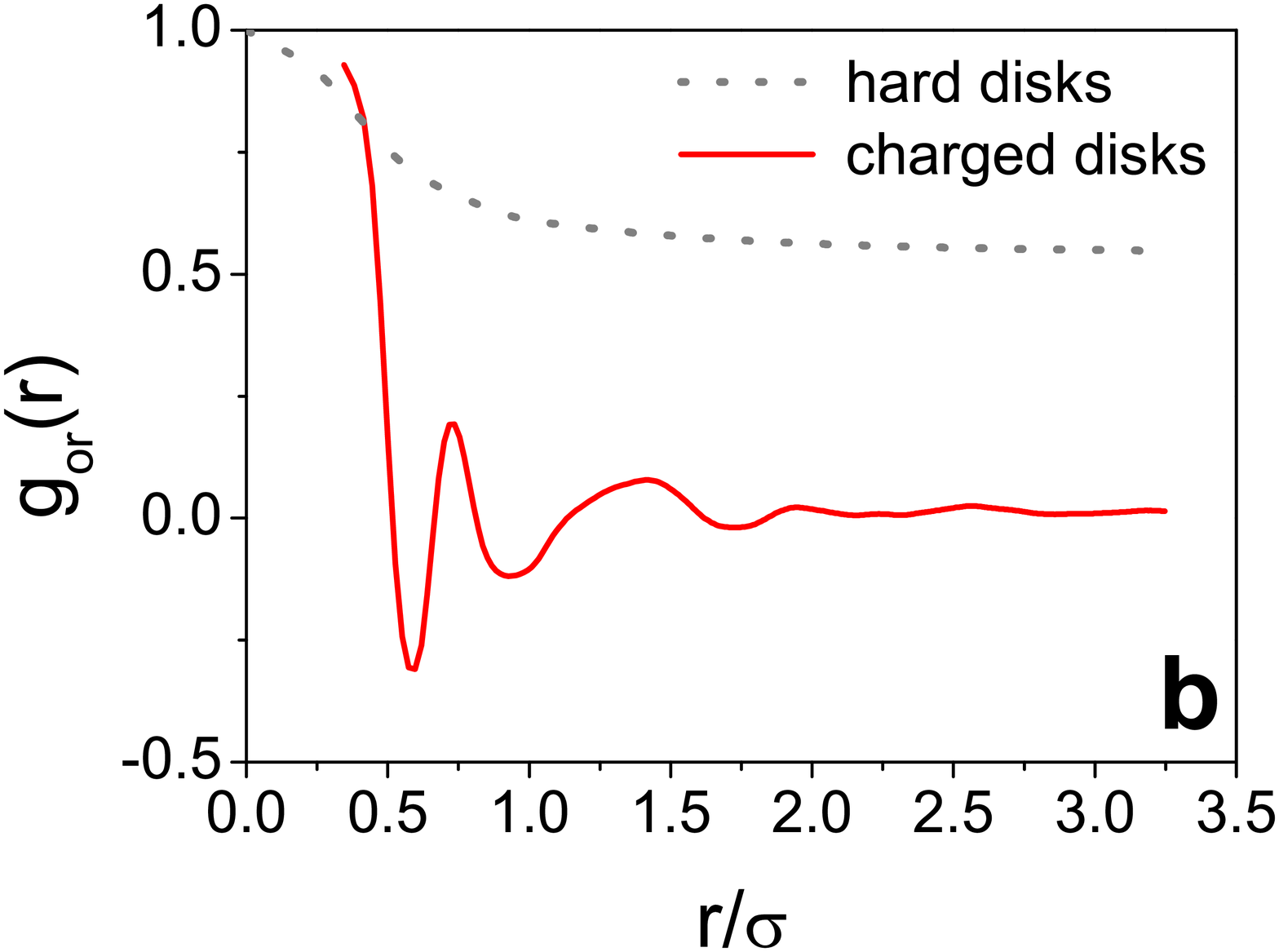}
\includegraphics[scale=0.2]{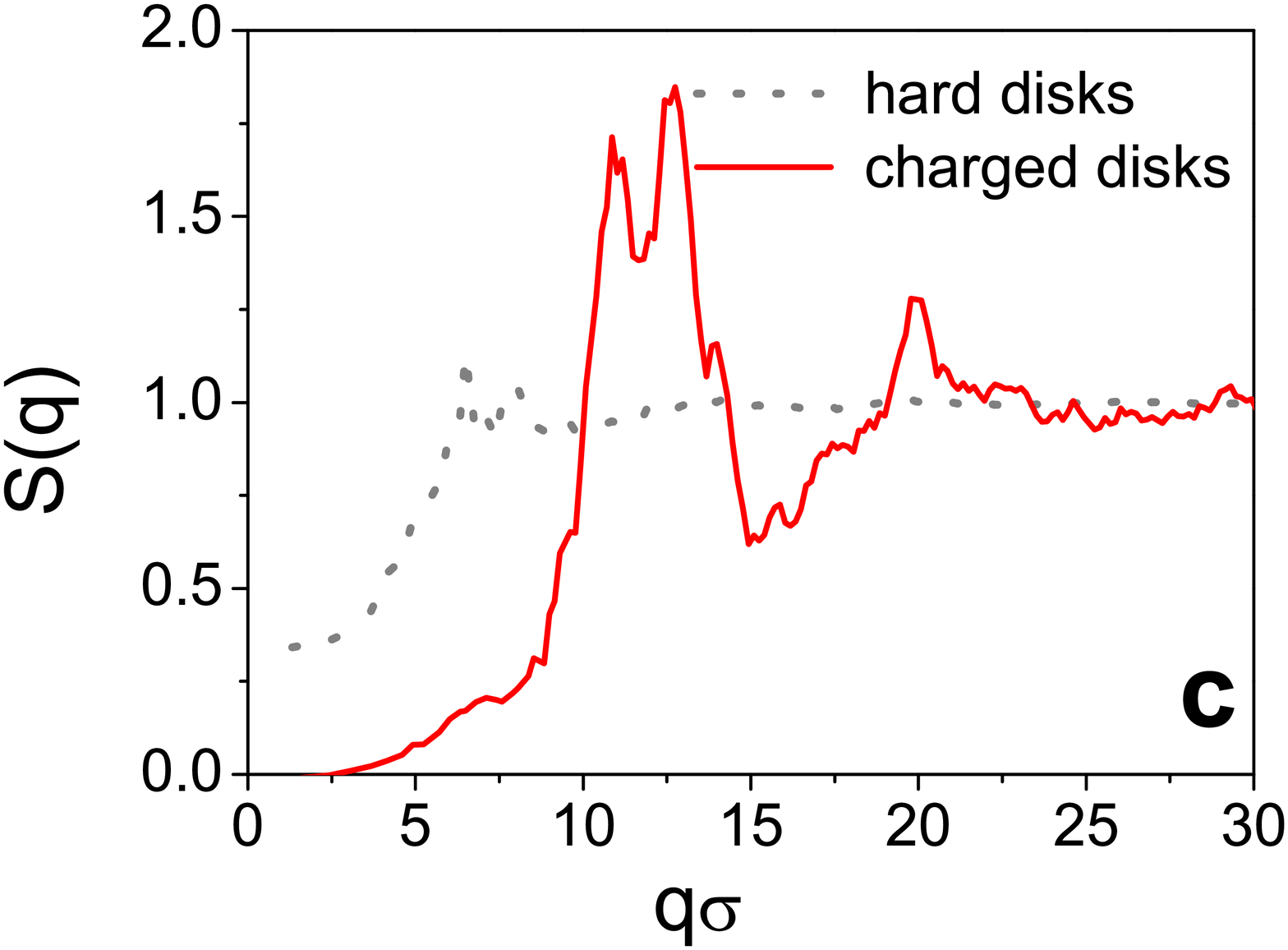}
\includegraphics[scale=0.2]{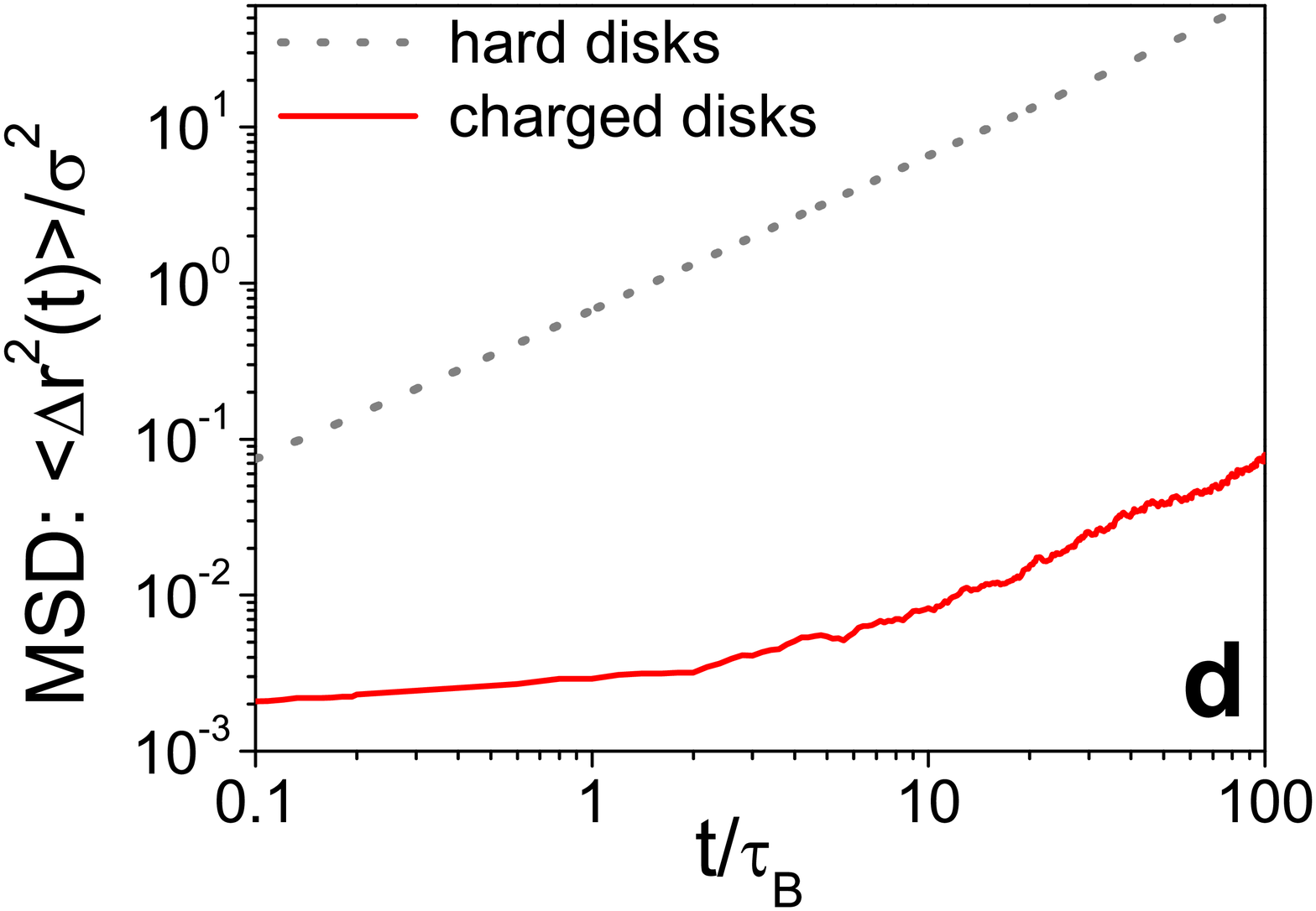}
\includegraphics[scale=0.2]{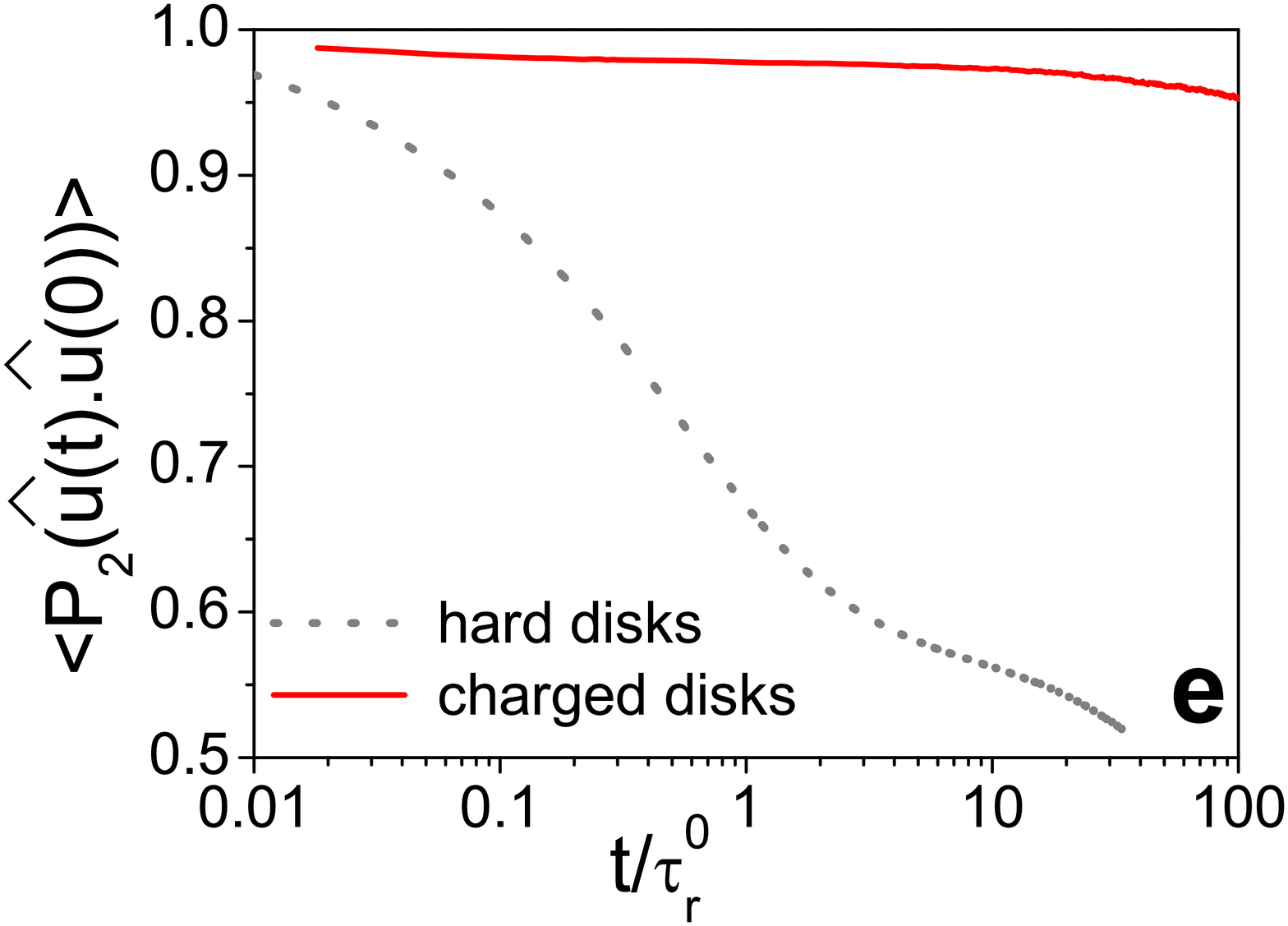}
\includegraphics[scale=0.2]{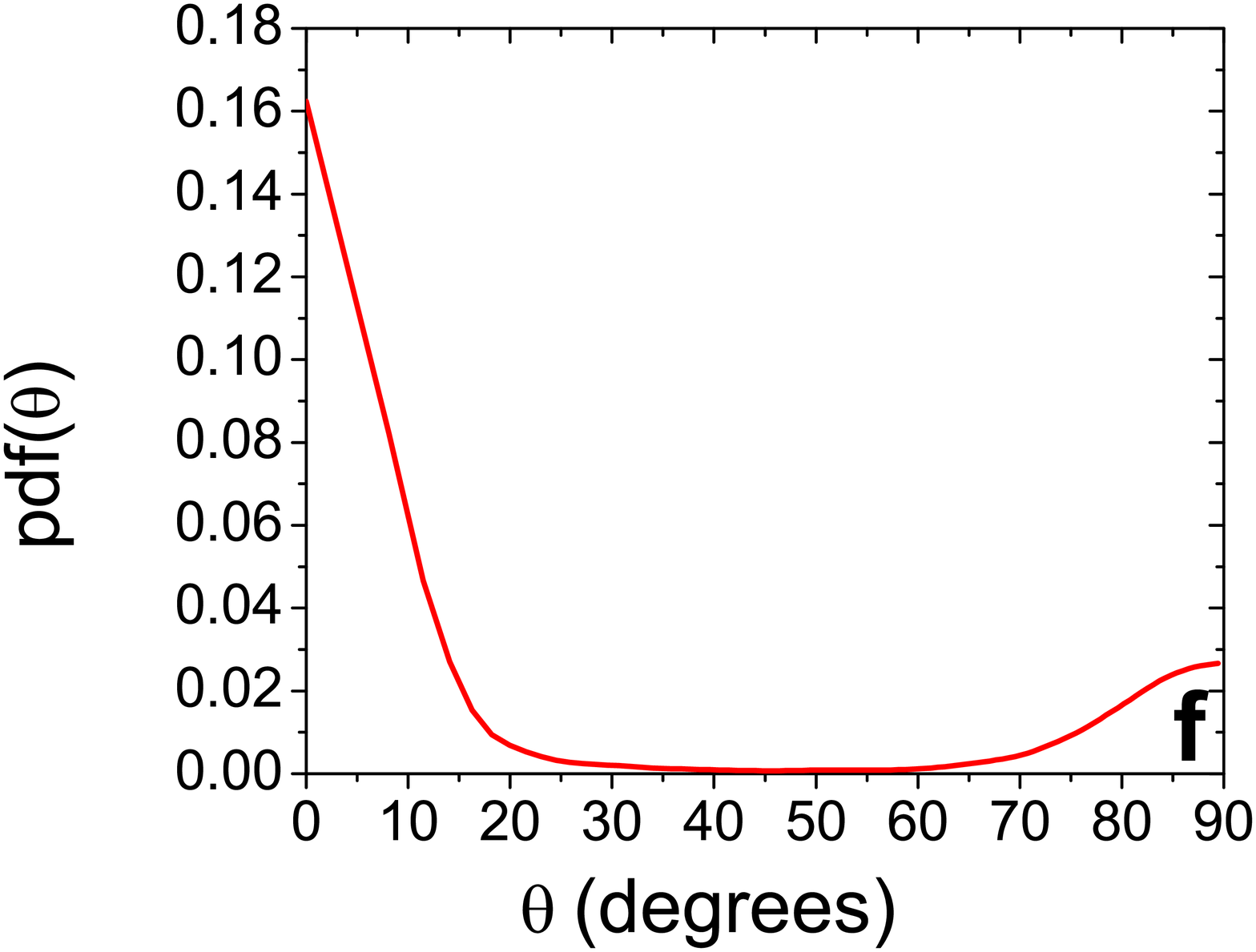}
\caption{ a) The  radial pair distribution function $g(r)$, b) the orientational pair distribution function $g_{or}(r)$ ,  c) the structure factor $S(q)$,
 d) mean squared displacement $\left\langle \Delta r^2(t) \right\rangle $,  e) orientational time correlation function
  $\left\langle P_2(\widehat{u}(t).\widehat{u}(0)) \right\rangle$ shown for  $\rho^*=5$ and $\kappa \sigma=4$, an \emph{intergrowth texture}. f) Angular probability distribution function  
  pdf$(\theta)$, for $\kappa \sigma=4, \rho^*=4.5$ where $\theta$ is the angle of the orientation vector of each disk with respect to the director of the nematic region. }
\label{fig9}
\end{figure}

The dynamics of charged disks in an example of intergrowth texture  are shown in Figs. \ref{fig9}d and \ref{fig9}e.  We find that both 
 translational and rotational dynamics  are significantly decreased  pointing to the effect of electrostatic interactions in  relatively dense suspensions.

\item \textbf{Nematic stacks} \\
Having discussed the configurations with vanishing order parameters,  we now focus on the orientationally ordered phases.  Nematic stacks is the simplest orientationally ordered state with  $S>0.4$ that appears at  moderate to high ionic strengths for densities in the range $ 5<\rho^* <6 $.
The non-zero value of $S$ is  also reflected  in $g_{or}(r)$  which shows a non-decaying plateau for large distances as presented in
 Fig. \ref{fig10}b. However, the structure of the observed nematic phase  is different from  that of hard disks, as most of
 the disks are found to be in stack-like arrangements; see Fig. \ref{phase}.  This difference in structure from that of hard disks is also reflected in positional pair correlation function Fig. \ref{fig10}a and structure factor Fig. \ref{fig10}c which are sharply peaked at short distances and mark the presence of short-ranged stack-like spatial correlations. The average stacking distance  obtained from $g(r)$  for the example shown at $\kappa \sigma=20$ and $\rho^*=5$ is $0.25 \sigma$. 
 We have examined the positional pair correlation functions in directions  parallel and perpendicular to the nematic director as   presented in Figs. \ref{fig10}d and e. We notice that there is no long-range positional order in any of the directions, although a local ordering within the first few  neighbors exists, an evidence of the presence of stacks. Furthermore, similarly to the case of random stacks,  we have extracted the distribution function of stack sizes as presented in Fig. \ref{pdf}b. Here also, the number of stacks
 decreases exponentially with size; however, the average size of stacks  is larger in the nematic phase than in the isotropic state of
 random stacks.

\begin{figure}[t]
\includegraphics[scale=0.2]{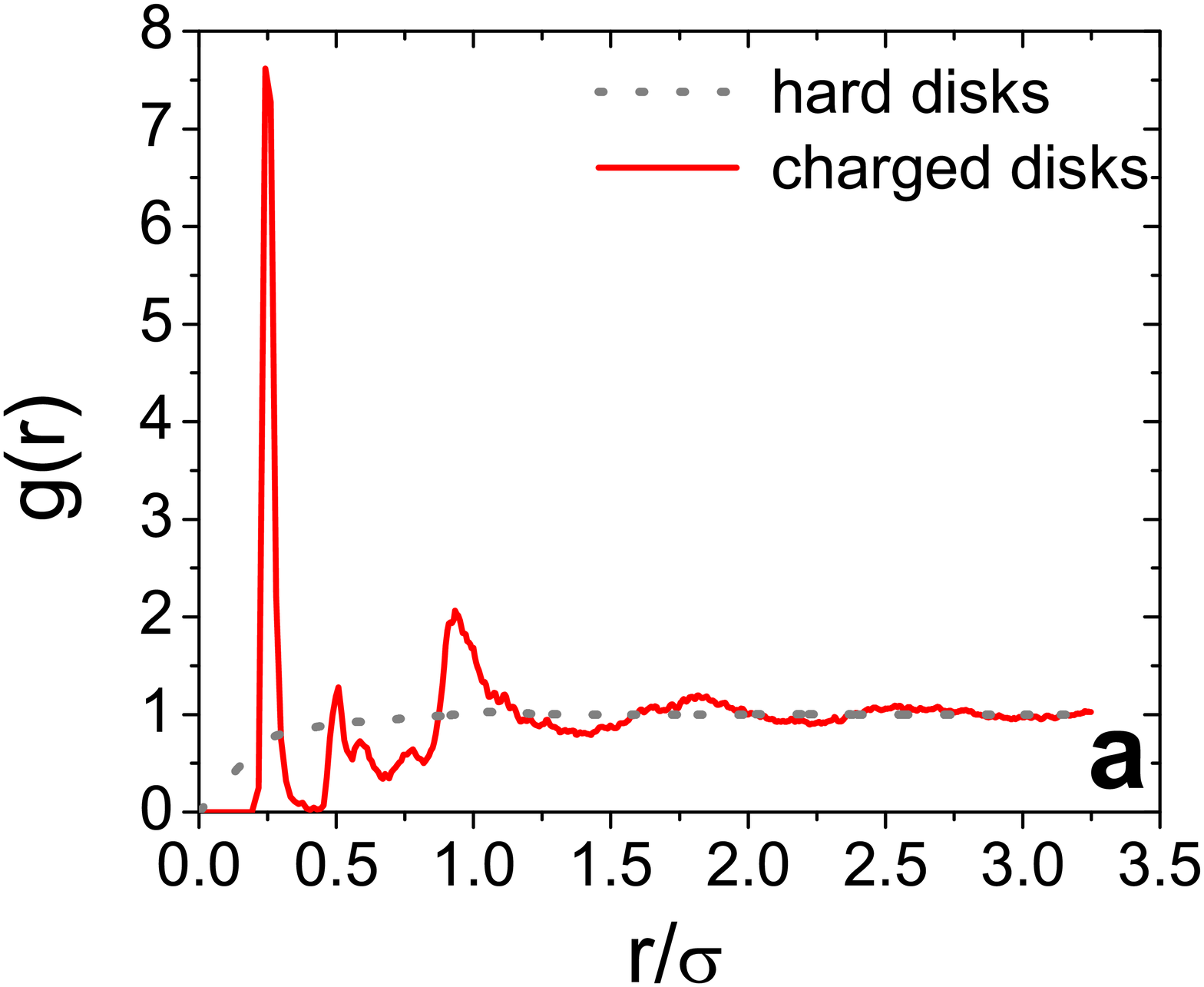}
\includegraphics[scale=0.2]{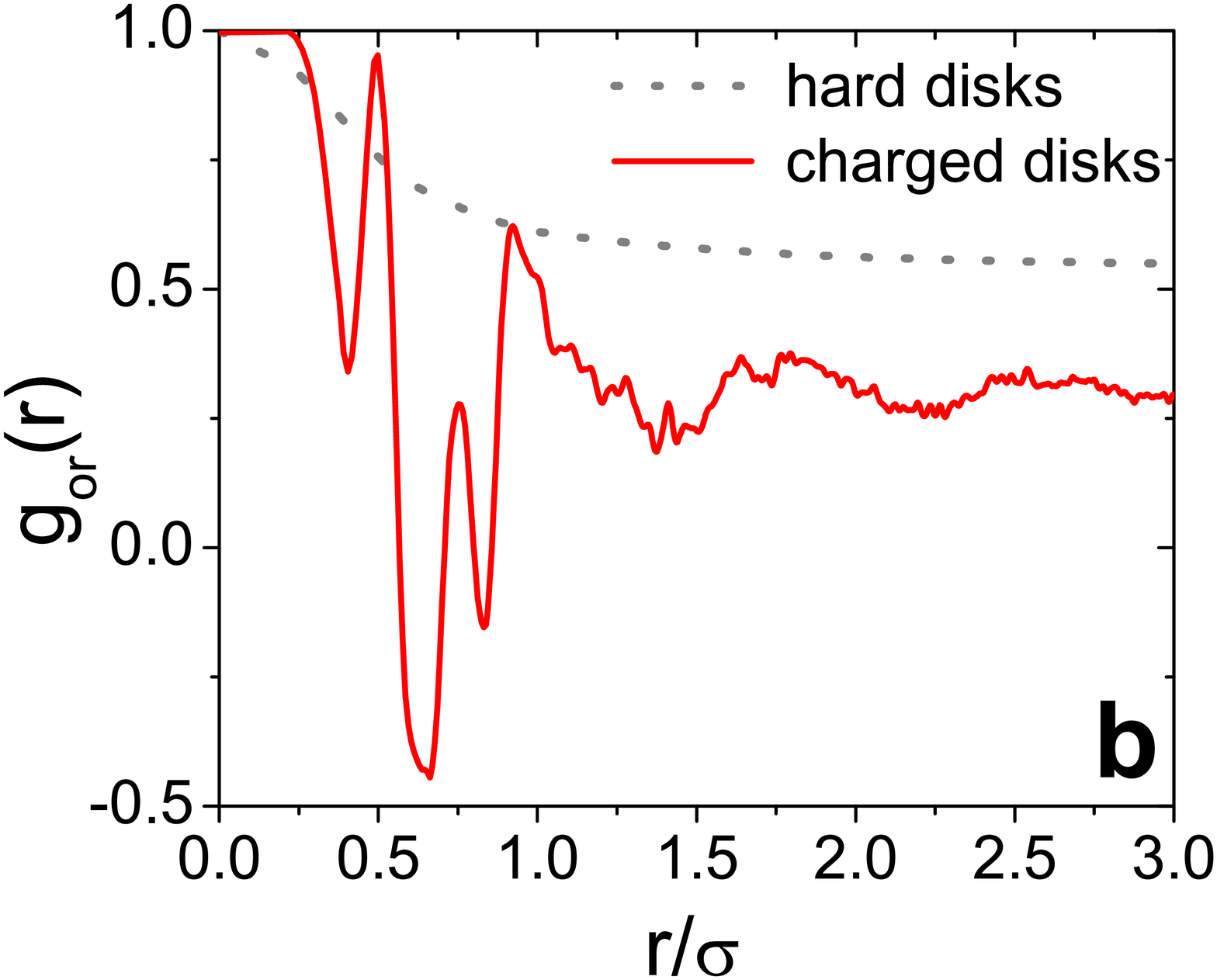}
\includegraphics[scale=0.2]{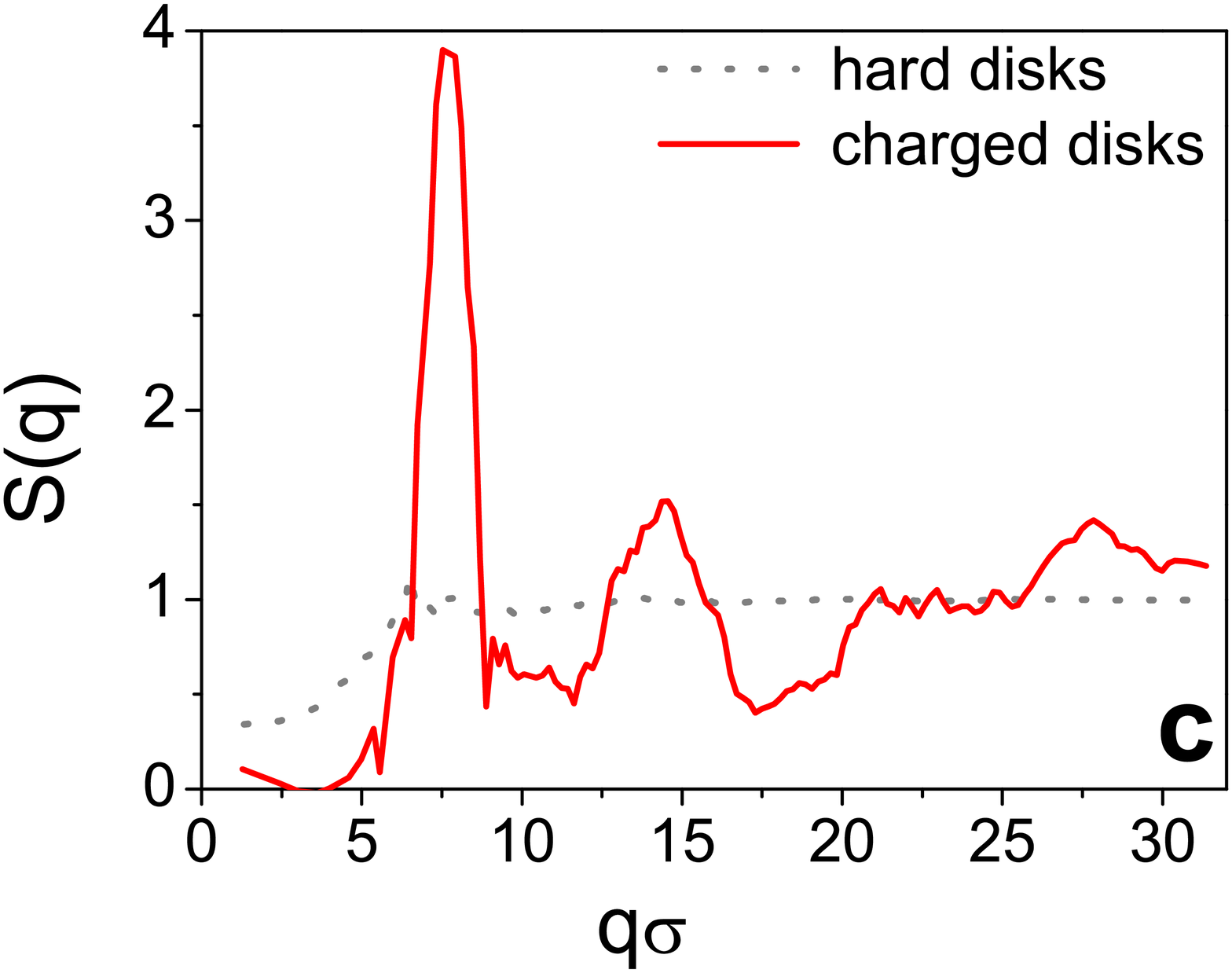}
\includegraphics[scale=0.2]{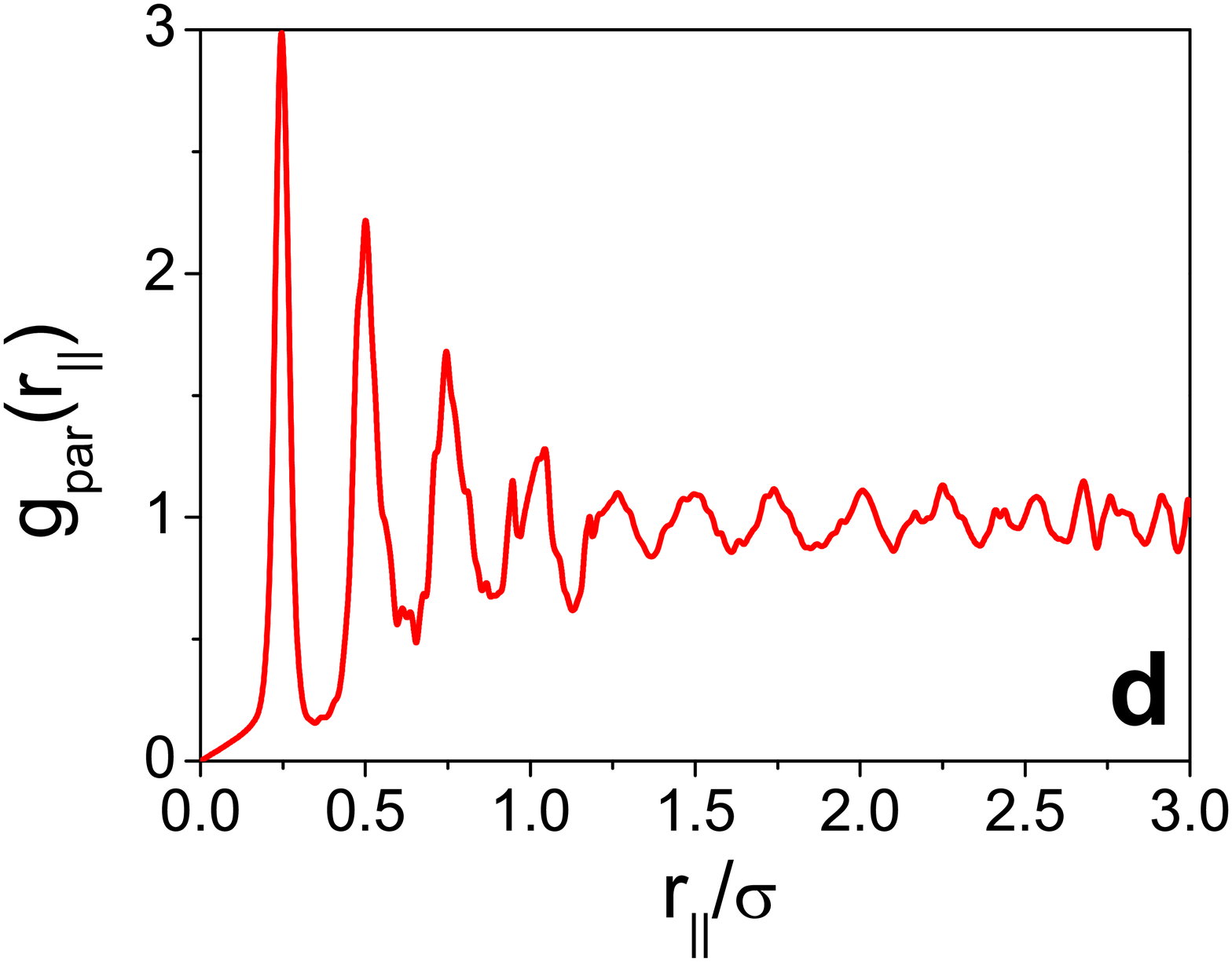}
\includegraphics[scale=0.2]{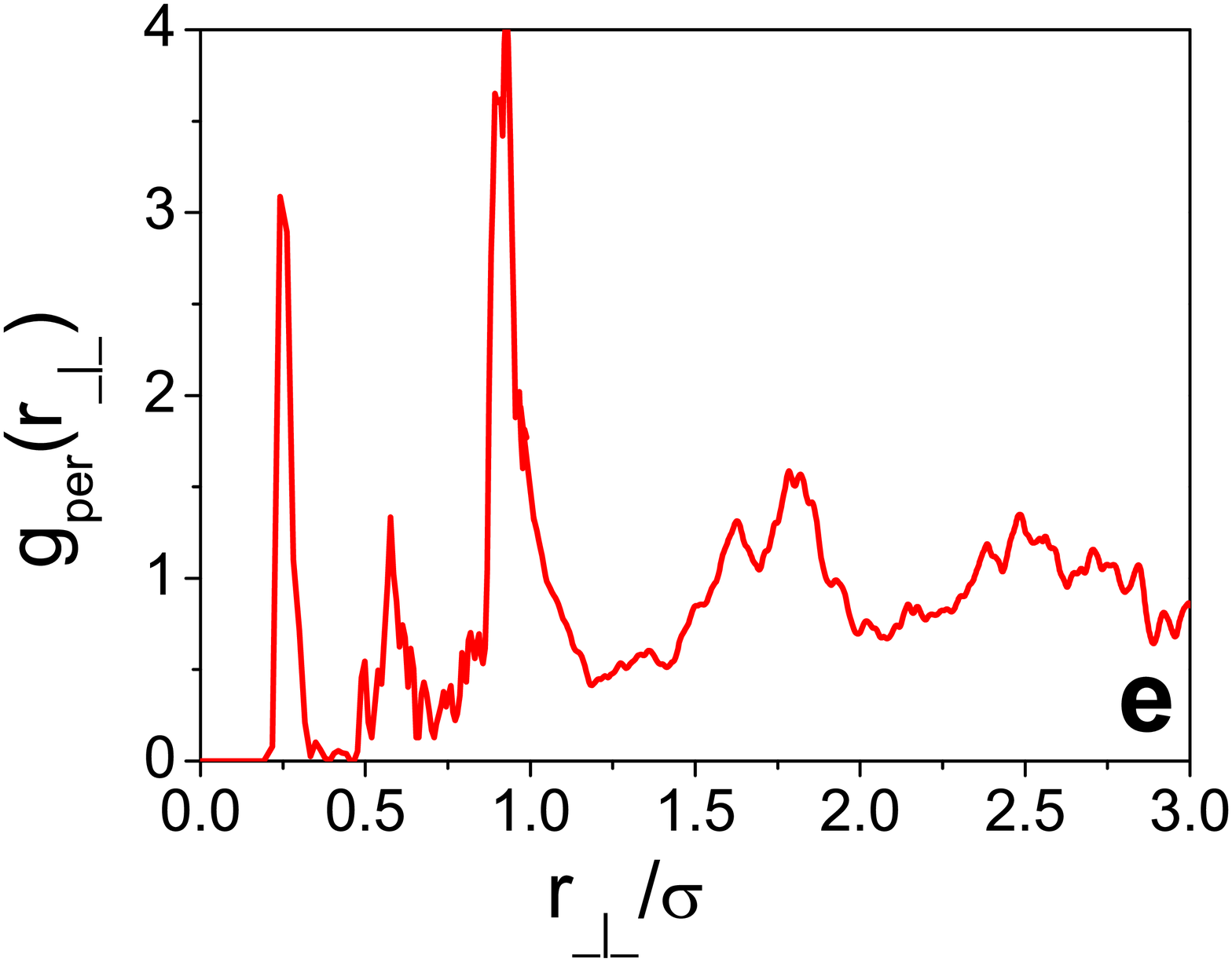}
\caption{ a) The  radial pair distribution function $g(r)$, b) the orientational pair distribution function $g_{or}(r)$, c) the structure factor $S(q)$,
 d) spatial correlation function in the direction parallel to the director e) spatial correlation function in the direction perpendicular to the director
shown for  $\rho^*=5$ and $\kappa \sigma=20$, corresponding to \emph{nematic stacks} structure.
}
\label{fig10}
\end{figure}

\item \textbf{Columnar hexagonal} \\
For larger densities, typically $ \rho^* > 6$, both orientational and positional
orders of the particles increase. Stacks of aligned disks become more organized and form  regular columns
 arranged on a hexagonal lattice (see Fig. \ref{phase}).  Our qualitative observations have been confirmed by investigating both positional and orientational pair correlation functions as presented in Fig. \ref{fig11} for an  example at $\kappa \sigma=10, \quad \rho^*=8$. The large value of plateau in $g_{or}$ reflects the large degree of orientational order at such high densities. The radial pair correlation function $g(r)$ has a strong peak at very small distances and integration over this peaks gives un an average number of nearest neighbors $N_{nn}= 2$ which is consistent with columnar arrangement of disks at small separations. 
 To examine the positional order of particles more closely, we inspect positional correlations in directions parallel and perpendicular to the director.  The pair correlation function in  direction parallel  to the director   $g_{par}$   in Fig. \ref{fig11}d shows almost equally spaced  peaks which  broaden with distance. This behavior represents a short-ranged liquid-like ordering of disks in the columns. The average distance between the disks in a column, i.e. intracolumnar spacing, is given by the position of the first peak, here $0.25 \sigma$.    
 
 Although the disks have  liquid-like order along the columns, i.e., in the parallel direction, they exhibit a strong crystalline order in the perpendicular directions as evidenced by the long-ranged positional order in $g_{per}$. Indeed, the positions of the successive peaks in $g_{per}$ are related as $a, a\sqrt{3}, 2a, a\sqrt{7},3a$, where $a$ is the lattice
 spacing of a 2D  hexagonal lattice. In the typical example, $\rho^*=8$
 and $\kappa \sigma=10$, shown in Fig. \ref{fig11}, $a=0.73 \sigma$
which is smaller than one particle diameter, indicating that the disks are  interdigitated, similar to the observed columnar
structures for hard  spherocylinders \cite{Marechal}. The hexagonal
nature of the  lattice is also inspected by computing
the hexagonal bond-orientational order parameter $q_6$, (see Appendix B for its  definition). For the given example,
we obtained  $q_6=0.97$.

%
\begin{figure}[h!]
\includegraphics[scale=0.2]{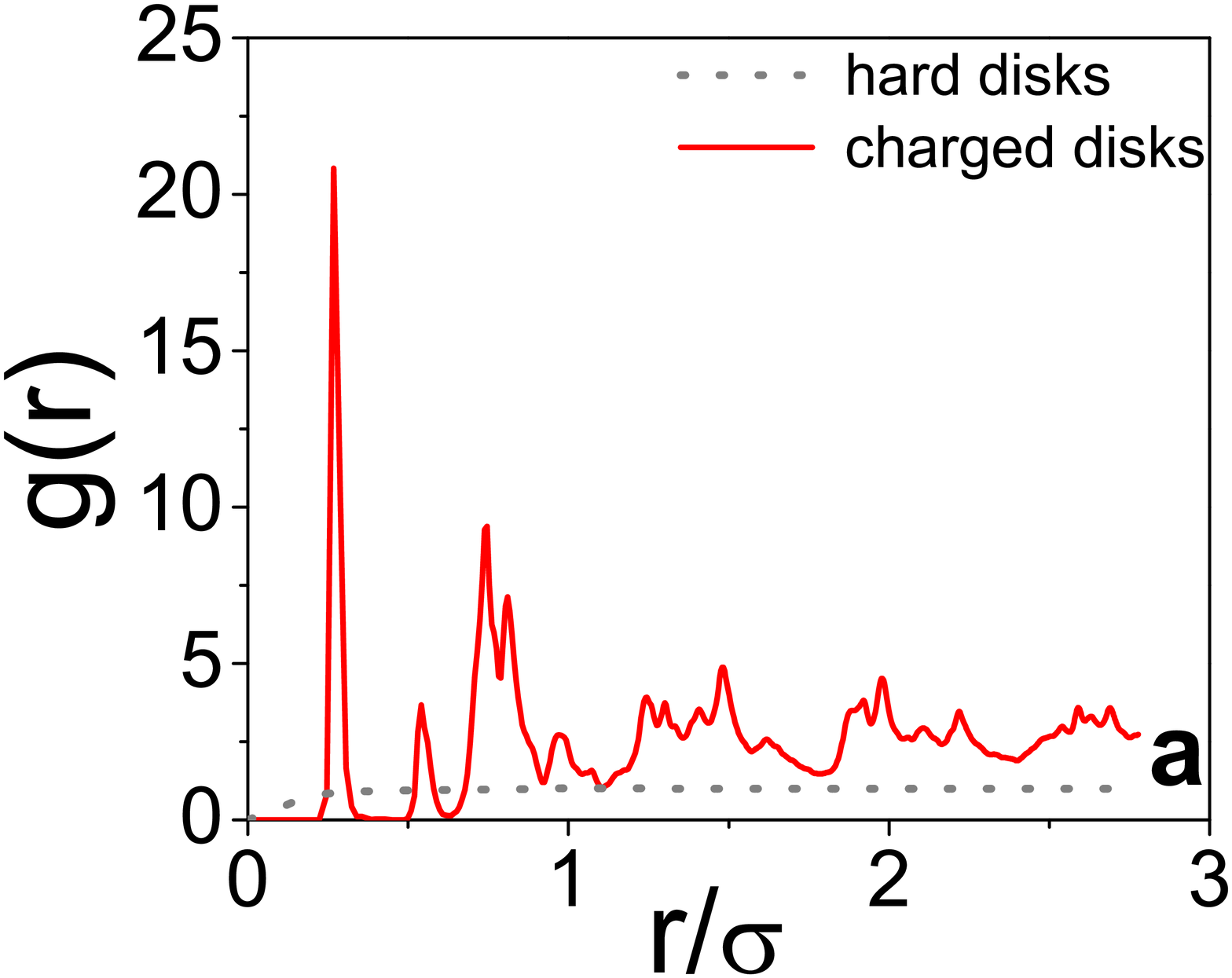}
\includegraphics[scale=0.2]{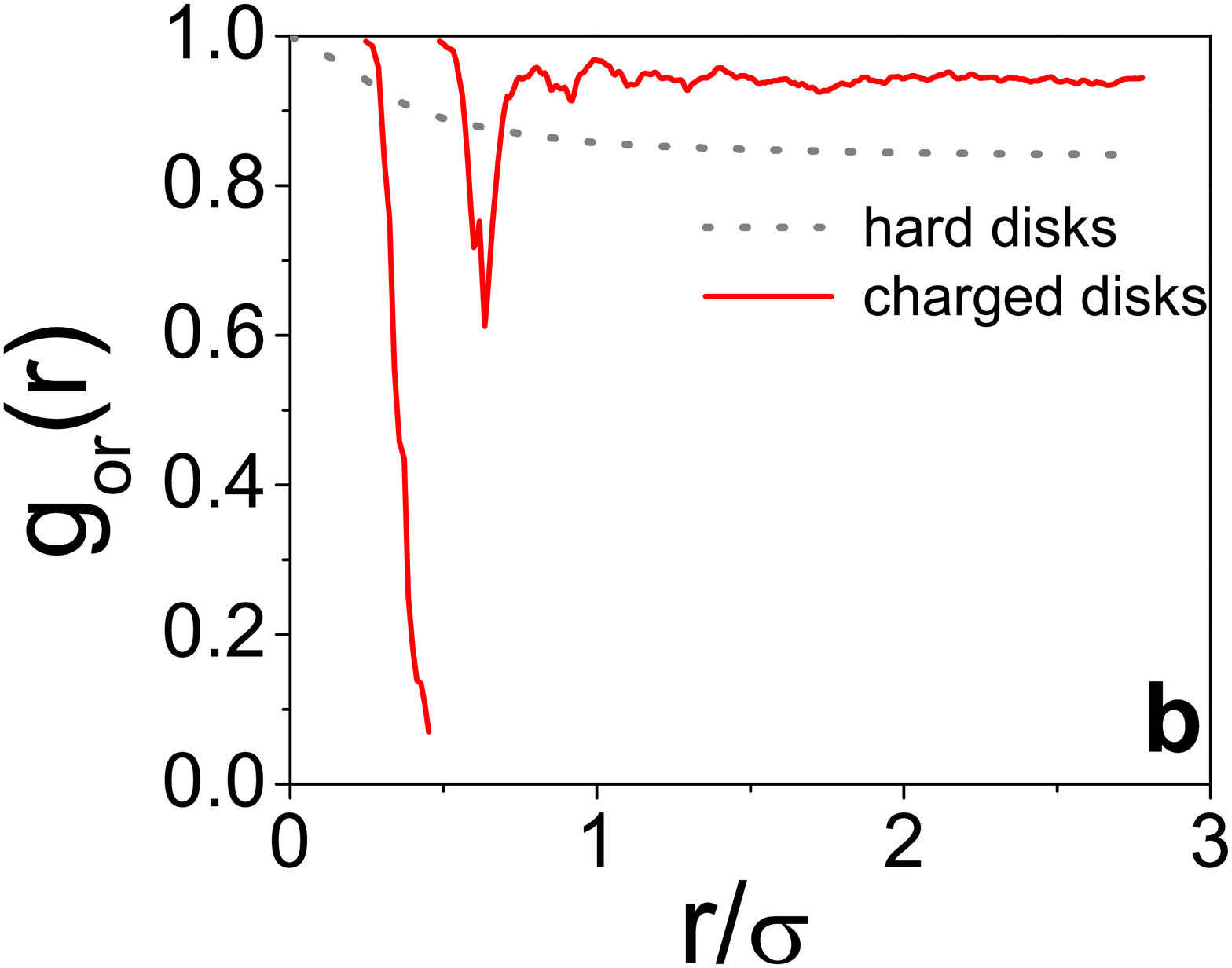}
\includegraphics[scale=0.2]{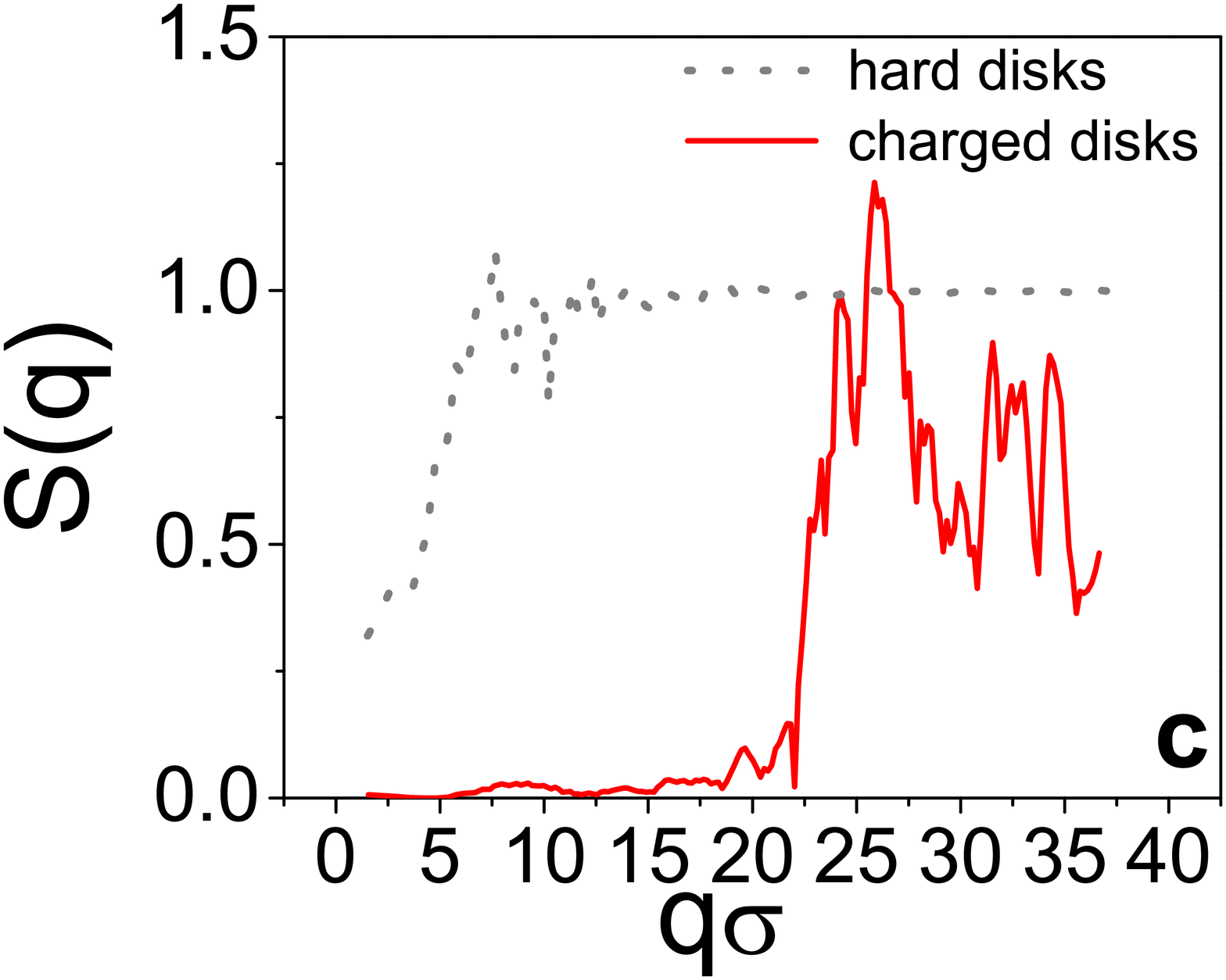}
\includegraphics[scale=0.2]{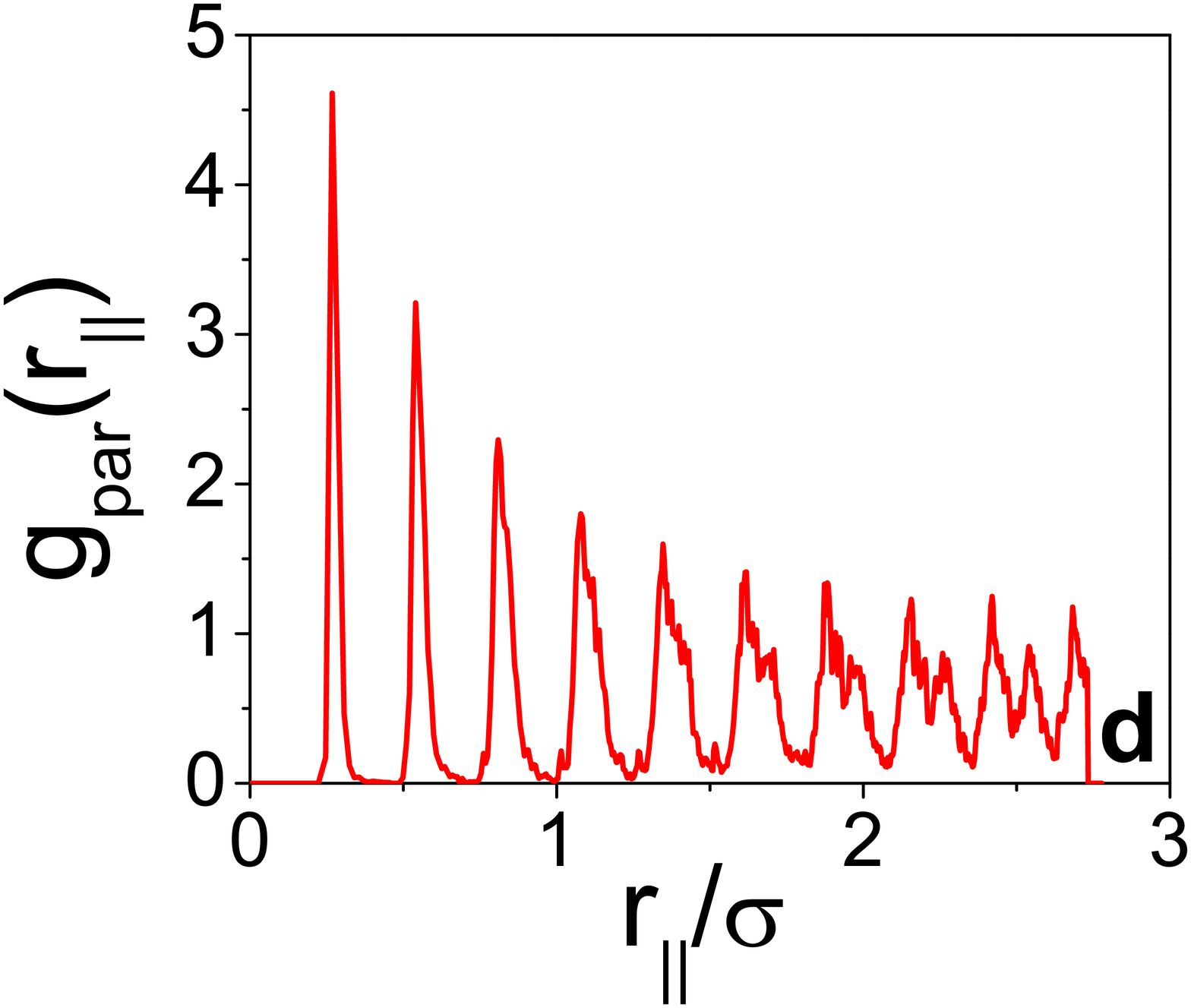}
\includegraphics[scale=0.2]{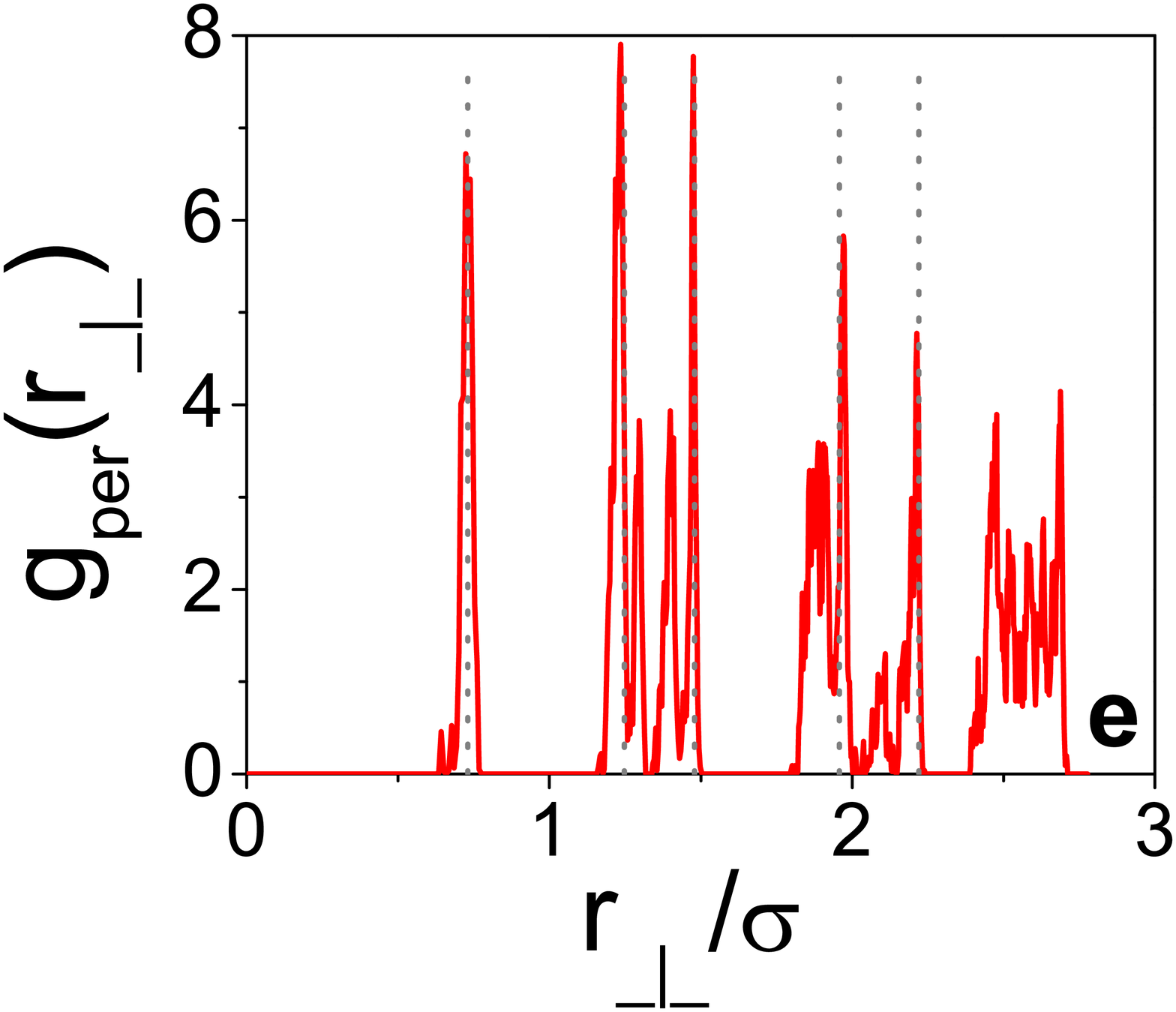}
\caption{ a) The  radial pair distribution function $g(r)$, b) the orientational pair distribution function $g_{or}(r)$, c) the structure factor $S(q)$
 d) spatial correlation function in the direction parallel to the director e) spatial correlation function in the direction perpendicular to the director
shown for   $\rho^*=8$ and $\kappa \sigma=10$, corresponding to a \emph{columnar hexagonal} structure.
}
\label{fig11}
\end{figure}


%
\begin{figure}[h!]
\includegraphics[scale=0.2]{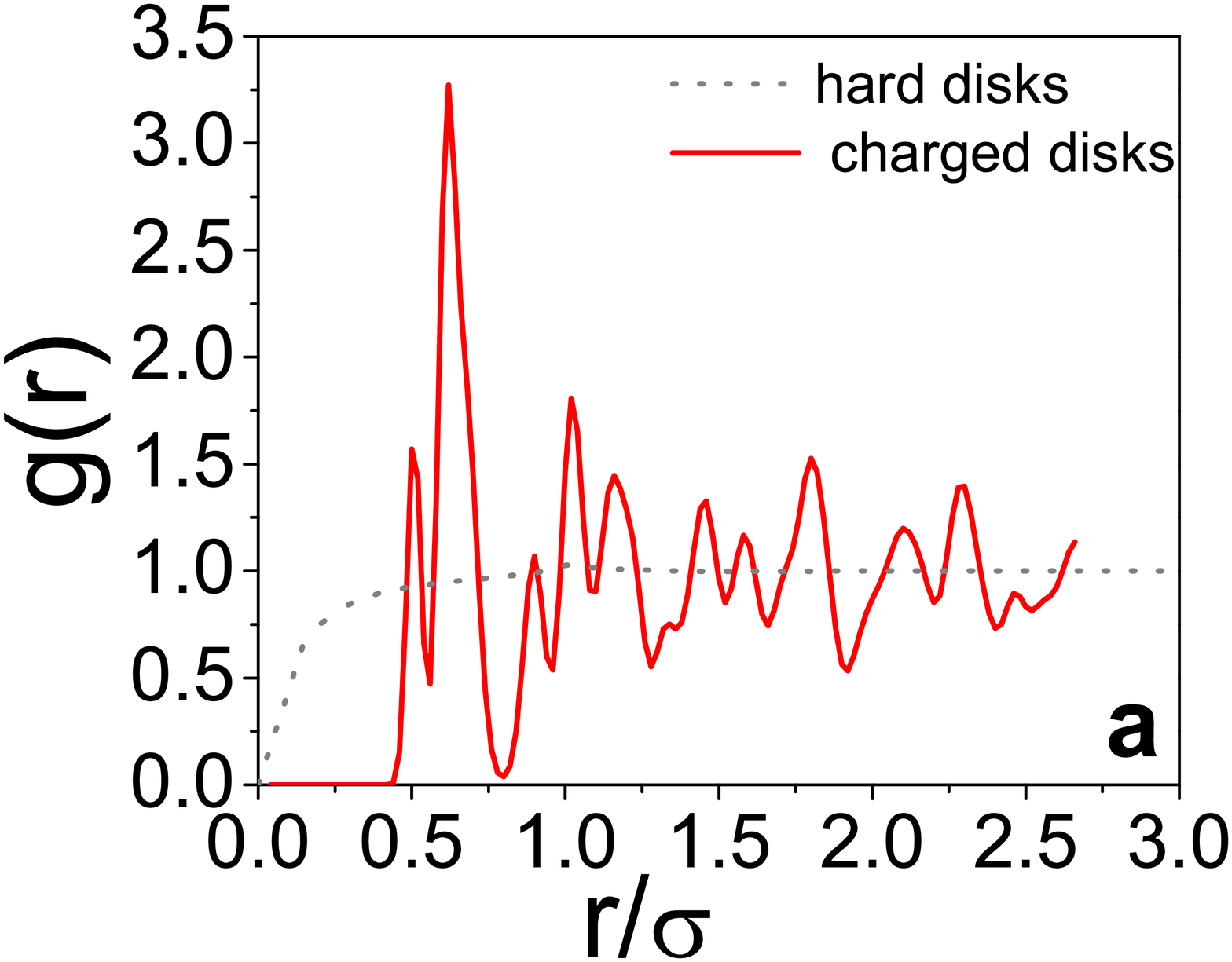}
\includegraphics[scale=0.2]{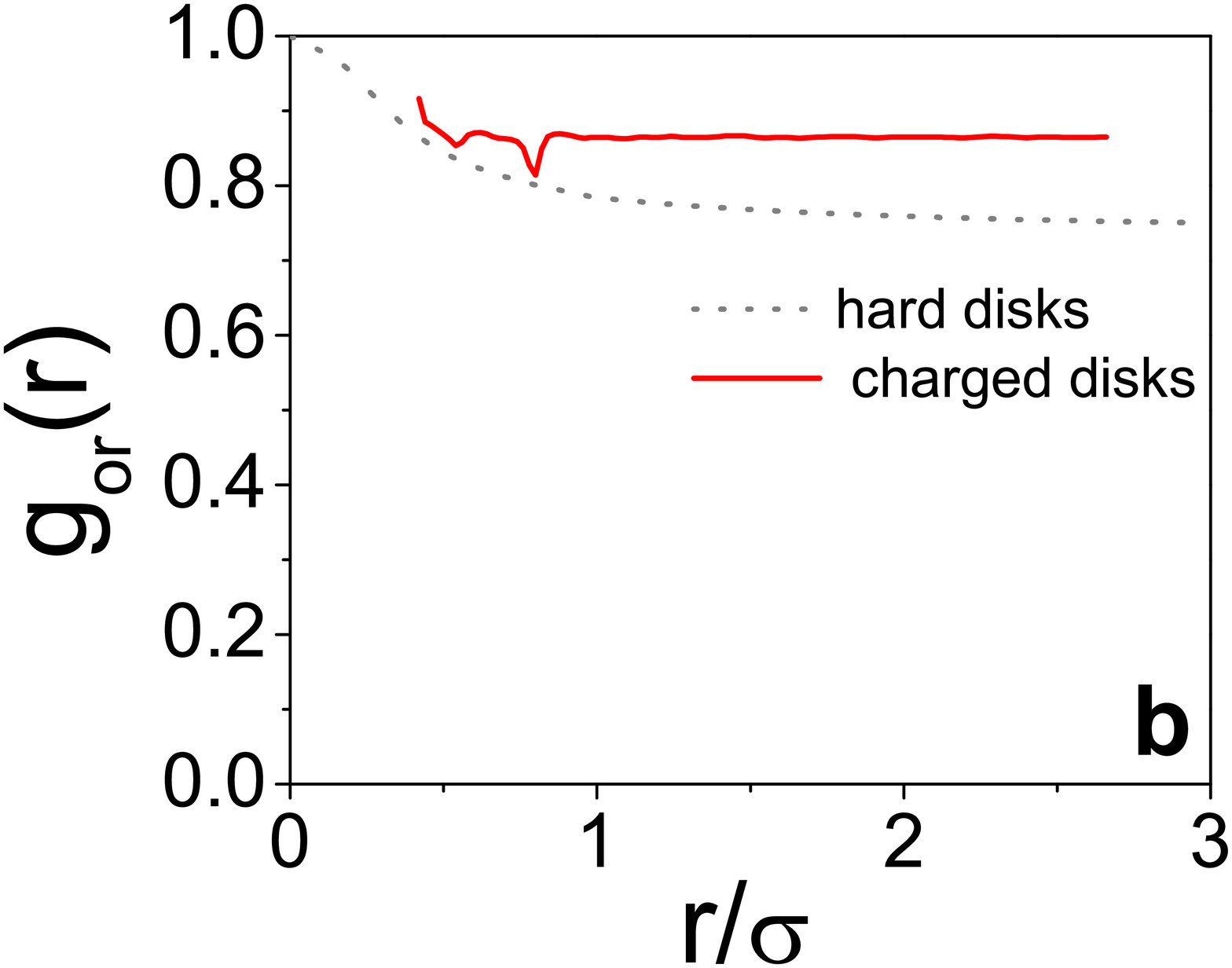}
\includegraphics[scale=0.2]{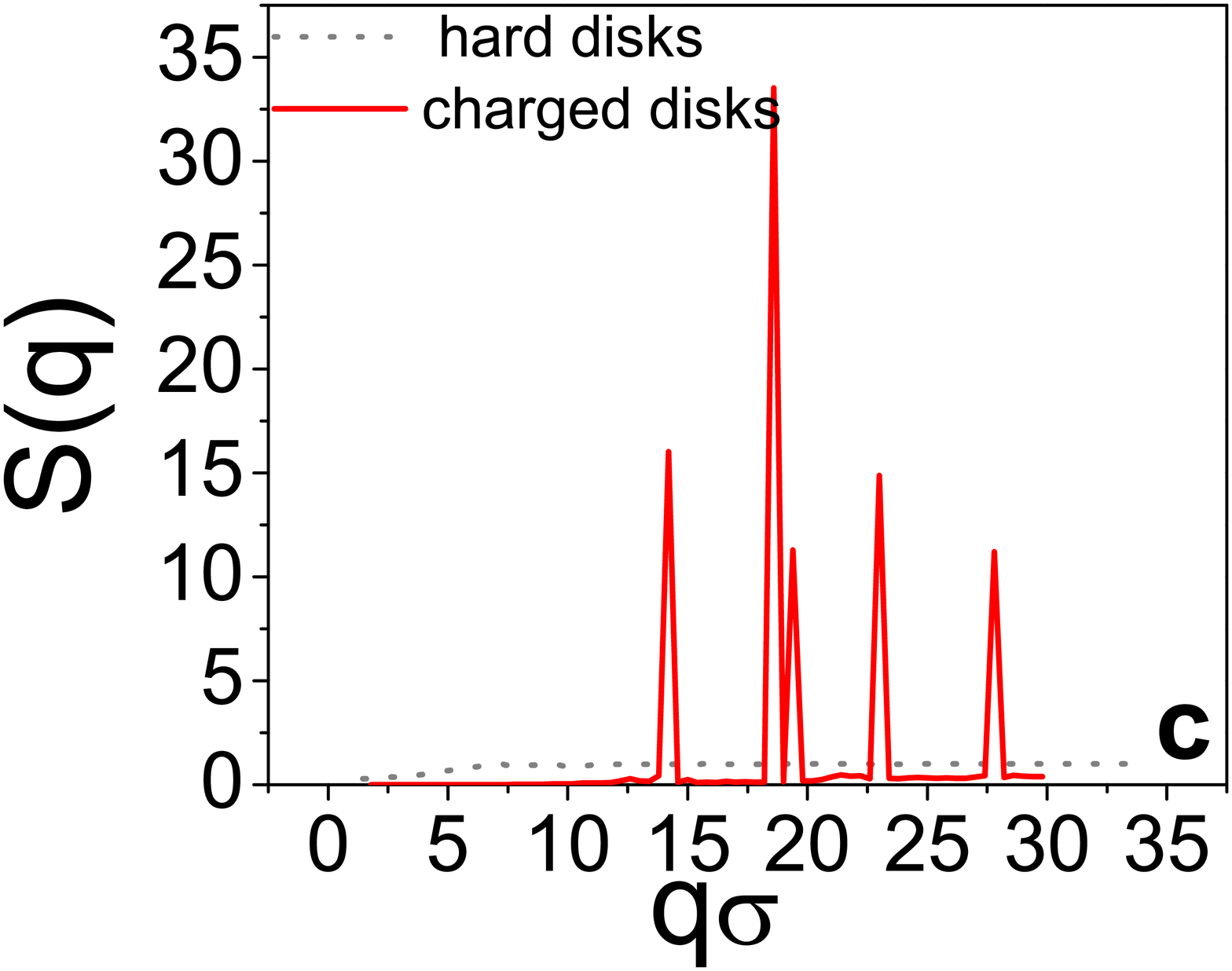}
\includegraphics[scale=0.2]{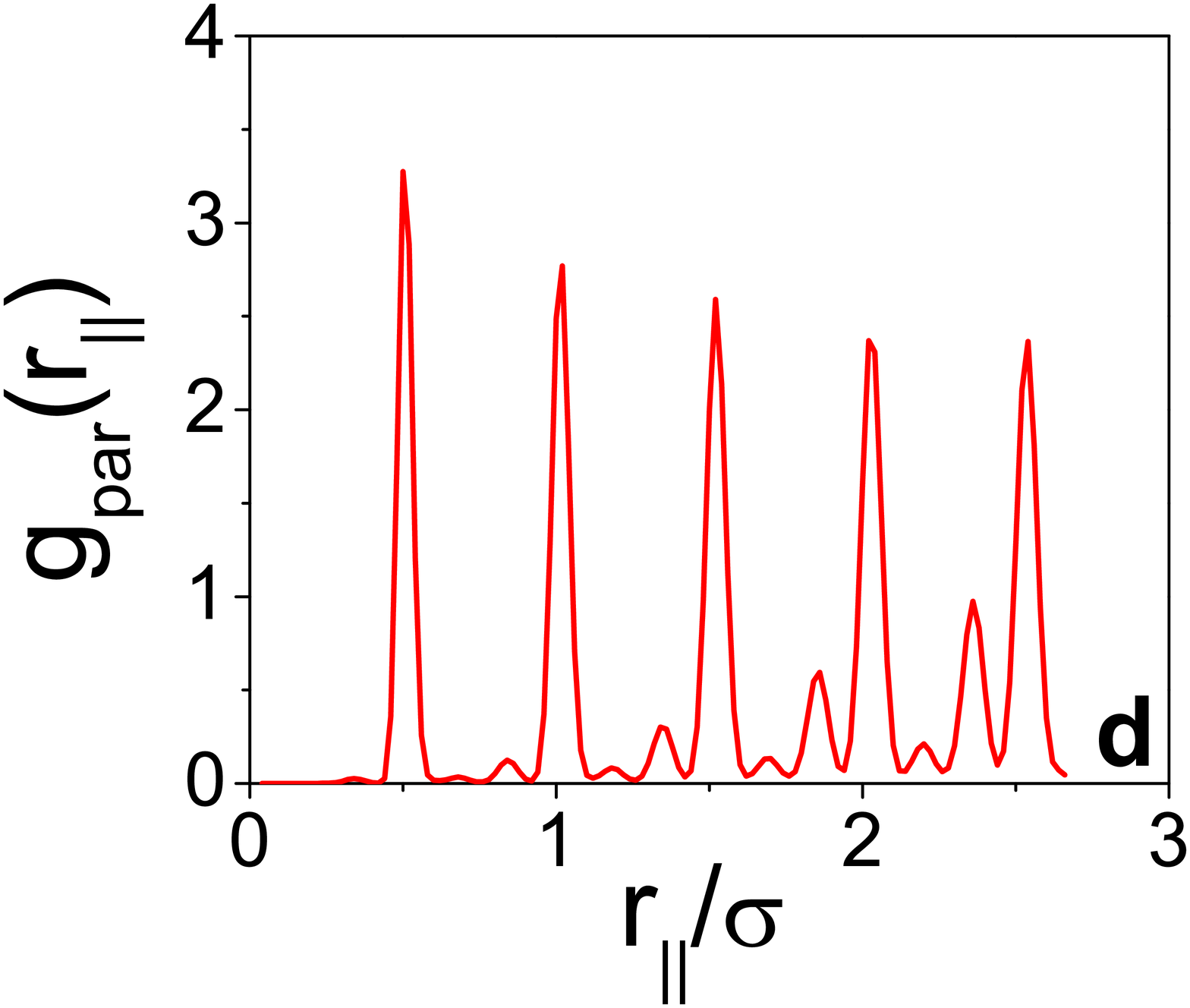}
\includegraphics[scale=0.2]{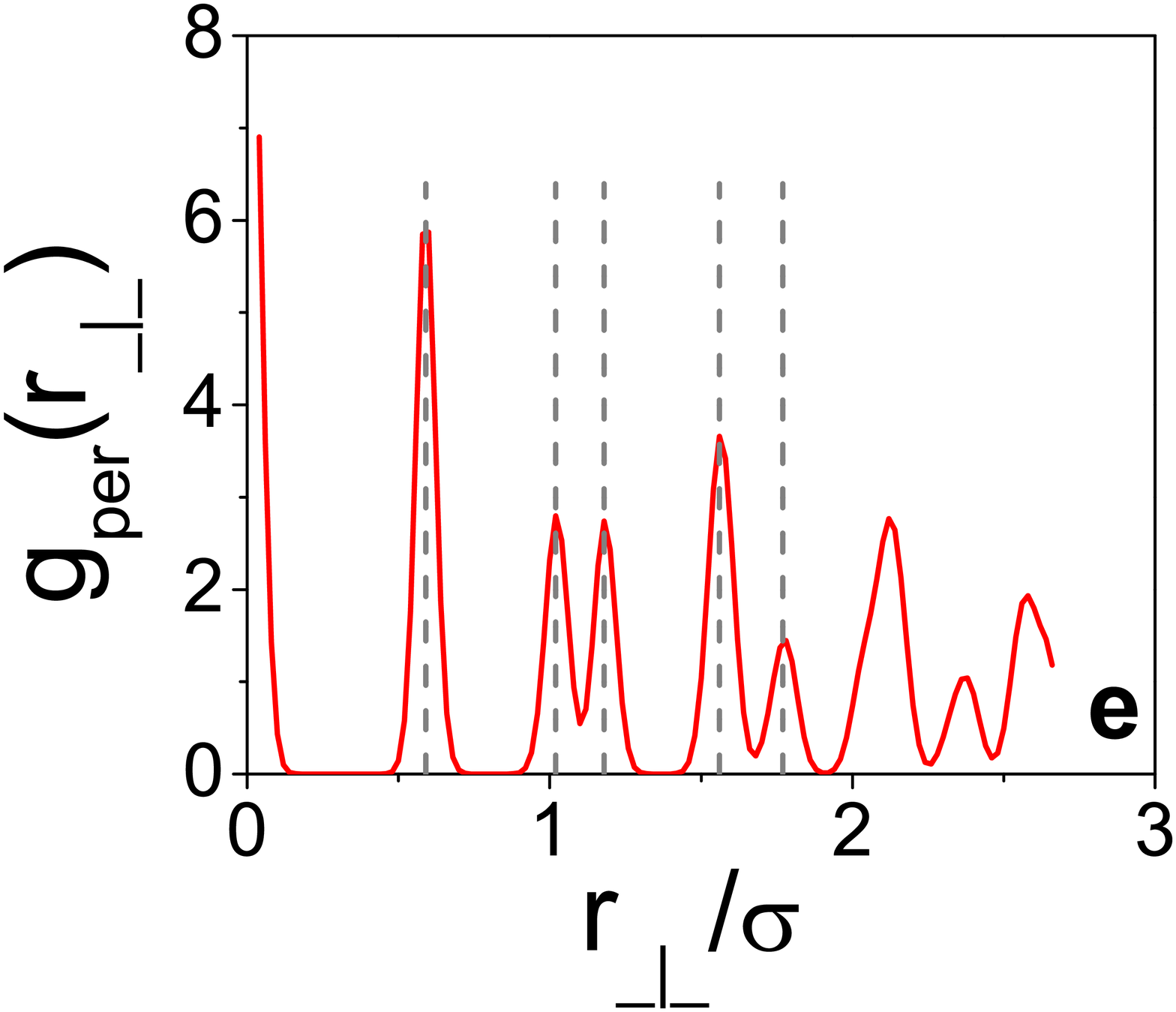}
\includegraphics[scale=0.2]{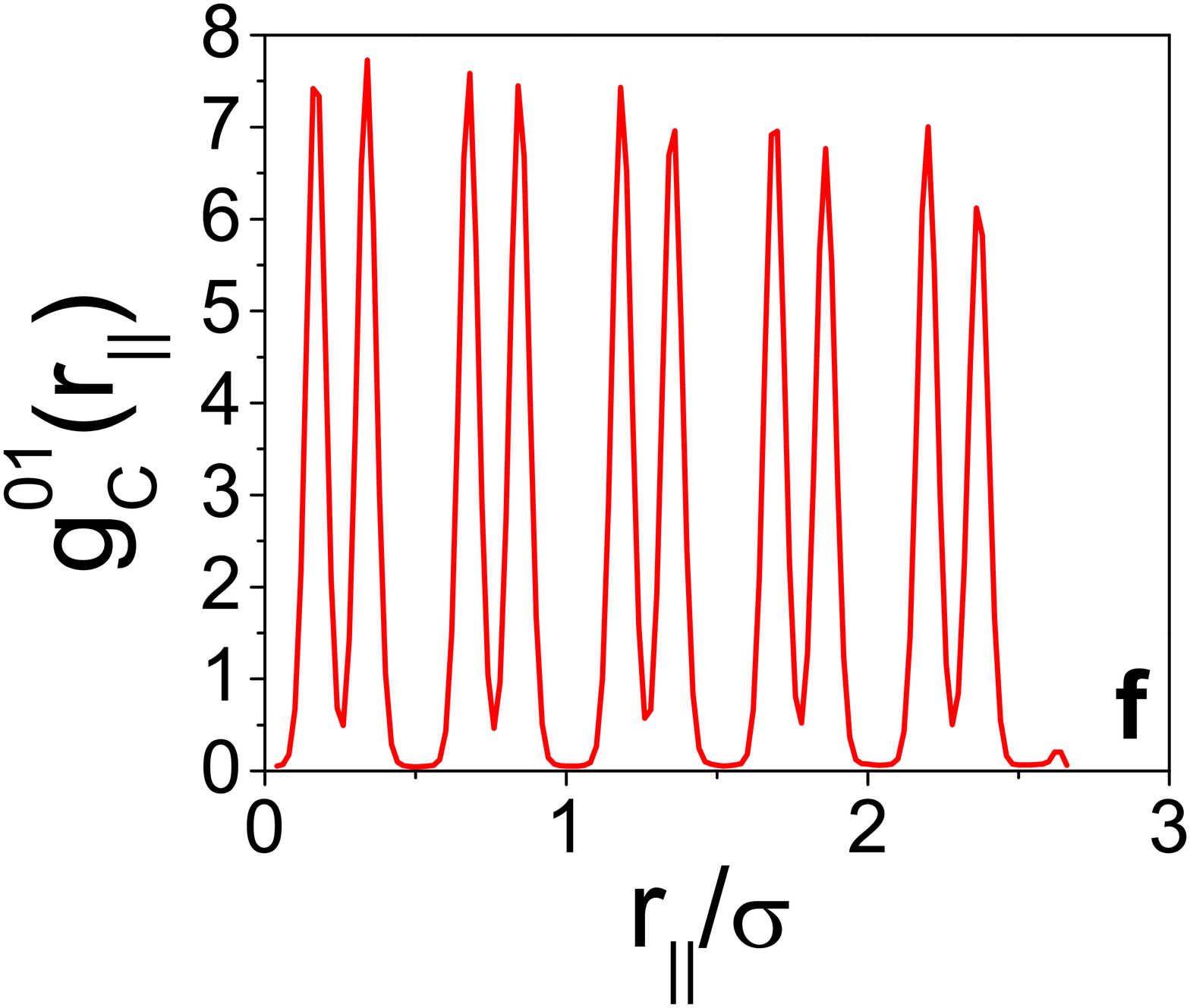}
\caption{ a) The  radial pair distribution function $g(r)$, b) the orientational pair distribution function $g_{or}(r)$, c) the structure factor $S(q)$
 d) spatial correlation function in the direction parallel to the director e) spatial correlation function in the direction perpendicular to the director f) intercolumnar pair correlation function
shown for   $\rho^*=6.5$ and $\kappa \sigma=2$, corresponding to a \emph{hexagonal crystal}.
}
\label{fig13}
\end{figure}
\item \textbf{Hexagonal crystal} \\
At  rather low ionic strengths,  at  $\kappa \sigma=2 $ and for $ \rho^*>6 $ , where the electrostatic  interaction potential is relatively long-ranged with  a maximum anisotropy of 20  \%, we observe a further
enhancement of positional ordering. In this region, not only stacks of aligned disks form hexagonal columns, but also the disks within the columns are arranged in an ordered crystalline manner. 
In Fig. \ref{fig13}, we have 
displayed the positional and orientational correlation functions, structure factor as well as intracolumnar $g_{par}$ and intercolumnar correlation function and pair correlation in the direction perpendicular to the 
director $g_{per}$ for an example of such a structure, i.e.,  $\rho^*=6.5$ and $\kappa \sigma=2$.  The strong degree of orientational ordering is demonstrated by a large value of the plateau in $g_{or}$. The orientational 
order parameter in this system is $S=0.93$.  As can be noticed from $g(r)$ and $S(q)$,  a strong degree  of positional order also exists. Similar to the columnar hexagonal phase, we observe a strong hexagonal ordering  in the
directions perpendicular to the director as confirmed by the position of successive peaks seen at distances $a, a\sqrt{3}, 2a, a\sqrt{7}, 3a $ with $a=0.59 \sigma$ for this example.
The lattice spacing smaller than one diameter, here again, points to the interdigitated nature of the columns.
 The regularly spaced peaks with the same width for $g_{par}$  demonstrate a strong degree of crystalline order of disks within  the columns. The intracolumnar spacing in this case is $0.5 \sigma$.  
 We also calculated the intercolumnar pair correlation functions for two adjacent columns $g_{01}(r_{||})$ with $0.5 \sigma < r_{\bot}< 0.7 \sigma$, similar to the one 
 obtained for hard spherocylinders \cite{Cutos}. The positions of the peaks give us the interdigitation displacement $0.2 \sigma$.
   A strong  positional correlation between  the disks in two adjacent columns confirms the crystalline nature of the phase, hence assessing that  a hexagonal crystal is formed.

\begin{figure}[h]
\includegraphics[scale=0.2]{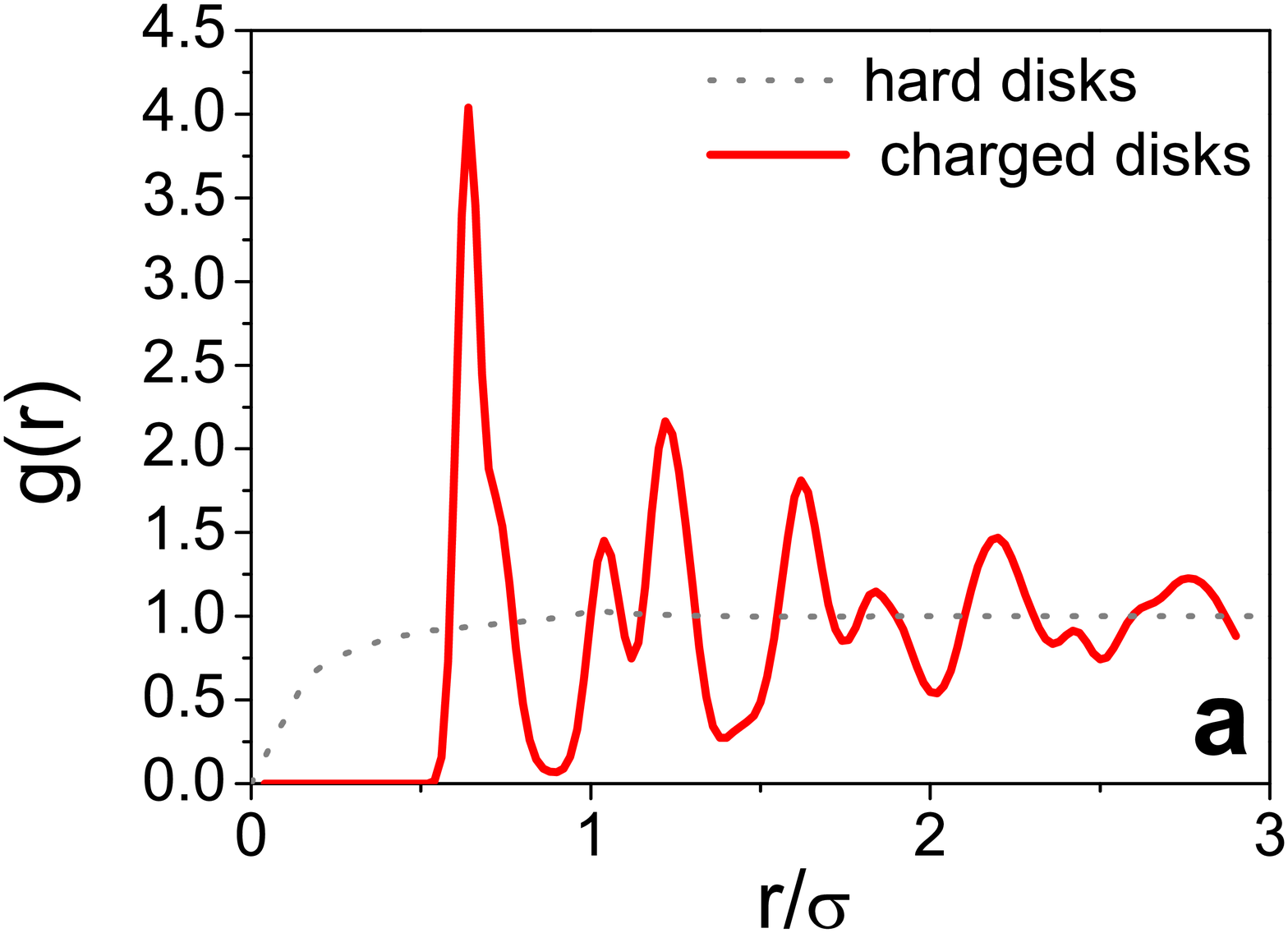}
\includegraphics[scale=0.2]{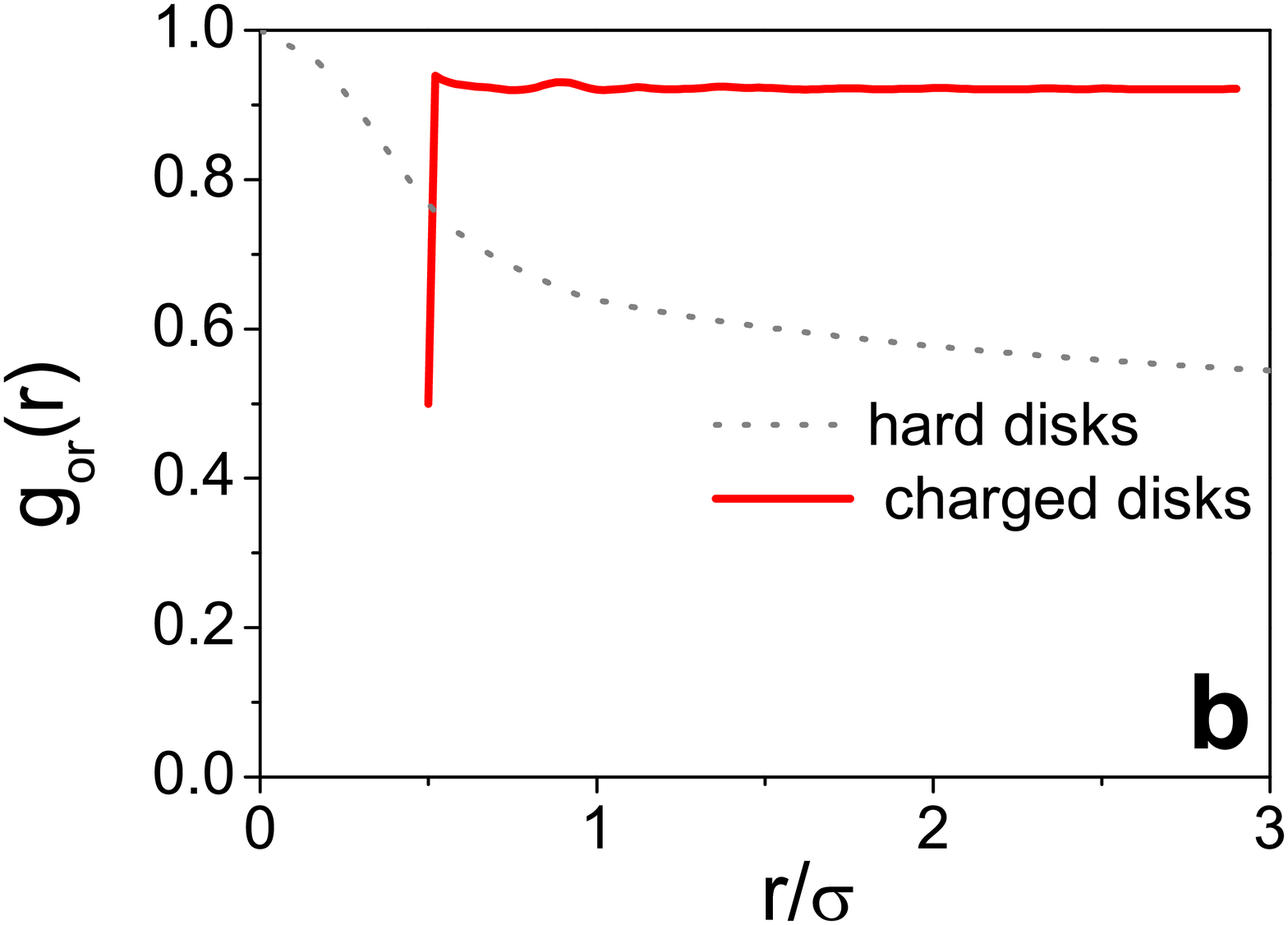}
\includegraphics[scale=0.2]{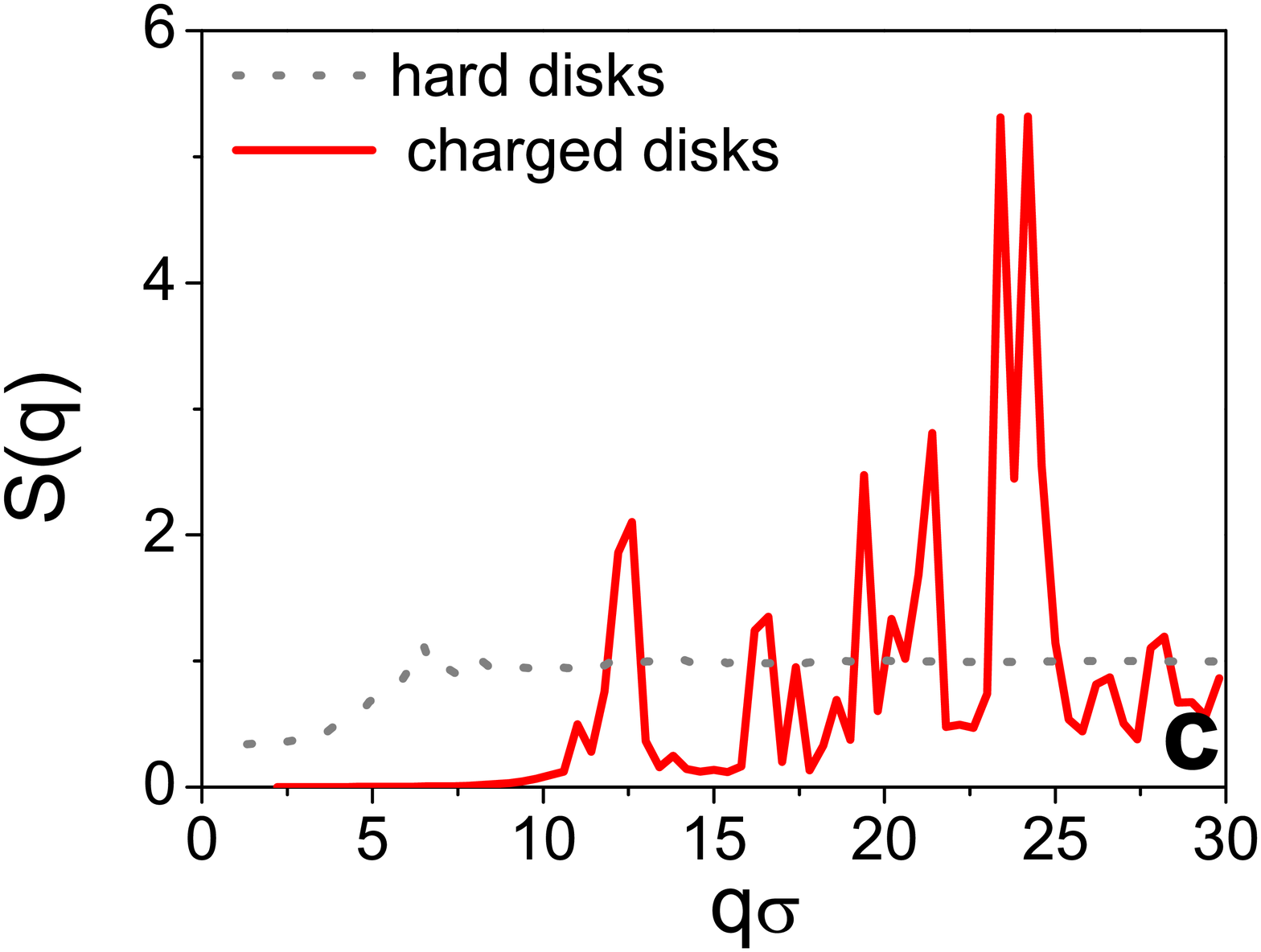}
\caption{ a) The  radial pair distribution function $g(r)$, b) the orientational pair distribution function $g_{or}(r)$, c) the structure factor $S(q)$   for  $\rho^*=5$ and  $\kappa \sigma=1$, 
corresponding to a \emph{BCC crystal}.}
 \label{fig14}
\end{figure}

\item \textbf{BCC crystal} \\
 At even lower ionic strengths where the interaction potential is
almost isotropic  ($\kappa \sigma=1$) and for densities $\rho^*> 3.5$,  another crystalline structure  appears which has a BCC-crystal  structure.  An example of  radial and orientational pair correlation functions and structure factor for  such a structure is shown in Figs. \ref{fig14}a,
\ref{fig14}b and \ref{fig14}c for $\rho^*=5$ and $\kappa \sigma =1$.  Both positional and orientational pair correlation functions as well as structure factor reveal a strong degree of positional and  orientational order. The peaks of $g(r)$  are found at separations  $ r/ \sigma=0.64,0.7, 1.04, 1.22, 1.62, 1.836,2.19$ which can be identified with peak positions for a BCC-crystal, {\it i.e.} $ a, 2/\sqrt{3}a, \sqrt{8}/\sqrt{3} a, \sqrt{11}/\sqrt{3} a, 2 a, 4/ \sqrt{3}a, \sqrt{19}/\sqrt{3} a$, and $2\sqrt{5} /\sqrt{3} a$.  Hence, from the positions of the peaks of $g(r)$, it is easily concluded that the structure is that of BCC crystal and the obtained orientational order parameter is $S=0.9$.

\end{itemize}
\subsection{\label{sec:charge}  The influence of effective  charge value on the structure}

\begin{figure}[h]
\includegraphics[scale=0.3]{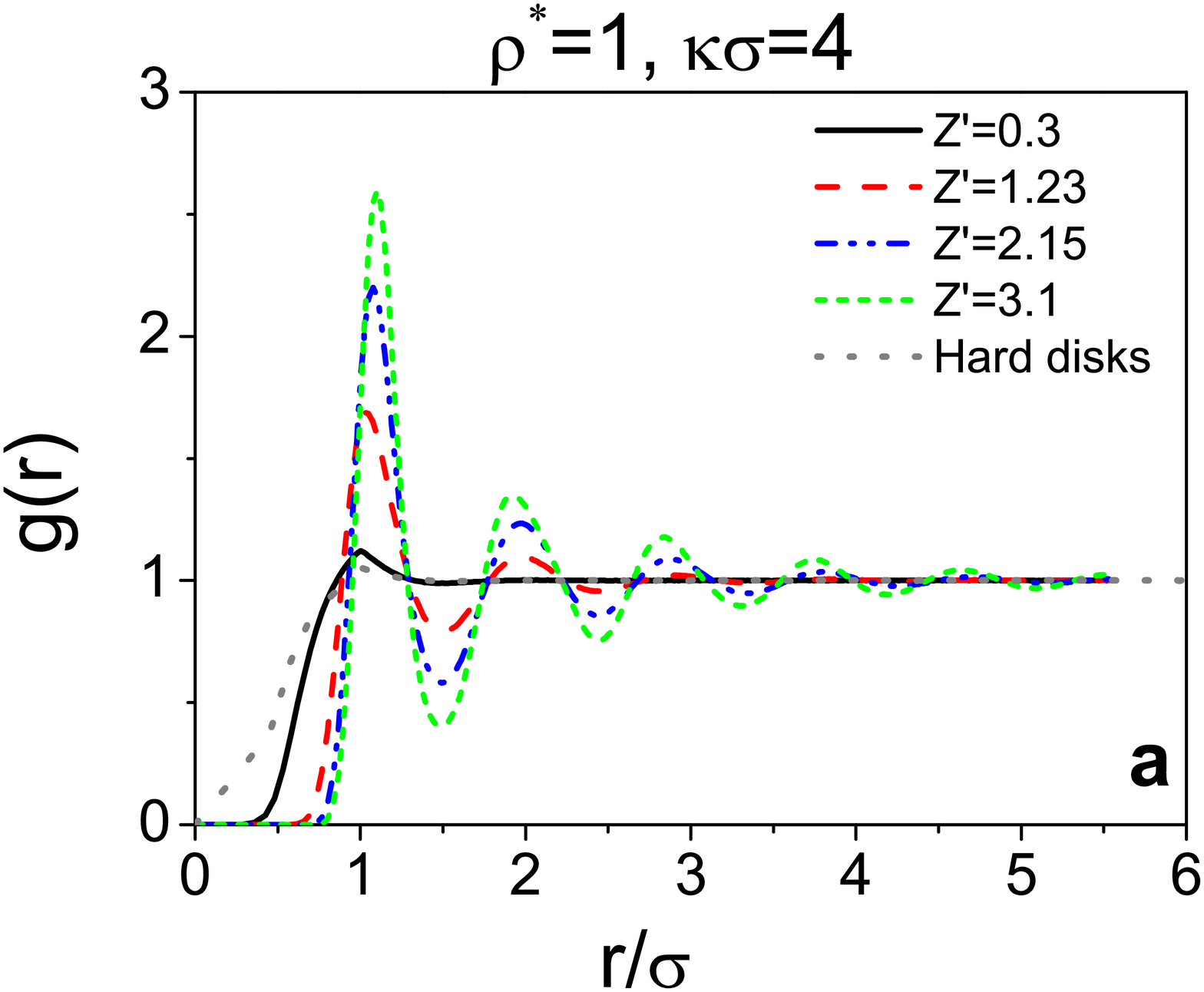}
\includegraphics[scale=0.3]{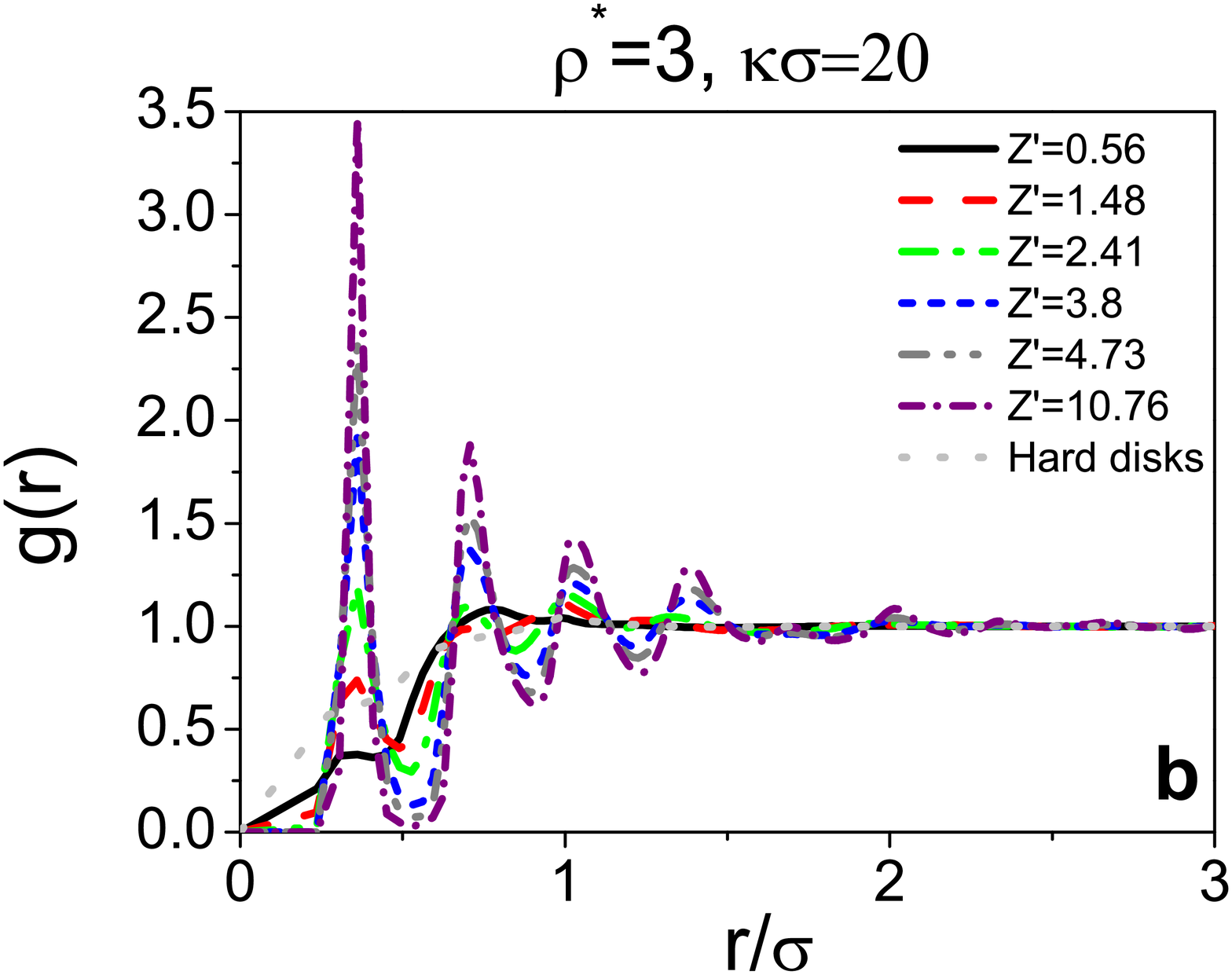}
\includegraphics[scale=0.3]{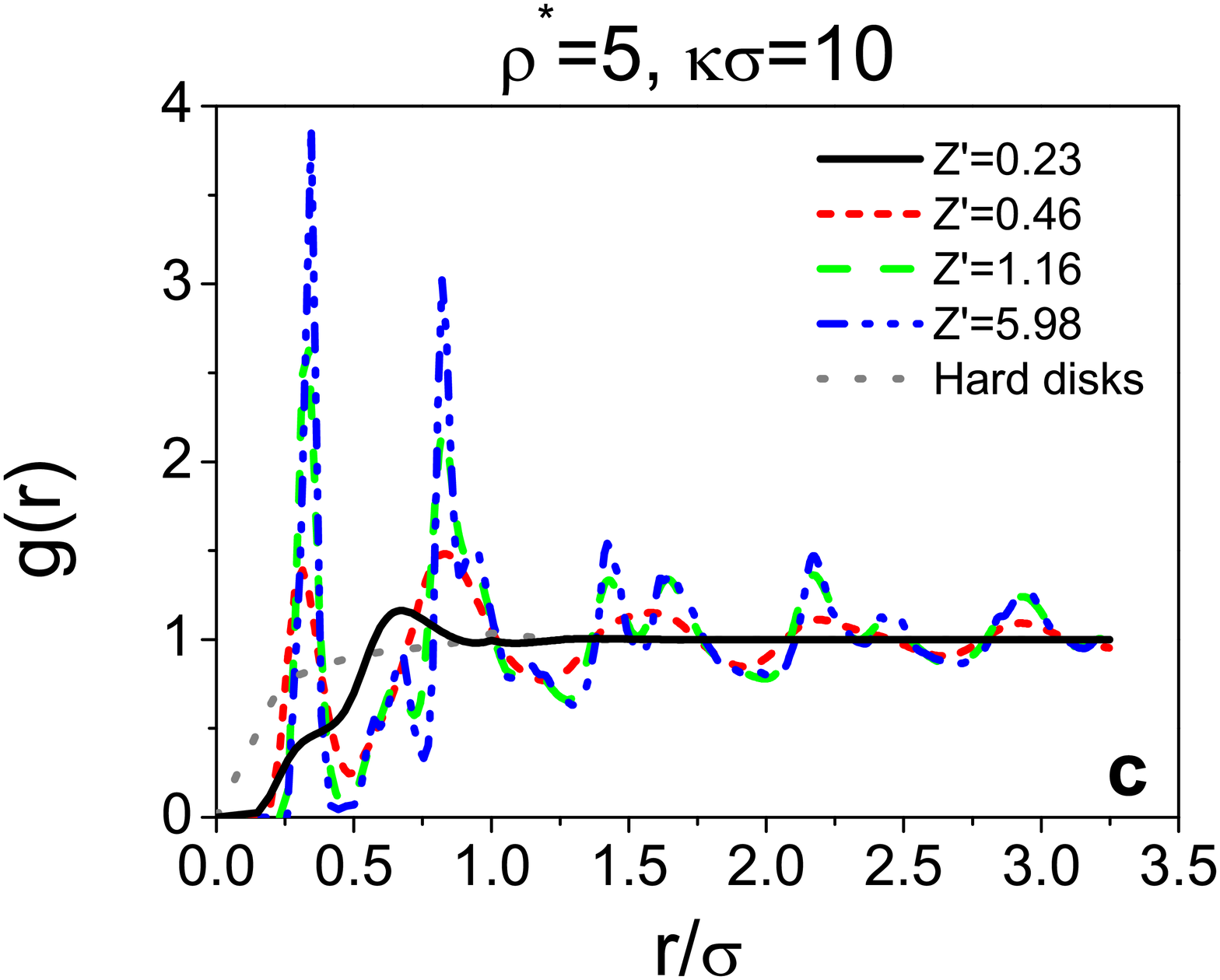}
\includegraphics[scale=0.3]{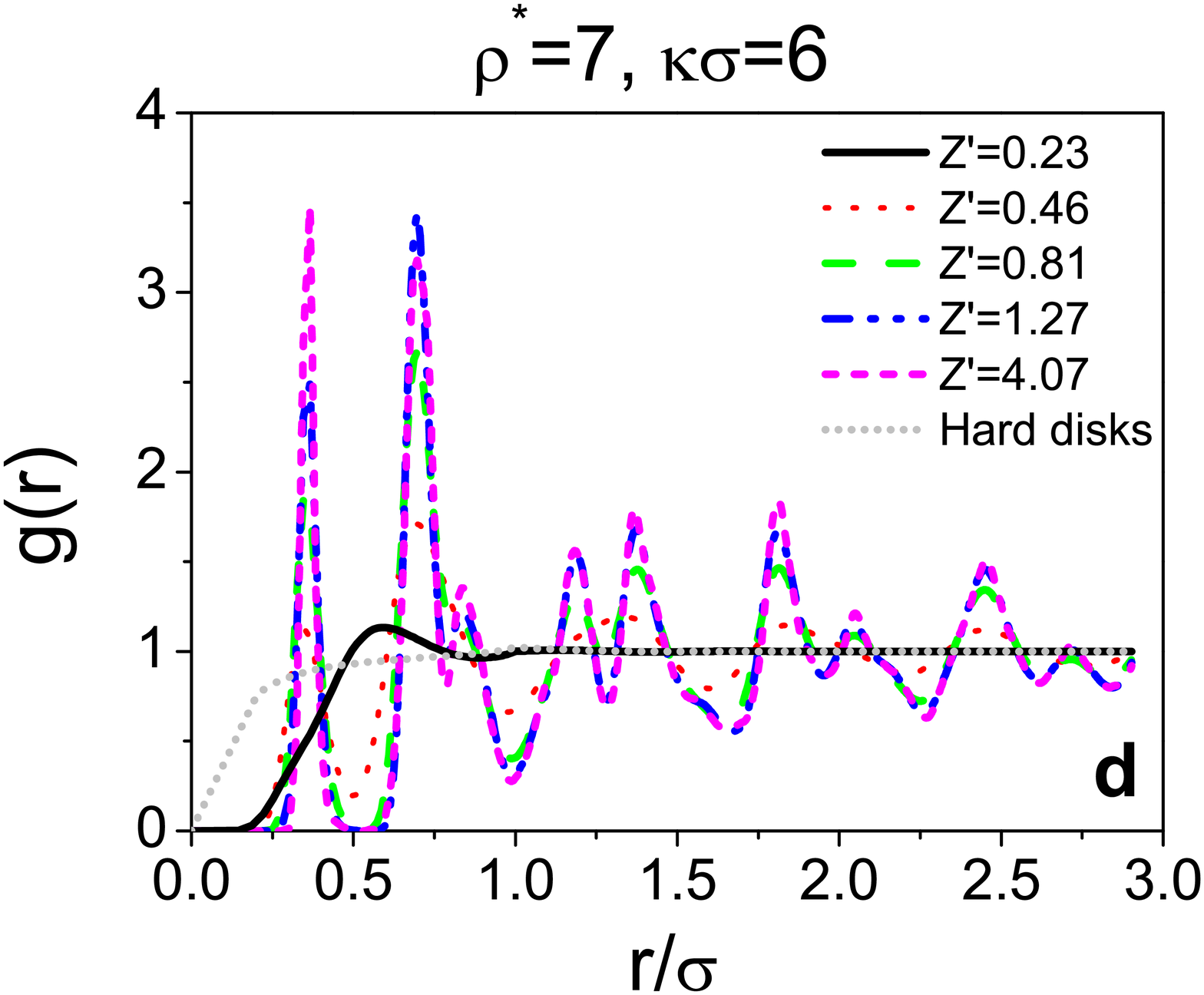}
\caption{ The  radial pair distribution function $g(r)$ shown for different values of $Z_{eff}$  for samples
 a) $\rho^*=1$ and $\kappa \sigma=4$ b)  $\rho^*=2$ and $\kappa \sigma=20$ c)$\rho^*=5$ and $\kappa \sigma=10$,
 d) $\rho^*=7$ and $\kappa \sigma=6$. The graphs are  complemented  with $g(r)$ of hard disks at the same density
 as the sample. The corresponding values of effective charges are shown in the legends, with $Z' = Z_{eff} \lambda_B/\sigma$.
}
\label{fig15}
\end{figure}

Having examined in detail the features of the various structures found for highly charged disks with
$Z_{eff}=Z_{eff}^{sat} (\kappa \sigma)$,  we now turn to the cases where $Z_{eff}< Z_{eff}^{sat}$. We present the results for the evolution of structure upon increasing charge
 for several densities and screening parameter values. As explained in Sec. \ref{sec:method},
  the effective charge values were varied from  relatively small values to $Z_{eff}^{sat} (\kappa \sigma)$
  for each $\kappa \sigma$ and density starting from an initial configuration of equilibrated hard disks.
   In Fig. \ref{fig15}, we  show the evolution of the radial pair distribution function $g(r)$
with the effective charge for two isotropic structures, i.e., the isotropic fluid and the
random stacks and two examples of liquid-crystalline
phases, i.e. nematic stacks and columnar hexagonal structure.
The general trend that we observe from all these examples is that upon increase of the effective charge, and
consequently of the strength of repulsion, the particles, as expected, are  excluded from the proximity of each other, i.e.,
 the shortest distance for which $g(r)$ is different from zero is shifted towards larger values.
The second important effect is the appearance of  peaks at well-defined positions in $g(r)$, different from
the peak of $g(r)$  for hard disks  at $r/\sigma=1$,  that become sharper with an increase of
the effective charge. We also observe that  for large densities, typically $\rho^*>4$, the structural changes
are less pronounced upon further increase of charge for  $ Z_{eff} \lambda_B/\sigma> 2.3 $.

\begin{figure}[h]
\includegraphics[scale=0.25]{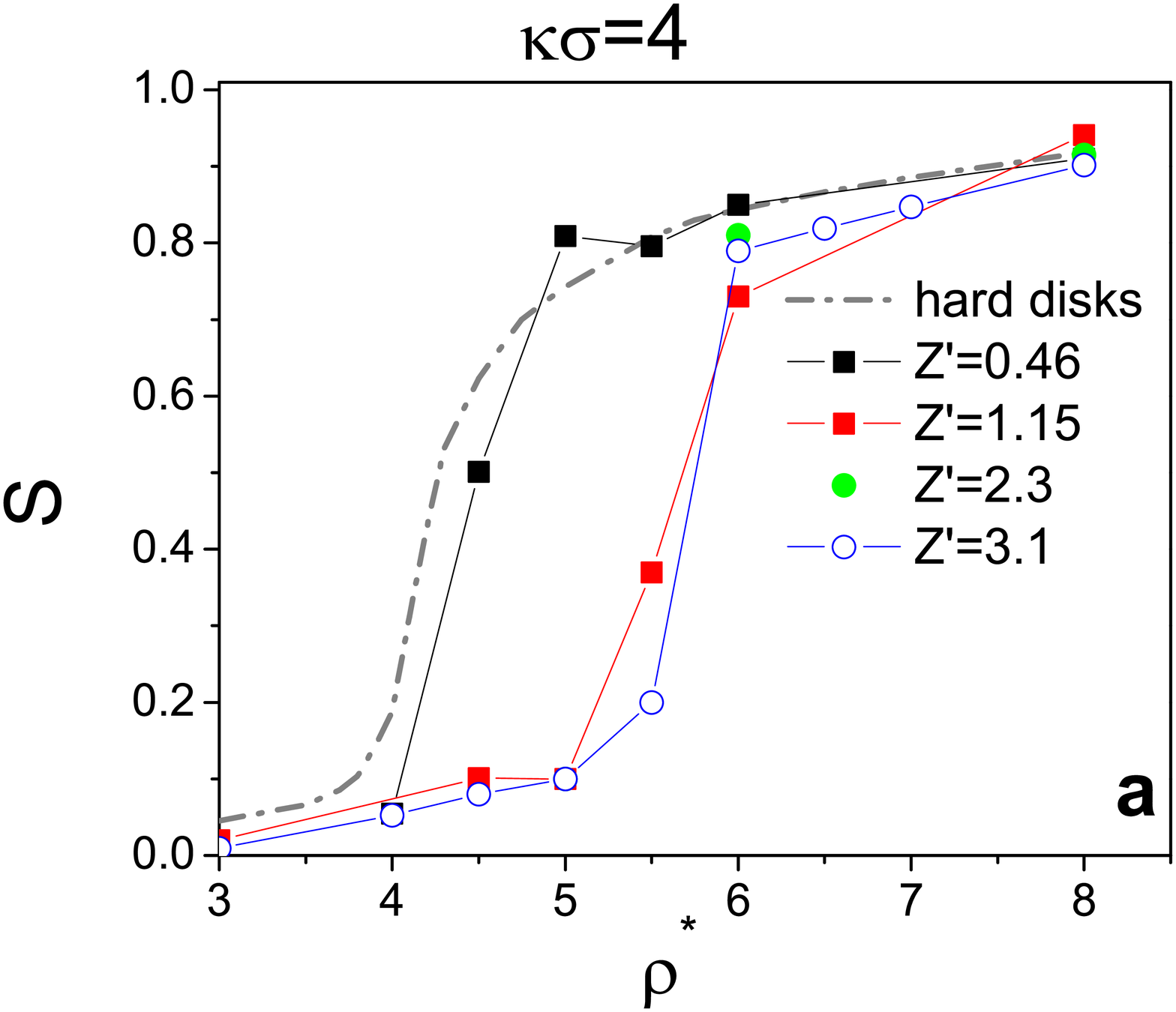}
\includegraphics[scale=0.25]{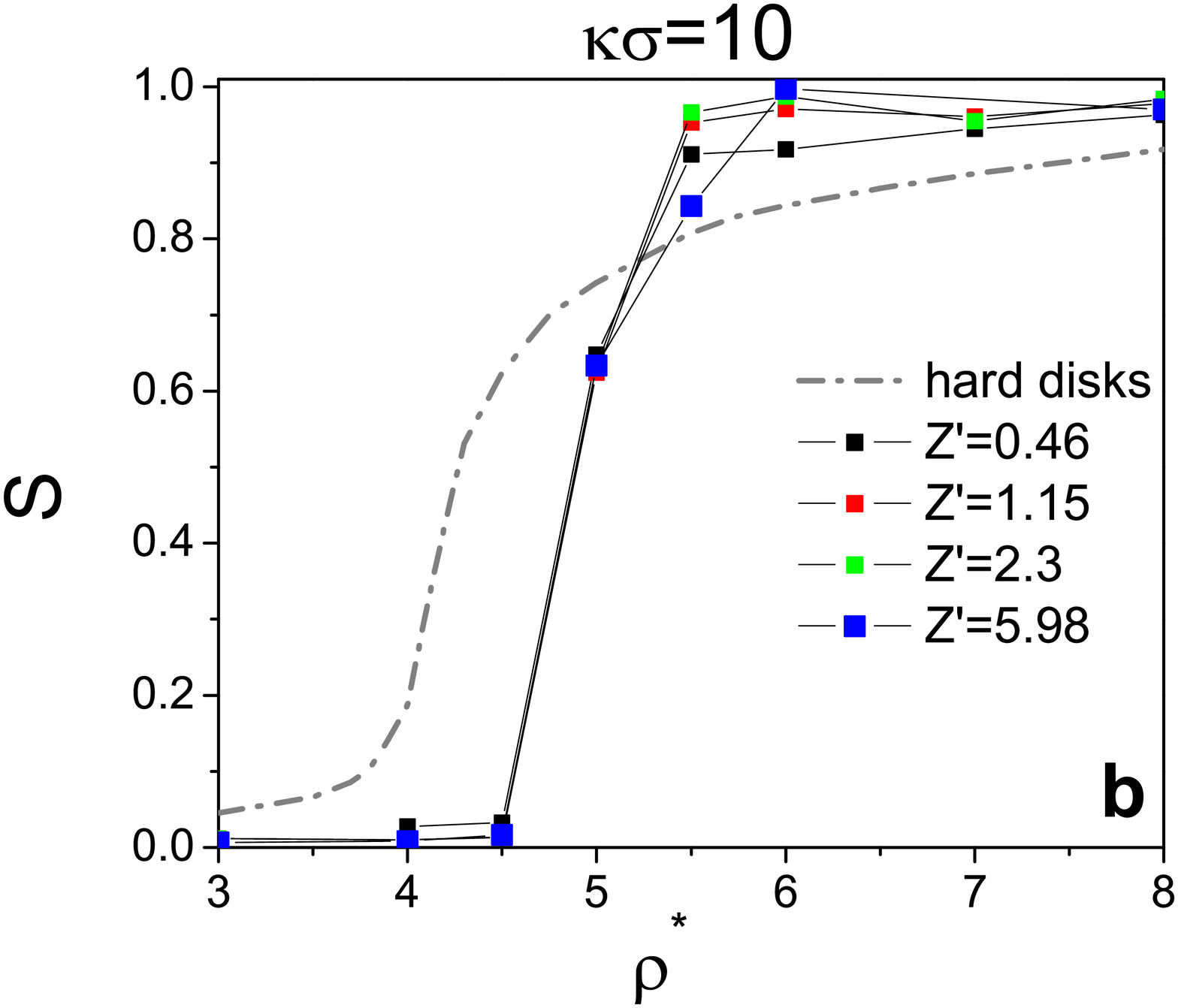}
\caption{ The nematic order parameter as a function of density for several values of effective charge
(shown in the legend) for a) $\kappa \sigma=4$ and b) $\kappa \sigma=10$.
}
\label{fig16}
\end{figure}

We have also investigated the dependence of the nematic order parameter on $Z_{eff}$, as depicted in Fig. \ref{fig16} for two different values of $\kappa \sigma$. In  the structures with vanishing
$S$, the  nematic order parameter did not change with increase of charge for $\rho^*< 4$ .
For  the orientationally ordered phases,  again we notice  that for  $ Z_{eff} l_B/ \sigma> 1.2 $, the
nematic order parameter is not very sensitive to the value of $Z_{eff}$. This is an important finding, because although the strength of interactions depends on the value of 
effective charge and size of particles, for sufficiently large  effective charge values the observed structure is not a sensitive function of particle size. 
Therefore, we argue that although our simulations were performed for $\sigma/\lambda_B=43$ , the obtained phase diagram for  $Z_{eff}^{sat}$ are relevant for other sizes as well.

%

%

\section{ Discussion and concluding remarks }
\label{sec:conclusion}

We have studied  the structure and dynamics of charged disks interacting with purely repulsive electrostatic forces.  The electrostatic interactions in our model are accounted for via a two-body effective potential 
between disks, that is obtained from Poisson-Boltzmann theory. We find that  anisotropic electrostatic interactions  modify the disks arrangements and
their dynamics immensely and lead to a 
rich phase behavior. The  phases uncovered not only include the isotropic fluid and nematic phases observed for infinitely thin disks, but also comprise 
 columnar  hexagonal liquid crystalline,  hexagonal, BCC crystalline phases  and  a novel phase of intergrowth texture. One  germane feature of our phase diagram is 
 the non-monotonic behavior of orientational disorder-order transition with ionic strength $ \propto (\kappa \sigma)^2$ which results from the opposing effects of the decreasing potential range and the increasing amplitude of the
anisotropy function, upon increasing ionic strength.  Another important aspect of our study is the investigation of dynamical signatures of orientationally disordered states. 
We find that both translational and
rotational dynamics slow down remarkably already in the isotropic phases at densities somewhat below the density of the orientational order-disorder transition.

 Before discussing our results in the light of experimental data, a comment is in order. Our phase diagram highlights the importance of   intrinsic anisotropy of the electrostatic potential between charged platelets and shows that the phase diagram of charged disks can not be obtained by a simple rescaling of particles dimensions from the phase diagram of
 hard spherocylinders \cite{Marechal} or uniaxial ellipsoidal particles \cite{spheroids}.  Nevertheless in the limit of very low ionic strengths where the potential is  isotropic, we observe the same trend as ellipsoidal particles
 with  aspect ratios close to unity where the system goes from an isotropic fluid to a plastic crystal and finally an orientationally ordered crystal with FCC structure upon increase of density albeit with a different 
 symmetry than the one observed for charged disks (BCC symmetry).  For high ionic strengths, our phase diagram resembles the phase diagram of oblate spherocylinders \cite{Marechal} at small aspect ratios $ L/ \sigma$ where a transition from isotropic fluid to nematic phase and
 subsequently to a columnar hexagonal phase occurs. However, unlike the case of charged spheres where the electrostatic interactions lead simply to a shift of phase boundaries, here we can not deduce the phase behavior of charged
 disks from hard anisotropic colloids with similar symmetry. New phase of intergrowth texture arises as a result of original anisotropy of the potential that highlights the competing effects of anisotropy in shape and interactions.
 
Our phase diagram shows a good agreement with  the  phase behavior of  most extensively studied charged colloidal platelets: Gibbsite, and Beidellite systems for which the 
repulsive interactions are predominant  \cite{Scirep}. Gibbsite and Beidellite both show an I/N transition
for a wide range of ionic strengths, and platelet stacking is often observed in the nematic phase \cite{Gibphase,Beidellite}. Moreover, Gibbsite suspensions
also display a columnar hexagonal liquid-crystalline phase. These features are well captured in our simulations. We provided a direct comparison of our phase diagram
with  experimental phase diagrams by replotting them as a function of 
density and $\kappa \sigma$   \cite{Scirep} and  argued that, in spite of the presumably important effects of polydispersity, the two phase diagrams agree quite well
 for overlapping regions. Of course a more detailed and extended experimental study might shed  some more light on this issue and taking into account the effects of polydispersity and other specific interactions in MC simulations would possibly lead to a closer agreement. Another evidence of predominance of repulsive electrostatic interactions on phase behavior of Gibbsite  comes from measurements of  long-time translational and rotational diffusion of isotropic suspensions of Gibbsite in DMSO \cite{gibdynamics}.  In  good agreement with our findings, the experiments demonstrate a slow dynamics for  both translational and  rotational diffusion
 at densities well below the orientational disorder-order transition.  These results give us a hint why we observe arrested states for some of the charged platelet systems such as Laponite and Montmorillonite at very low volume fractions. 

Laponite is another system of charged platelets which has attracted a lot of attention because of its ability to form aging gel/glassy states at  low 
volume fractions ($\phi <0.02$)\cite{Laponite,Sara-PRE2008,Wigner}  well below the I/N transition. The range of densities and ionic strengths 
investigated in experiments corresponds to $\rho^* < 1$ and \cite{Sara-PRE2008} $1 \leq \kappa \sigma <20$. At such low densities, in our phase diagram, we observe 
a solid-like state (plastic crystal) only at small $\kappa \sigma=1$. Interestingly, exploring the response of our plastic crystal,  we observe a slow dynamics with a two-step relaxation similar to what is seen for Laponite suspensions \cite{Abou}. This state could presumably only be glassy if polydispersity were considered as seen with a bidisperse system of  point-like particles interacting with Yukawa potential \cite{Zaccarelli,LiLi}. Indeed, a Wigner glass at very low ionic strengths $I < 10^{-4}$ has been reported for Laponite suspensions \cite{Levitz,Wigner}. Upon further increase
of ionic strength, we observe an ergodic repulsive isotropic liquid in the low-density region of our phase diagram which is at equilibrium unlike the aging Laponite suspensions.  Therefore, our model system only based on repulsive interactions cannot fully explain the phase behavior of Laponite for larger ionic strengths, although 
 we find that onset of slow dynamics shifts towards lower densities \cite{Scirep}   upon increase of ionic strength, similar to the trend observed for Laponite suspensions \cite{Sara-PRE2008}.  
This points to the possible relevance of attractions \cite{Ruzicka} or other specific interactions in this system that need to be incorporated  in our modeling  to reproduce the aging phenomenon observed for Laponite suspensions.

\begin{figure}[h]
\includegraphics[scale=0.25]{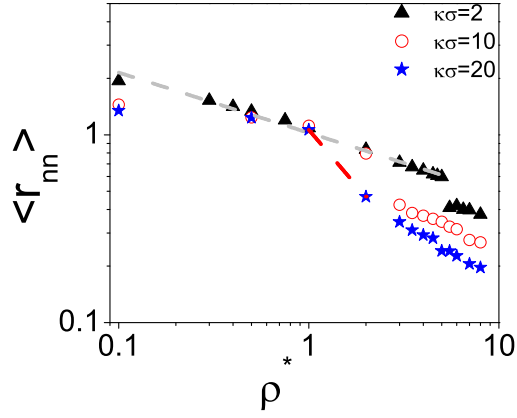}
\caption{Nearest neighbor distance $\langle r_{nn}\rangle$ as a function of density plotted for several values of $\kappa \sigma$ as shown in the legend. In the low density regime $\langle r_{nn}\rangle \propto \rho^{-1/3}$ indicated by the gray dashed line.
}
\label{figNN}
\end{figure}

As another test of our model against experiments, we extracted the nearest-neighbor distance $\langle r_{nn}\rangle$ from the position of the first peak of $g(r)$. 
This quantity is a measure of average interparticle distance. It has been suggested that organization of highly anisotropic particles such as disks can be understood based on simple geometrical 
considerations \cite{baravian}  and for disks,  the interparticle distance at low densities (volume fractions) should scale as $\rho^{-1/3}$ and as $\rho^{-1}$ at high densities. 
In Fig. \ref{figNN}, we have plotted $\langle r_{nn}\rangle$ as a function of density for several values of $\kappa \sigma=2, 10 $ and 20. Interestingly, in the low density regime 
$\langle r_{nn}\rangle$ scales as $\rho^{-1/3}$, however, the high density regime does not follow $\rho^{-1}$ and the exponent obtained for high densities is a function of $\kappa \sigma$, 
its value being $-0.24, -0.46$
and $-0.62$  for $\kappa \sigma=2, 10 $ and 20, respectively. This difference in the high density regime presumably arises from the fact that 
the experimental systems of Gibbsite and Beidellite are not equilibrated, and are trapped in arrested states.

In conclusion, numerical simulations of charged platelets, interacting with an effective potential of the
anisotropic screened Coulomb form,  display a rich phase diagram that captures the generic features of  charged colloidal platelet systems.
This study  provides us with an insight about the influence of electrostatics on slowing down of dynamics, upon varying density and ionic strength.
The required procedure of slow charge increase, at moderate and high densities, echoes the difficulty for
experimental systems to reach thermodynamic equilibrium. Better agreement with experiments could arguably be achieved by
considering polydispersity, platelet flexibility, van der Waals attractions, and interactions specific to each system such as solvation forces.
The features brought to the fore here, though, can be considered as the non-specific effects pertaining to charged colloidal
 platelets. \\




\appendix

\section{Ewald sum for long-range Yukawa potential}
 \label{appendixA}
 
 The Ewald sum for the energy of the Yukawa interaction
\begin{equation}
v_{ij}(r) =  \left\{ \begin{array} {l@{~~~~~~~~}l}
                             \infty  & \mbox{overlap} \\
                \displaystyle  \frac{e^{-\lambda r}}{r}
                                     & \mbox{no \ overlap} \\
                             \end{array}
                             \right.
\label{potential2}
\end{equation}
is given by \cite{salin:00,caillol:07}:
\begin{eqnarray}
\label{I2}
U =&& \frac{1}{2}  \sum_{i=1}^N \sum_{j=1}^N \sum_{\vec{n}}^{'}
  \left[ {\rm erfc} \Big(\eta |\vec{r}_{ij}+ L\vec{n}|+ \displaystyle
  \frac{\lambda}{2 \eta} \Big) e^{\lambda  |\vec{r}_{ij}+ L\vec{n}|} \right.
  \nonumber \\
&& \left. + {\rm erfc} \Big(\eta |\vec{r}_{ij}+ L\vec{n}|-  \displaystyle
  \frac{\lambda}{2 \eta} \Big) e^{-\lambda |\vec{r}_{ij}+ L\vec{n}|} \right]
 \Big / 2  |\vec{r}_{ij}+ L\vec{n}| \nonumber \\
&&  \frac{2\pi}{V}  \sum_{i} \sum_{j \ne i} \sum_{\vec{k} \neq 0}^\infty \displaystyle
 \frac{e^{- (\vec{k}^2 +\lambda^2) / 4 \eta^2}}{\vec{k}^2 +\lambda^2}
   \displaystyle \exp[i\vec{k} \cdot \vec{r}_{ij}] + \frac{2\pi}{V}  \displaystyle
 \frac{e^{- \lambda^2 / 4 \eta^2}}{\lambda^2}. \nonumber \\
&&  +  U_{self}
\end{eqnarray}

In the limit $\lambda \to 0$ the divergence in the third term in Eq.\
(\ref{I2}) is avoided by assuming a uniform background of opposite
charge.
This amounts to subtracting a term  $ \displaystyle \frac{4\pi}{\lambda^2 V}$.

The self term is given by
\begin{eqnarray}
\label{I3}
U_{self}/N =&& \frac{1}{2}  \sum_{\vec{n} \neq 0}
  \left[ {\rm erfc} \Big(\eta L |\vec{n}|+ \displaystyle
  \frac{\lambda}{2 \eta} \Big) e^{\lambda L |\vec{n}|} \right.
  \nonumber \\
&& \left. + {\rm erfc} \Big(\eta L |\vec{n}|-  \displaystyle
  \frac{\lambda}{2 \eta} \Big) e^{-\lambda L |\vec{n}|}
  \right] \Big /2 L |\vec{n}| \nonumber \\
&&  \frac{2\pi}{V} \sum_{\vec{k} \neq 0}^\infty \displaystyle
 \frac{e^{- (\vec{k}^2 +\lambda^2) / 4 \eta^2}}{\vec{k}^2 +\lambda^2}
+ \frac{2\pi}{V} \displaystyle
 \frac{e^{- \lambda^2 / 4 \eta^2}}{ \lambda^2}
  \nonumber \\
&& -\frac{\eta}{\sqrt \pi}  {e^{- \lambda^2 / 4 \eta^2}}
+ \frac{\lambda}{2} {\rm erfc} \Big( \displaystyle \frac{\lambda}{2
  \eta} \Big).
\end{eqnarray}
Again, in the presence of a uniform background,  a term
$\displaystyle \frac{4\pi}{\lambda^2 V }$ has to be subtracted.

In Eq.\ (\ref{I2}) $\vec{r}_{ij}=\vec{r}_j-\vec{r}_i$,
$L=$ is the box length (assuming a cubic box) and $\rm{erfc}$ denotes the
complementary error function. The prime in the sum over
$\vec{n} = (n_x,n_y,n_z)$, with $n_x$,$n_y,n_z$ integers, restricts it to $i \neq
j$ for $\vec{n}=0$. The parameter  $\eta$ controls the convergence of
the direct and Fourier space sums in (\ref{I2}).
The sum in reciprocal space extends
over all lattice vectors $\vec{k} = 2 \pi  \vec{n}/L$.
If the $\vec{k}$-space contribution in the self energy is added to the
reciprocal space term in Eq.(\ref{I2}), the former can be written
\begin{equation} \label{I4}
 \frac{2\pi}{V} \sum_{\vec{k} \neq 0}^\infty \displaystyle
 \frac{e^{- (\vec{k}^2 +\lambda^2) / 4 \eta^2}}{\vec{k}^2 +\lambda^2}
  F(\vec{k})F^*(\vec{k})
\end{equation}
where
\begin{eqnarray}
\label{I6}
F(\vec{k}) &=& \sum_{i=1}^N  \displaystyle \exp[i\vec{k} \cdot \vec{r}_i],
\end{eqnarray}
and $F^*(\vec{k})$ denotes the complex conjugate.

In the present model, the interaction between two disks depends also on
the orientation of each disk with respect to the interparticle distance.
In that case the reciprocal space term in the energy cannot be factorized 
as in Eq.\ (\ref{I4}), but involves a triple sum over $i$, $j$, and  $k$,
which would lead to prohibitively large computation times.
In our
calculations, we adopted a value of $\eta L=5$ large enough that only terms with $\vec{n}=0$ need to be retained in
Eq.\ (\ref{I2}) but small enough that  the energy  from the
direct term is the dominant contribution.  We then neglected the Fourier
space term.
Of course, in the case of an angle-averaged potential the energy can be
calculated exactly. We verified in the case of the isotropic potential
that neglect of the Fourier space term had no measurable effect on the
structural
properties.
\section{ Computed quantities}
\label{appendixB}
\begin{itemize}
\item \textbf{Radial pair distribution function} $g(r)$:  orientationally averaged pair distribution function that describes the correlation between the centers of disks whose distance is  $r$ .
\begin{equation}
g(r)=\frac{1}{4 \pi r^2 N \rho} \left\langle \sum_{i=1}^N  \sum_{j\neq i}^N \delta (r-r_{ij}) \right\rangle
\end{equation}
The average number of nearest neighbors for a disk $N_{nn}$ can be obtained from the integral of the radial distribution function over  its first peak
as  $N_{nn}=4 \pi \rho \int_{r_{min1}}^ {r_{min2}} g(r) r^2 dr $

   \item \textbf{Structure factor} $S(q)=\frac{1}{N}  \left\langle \sum_{i,j=1} ^N \exp(i\vec{q} \cdot (\vec{r}_i-\vec{r}_j)\right\rangle$ that is  related to
     the Fourier transform of $g(r)$ as  $S(q)=1+\rho \int d^3r (g(r)-1)\exp(i \vec{k}\cdot \vec{r})$   .
\item \textbf{Orientational pair distribution functions} $P_2(r)$ and  $P_4(r)$  are defined  as  the statistical average of the second and
fourth Legendre polynomial of azimuthal angle between the normal vectors of two disks, $\cos \theta= \hat{u}_i . \hat{u}_j $
\begin{eqnarray}
P_2(r)&=&\left\langle \frac{1}{2} (3 \cos^2 \theta (r)-1) \right\rangle \\
\end{eqnarray}
\item \textbf{Nematic order parameter} $S$: is defined as the average of the largest eigenvalue of the second-rank tensor
\begin{equation}
Q_{\alpha \beta}= \frac{1}{N} \sum_{i=1} ^{N}(3 u_{\alpha}^i u_{\beta}^i-\delta_{\alpha \beta})/2
\end{equation}
\end{itemize}
The normalized eigenvector corresponding to the largest eigenvalue of $Q_{\alpha \beta}$ gives us
  the director  $\hat{n}$ that characterizes the direction of  dominant orientation.\\ \\
\textbf{Quantities specific to orientationally ordered phases}\\
These  quantities are computed only for the samples with $S  \geq 0.4$

\begin{itemize}
\item \textbf{Perpendicular pair-correlation function}: $  g_{per}(r_{\bot})$  is calculated for all pairs of disk centers whose   distance perpendicular to
the director $\hat{n}$  lies on a cylinder of radius $r_{\bot}$ and thickness $2l_t$
with its  axis parallel to the director $\hat{n}$.  $l_t$ was taken to be $0.01 \sigma$ in our simulations.
\begin{equation}
g_{per}(r_{\bot})=\frac{1}{2 \pi l_t  r_{\bot} N \rho} \left\langle \sum_{i=1}^N  \sum_{j \neq i}^N \delta (r_{\bot}-|\vec{r}_{ij} \times \hat{n}|) \right\rangle
\end{equation}

\item \textbf{ Parallel (columnar) pair-correlation function}: $g_{par}(r_{||})$  is  calculated for all pairs of disk centers inside a cylindrical
volume parallel to $\widehat{n}$ with radius $R=\sigma/2$.
\begin{equation}
g_{par}(r_{||})=\frac{1}{ \pi R^2  N \rho} \left\langle \sum_{i=1}^N  \sum_{j \neq i}^N \delta (r_{||}-|\vec{r}_{ij}.\hat{n}|) \right\rangle
\end{equation}
\item \textbf{Hexagonal bond-orientational order parameter}: $q_6$ provides  us  the information about  arrangement of disks
in an hexagonal bidimensional lattice  within a sheet of thickness $l_t$ perpendicular to the nematic director
\begin{equation}
q_6=\left\langle \frac{ \sum_{i=1}^{N-1}  \sum_{j > i}^N P_{ij} \exp(i6 \theta_{ij})}{ \sum_{i=1}^{N-1}  \sum_{j > i}^N P_{ij}}  \right\rangle
\end{equation}
where
\begin{equation}
P_{ij}=1 \quad if \quad  r_1 \leq r_{\bot ij} \leq r_2 \quad and \quad 0 \leq r_{|| ij} \leq l_t
\end{equation}
\end{itemize}
$\theta_{ij}$ is the angle between the vector $\vec{r}_{ij}$ joining the centers of  $i$th and $j$th particles and a fixed axis perpendicular to
the director and  $r_1$ and $r_2$ give the lower and upper limits of the first peak of $g_{per}(r_{\bot})$.
Thus $q_6=1$ correspond to a perfect hexagonal order in a sheet of thickness $l_t$  and $q_6=0$ would mean  absence of any hexagonal order.\\ \\

\textbf{Dynamical quantities}\\

\begin{itemize}
\item \textbf{Mean-squared displacement} (MSD):
\begin{equation}\label{D1}
\left\langle \Delta r^2(t) \right\rangle=1/N \sum_{i=1}^{N}  \left\langle |\vec{r}_i(t)-\vec{r}_i(0)|^2 \right\rangle,
\end{equation} 
where $t$ 
 is  obtained in terms of the number of MC steps.

\item \textbf{Self-intermediate scattering function} is defined as
\begin{equation}\label{D2}
  F_s(q,t)=\frac{1}{N}  \left\langle \sum_{i=1} ^N \exp(i\vec{q} \cdot
(\vec{r}_i(t)-\vec{r}_i(0))\right\rangle
\end{equation}

\item \textbf{ Orientational  time correlation functions}  characterize the dynamics of orientational degrees of freedom of disks and are defined in terms of
Legendre polynomials:
\begin{equation}\label{D3}
   \left\langle P_l(\widehat{u}_i (t)\cdot \widehat{u}_i(0) ) \right\rangle.
\end{equation}

\end{itemize}
\begin{acknowledgments}
We wish to acknowledge the  support of Foundation Triangle de la Physique and IEF Marie-Curie fellowship. We are also grateful to
A. Maggs, H. Tanaka and  H. H. Wensink for fruitful discussions.
\end{acknowledgments}

\bigskip

\textbf{\large{List of symbols}}\\
\begin{itemize}
\item $ N $: number of particles in the simulation box of size $L$;
\item $ \sigma=2R $: disk diameter;
\item $ n $: number of Monte Carlo cycles, where a cycle is defined as one trial move for translational
                  and rotational displacements of the N particles;
\item $ \rho \equiv N/L^3  $:  number density of  disks; $\rho^*= \rho \sigma^3  $ is the dimensionless density ;
\item $ Q_s $: surface charge density of disks;
\item$\lambda_B \equiv e^2/ k_B T$ : Bjerrum length;
\item $ Z_{eff} $: effective charge of disks seen from far distance and $Z^{\prime}=Z_{eff} \lambda_B/\sigma$;
\item $ \kappa\equiv 1/l_{Debye} $: screening parameter defined as $\kappa^2=4 \pi \lambda_B \sum _i n_i z_i^2$ where $n_i$ and $z_i$ are the density
of ions of type $i$ and their corresponding charge, respectively;

\item $ U_{12} $: the pair interaction between two charged disks;

\item $ \widehat u_i $: the orientation of  $i$th disk (normal to the plate) ;
\item $ \theta_i $: the angle of orientation of  $i$th disk with a line that connects the center of this disk to the center of another disk;
\item $S$: orientational (nematic) order parameter;
  \item $ q_6 $: hexagonal bond-orientational order parameter;
\item $ \tau_B \equiv \sigma^2/ (6 D^t_{0}) $: Brownian time-scale, the time  required for diffusing over a distance equal to particle diameter;
 \item $\tau_{0}^{r}\equiv1/(2 D_{0}^r) $:  the first order  relaxation time of an isolated particle, for disks $ \tau_{0}^{r}=2/3 \tau_B$  ;
  \item $ q $: modulus of the scattering vector;
   \item $ F_s(q,t) $: the self-intermediate scattering function;
 \item $\langle r_{nn}\rangle$: nearest-neigbor distance.


\end{itemize}


\begin{thebibliography}{99}


    \bibitem{Beidellite} E. Paineau, K. Antonova, C. Baravian, I. Bihannic, P. Davidson, I. Dozov, M. Imp\'{e}ror-Clerc, P. Levitz, A Madsen, F Meneau, L. J. Michot,  \textit{J. Phys. Chem. B} \textbf{113}, 15858 (2009). 
     \bibitem{nontronite}  L.J.  Michot,   I. Bihannic,   S. Maddi,   S.S., Funari, C. Baravian, P.  Levitz,  and  P.   Davidson,   \textit{Proc. Nat. Acad. Sci. USA} \textbf{103}, 16101 (2006).
     
     \bibitem{Laponite} B. Ruzicka, and  E. Zaccarelli,  \textit{Soft Matter} \textbf{7}, 1268 (2011).
      \bibitem{Sara-PRE2008} S. Jabbari-Farouji, H.  Tanaka,  G. H. Wegdam, and D. Bonn,
\textit{Phys. Rev. E.} \textbf{78}, 061405 (2008).

    \bibitem{bentonite}  J.-C. P. Gabriel, C.  Sanchez, and  P. Davidson, 
  \textit{J. Phys. Chem.} \textbf{100}, 11139 (1996).
        
        
       \bibitem{clay1} A.  Bakk, J.O. Fossum, G.J. da Silva, H.M. Adland, A. Mikkelsen, and A. Elgsaeter,   \textit{Phys. Rev. E} \textbf{65}, 021407 (2002).
\bibitem{clay2} E.  DiMasi,   J.O. Fossum, T. Gog,, and Venkataraman, C.  \textit{Phys. Rev. E} \textbf{64}, 220405 (2001).
       
  \bibitem{clay3} N. Miyamoto, H. Iijima, , H. Ohkubo, and Y. Yamauchi,   \textit{Chem. Commun.}, \textbf{46}, 4166 (2010).
  
  
       \bibitem{zirconium}   D. Sun,  H.-J. Sue,    Z. Cheng,  Y.  Martinez-Ratun,  and  E.  Velasco,  \textit{Phys. Rev. E} \textbf{80}, 041704 (2009).

   \bibitem{Gibphase}  M. C. D. Mourad, D. V. Byelov,  A. V.  Petukhov,  D. A.  Matthijs de Winter,  A. J. Verkleij,  and   H. N. W. Lekkerkerker, 
 \textit{J. Phys. Chem. B} \textbf{113}, 11604  (2009).

     \bibitem{gib2008}M. C. D. Mourad, D. V. Byelov, A. V. Petukhov, and H. N. W. Lekkerkerker, \textit{J. Phys.: Condens. Matter} \textbf{20}, 494201 (2008).
    
\bibitem{Gibbsite} D. van der Beek, H. N. W.  Lekkerkerker,    \textit{Langmuir}  \textbf{20}, 8582 (2004).
\bibitem{Lekker} F. M. van der Kooij, K. Kassapidou, and H.N.W. Lekkerkerker, \textit{Nature}, \textbf{406}, 868  (2000).
\bibitem{crys1} S.Y. Liu, J. Zhang, N. Wang, W.R. Liu, C.G. Zhang, D.J. and Sun,  Chem. Mater., 15, 3240 (2003).
  \bibitem{crys2} A.B.D.  Brown, and A.R. Rennie,  \textit{Phys. Rev. E} \textbf{62}, 851 (2000).

  \bibitem{crys3} A.F.Mejia, Z. Cheng and M.S. Mannan, \textit{Phys. Rev. E} \textbf{85}, 061708 (2012).
  

\bibitem{nanosheets} Z. Liu, R. Ma, M. Osada, N. Iyi, Y. Ebina, K. Takada, and T. Sasaki    \textit{J. Am. Chem. Soc.} \textbf{128} (14), 4872 (2006)
\bibitem{nanosheets1} D. Yamaguchi, N. Miyamoto, T. Fujita, T. Nakato, S. Koizumi, N. Ohta, N. Yagi, and T. Hashimoto 
\textit{Phys. Rev. E} \textbf{85}, 011403 (2012).


\bibitem{nano1} N. Miyamoto, and T. J. Nakato,  \textit{Phys. Chem. B}, \textbf{108}, 6152  (2004).
\bibitem{nano2}  N. Miyamoto, N. and T. Nakato, \textit{Langmuir}, \textbf{19}, 8057 (2003).
\bibitem{nano3}  T. Nakato,  and   N. Miyamoto, \textit{J. Mater. Chem.}, \textbf{12}, 1245  (2002).
\bibitem{nano4}  J.C.P. Gabriel, F. Camerel, B.J. Lemaire, H. Desvaux, P. Davidson, P. Batail, \textit{Nature}, \textbf{413}, 504 (2001).




\bibitem{Scirep} S. Jabbari-Farouji, J.-J. Weis, P. Davidson, P. Levitz, and E. Trizac, \textit{Scientific Reports} \textbf{3}, 3559 (2013).

\bibitem{Trizac} R. Agra, E. Trizac, L. Bocquet, \textit{Eur. Phys. J.} \textbf{E }15, 345 (2004).
\bibitem{Carlos} C. \'{A}lvarez and G. T\'{e}llez, \textit{J. Chem. Phys} \textbf{133}, 144908 (2010).
\bibitem{Rowan00} D. G.  Rowan, J.-P. Hansen, E. Trizac, 
\textit{Mol. Phys.} {\bf 98}, 1369 (2000).

\bibitem{Dijkstra} A-P. Hynninen and M. Dijkstra, \textit{Phys. Rev. E.} \textbf{68}, 021407 (2003).
\bibitem{Wensink} L. Morales-Anda,  H. H. Wensink,  A.   Galindo, and  A.
Gil-Villegas,  \textit{J. Chem. Phys.} \textbf{132}, 034901 (2012).

\bibitem{TrBA02} E. Trizac,   L. Bocquet,   M. Aubouy, 
\textit{Phys. Rev. Lett.} {\bf 89}, 248301 (2002).


\bibitem{BoTA02}
L. Bocquet,   E. Trizac,   M. Aubouy, \textit{J. Chem. Phys.} {\bf 117}, 8138 (2002).




\bibitem{Allen} M. P. Allen and D. J. Tildesley, \textit{Computer Simulation of Liquids} (Oxford University Press, Oxford, 1987).
\bibitem{Frenkel} D. Frenkel and B. Smit, \textit{Understanding Molecular Simulation: From Algorithms to Applications} (Academic Press, 2001), 2nd ed.


\bibitem{salin:00} Salin and J.M. Caillol, \textit{J. Chem. Phys.}  {\bf 113}, 10459 (2000).
\bibitem{SimuAnneal}S. Kirkpatrick, C. D.  Gelatt and M. P. Vecchi, \textit{Science} \textbf{20} (4598) 671 (1983).

\bibitem{DMCsara} S. Jabbari-Farouji and E. Trizac, \textit{J. Chem. Phys.}  \textbf{137}, 054107 (2012).

\bibitem{commentJJ}
Another point should be added. The free energy calculation route for this model, by thermodynamic integration,
would face the difficulty that there are no
convenient reference states for which the free energies are accurately
known. 

\bibitem{vanRoij} 
R. van Roij, M. Dijkstra, and J. P. Hansen, 
\textit{Physical Review E} {\bf 59}, 2010-2025 (1999).

\bibitem{Belloni}
L. Belloni, 
\textit{J. Phys.: Condens. Matt.} {\bf 12}, R549 (2000).

\bibitem{Dobnikar}
J. Dobnikar,   R. Castaneda-Priego,   H.H. von Gr\"unberg,   E. Trizac,  
\textit{New Journal of Physics} {\bf 8}, 277 (2006).

\bibitem{Denton} 
A. R. Denton,
J. Phys.: Condens. Matter {\bf 22}, 364108 (2010).



\bibitem{Daan-disks} R. Eppenga and D. Frenkel, \textit{Molecular Physics} \textbf{52}, 1303 (1984).
\bibitem{Harnau} L. Harnau,  Mol. Phys. 106, 1975 (2008).

\bibitem{chaikin} H. M. Lindsay and P. M. Chaikin, \textit{J. Chem. Phys.} \textbf{76}, 3774 (1982).
\bibitem{Abou} B. Abou, D. Bonn, and J. Meunier, \textit{Phys. Rev. E} \textbf{64}, 021510 (2001).

\bibitem{LivingPolymer}P. van der Schoot, \textit{Supramolecular Polymers}, 2nd ed. (CRC, 2005), Chap.
2, 77-106.
\bibitem{orGlass} Z. Zheng, F. Wang, and Y. Han, \textit{Phys. Rev. Lett.} \textbf{107}, 065702
(2011).
\bibitem{schilling} M. Letz, R. Schilling, and A. Latz, \textit{Phys. Rev. E} \textbf{62}, 5173 (2000).
\bibitem{antinem}     K. Sokalski and   Th.W. Ruijgrok, \textit{Physica A} \textbf{126}, 280 (1984).
\bibitem{antinem1}  I. A. Georgiou, P. Ziherl,  and G. Kahl, EPL 106, 44004 (2014).


\bibitem{Marechal}M. Marechal,  A. Cuetos, B. Martinez-Haya, and M. Dijkstra, \textit{J. Chem. Phys.} \textbf{134}, 094501 (2011).
\bibitem{Cutos}  A. and B. Martinez-Haya, \textit{J. Chem. Phys.} \textbf{129}, 214706 (2008).
\bibitem{spheroids} G. Odriozola, \textit{J.Chem. Phys.} \textbf{36}, 134505 (2012).



\bibitem{Levitz}P. Levitz, E. Lecolier, A. Mourchid, A. Delville and S. Lyonnard, \textit{Europhys. Lett.} \textbf{49} (5), 672 (2000).

\bibitem{Smectite}D. Kleshchanok, P. Holmqvist, J.-M. Meijer and H. N. W. Lekkerkerker, \textit{J. Am. Chem. Soc.} \textbf{134}, 5985 (2012).
\bibitem{gibdynamics} D.  Kleshchanok, M. Heinen, G. N\"{a}gele, and P. Holmqvist,  \textit{Soft Matter} \textbf{8}, 1584 (2012).
\bibitem{baravian} C. Baravian, L. J. Michot, E. Paineau, I. Bihannic, P. Davidson, M. Imp\'{e}ror-Clerc, E. Belamie and P. Levitz, \textit{Europhys. Lett.} \textbf{90} 36005 (2010).


\bibitem{caillol:07} J. M. Caillol, F. Lo Verso, E. Sch\"oll-Paschinger, and
  J.-J. Weis, \textit{Mol.Phys. } {\bf 105}, 1813 (2007).

  \bibitem{Wigner}  D. Bonn, D., H. Tanaka, G.  Wegdam, H.  Kellay, and J. Meunier,  \textit{Europhys. Lett.} \textbf{45}, 52--57 (1998).
\bibitem{Zaccarelli} E. Zaccarelli, S. Andreev, F. Sciortino and D. R. Reichman, \textit{Phys. Rev. Lett.} \textbf{100}, 195701 (2008).
\bibitem{Ruzicka}
B. Ruzicka,  E. Zaccarelli, L. Zulian,  R.  Angelini,   M. Sztucki,   A. Moussaid,  T.  Narayanan, and  F. Sciortino,
\textit{Nature Materials} \textbf{10}, 56 (2011).

\bibitem{LiLi} Li Li, L. Harnau, S. Rosenfeldt, and M. Ballauff, Phys. Rev. E 72, 051504 (2005).



\end{thebibliography}
\end{document}